\documentclass[11pt,a4paper]{article}
\usepackage{jheppub,subfigure}

\title{\boldmath Ratios of $W$ and $Z$ cross sections at large boson $p_T$ as a constraint on PDFs and background to new physics}
\author[a,1]{Sarah Alam Malik%
\note{Now at High Energy Physics Group, Blackett Laboratory, Imperial College, London, SW7 2BW, UK}}
\author[b,2]{and Graeme Watt%
\note{Now at Institute for Particle Physics Phenomenology, Durham University, Durham, DH1 3LE, UK}}
\affiliation[a]{Laboratory of Experimental High Energy Physics, The Rockefeller University,\\ 1230 York Avenue, New York, NY 10065, USA}
\affiliation[b]{Institut f\"ur Theoretische Physik, Universit\"at Z\"urich,\\ Winterthurerstrasse 190, CH-8057 Z\"urich, Switzerland}
\emailAdd{smalik@cern.ch}
\emailAdd{gwatt@physik.uzh.ch}

\abstract{We motivate a measurement of various ratios of $W$ and $Z$ cross sections at the Large Hadron Collider (LHC) at large values of the boson transverse momentum ($p_T\gtrsim M_{W,Z}$).  We study the dependence of predictions for these cross-section ratios on the multiplicity of associated jets, the boson $p_T$ and the LHC centre-of-mass energy.  We present the flavour decomposition of the initial-state partons and an evaluation of the theoretical uncertainties.  We show that the $W^+/W^-$ ratio is sensitive to the up-quark to down-quark ratio of parton distribution functions (PDFs), while other theoretical uncertainties are negligible, meaning that a precise measurement of the $W^+/W^-$ ratio at large boson $p_T$ values could constrain the PDFs at larger momentum fractions $x$ than the usual inclusive $W$ charge asymmetry.  The $W^\pm/Z$ ratio is insensitive to PDFs and most other theoretical uncertainties, other than possibly electroweak corrections, and a precise measurement will therefore be useful in validating theoretical predictions needed in data-driven methods, such as using $W(\to\ell\nu)$+jets events to estimate the $Z(\to\nu\bar{\nu})$+jets background in searches for new physics at the LHC.  The differential $W$ and $Z$ cross sections themselves, ${\rm d}\sigma/{\rm d}p_T$, have the potential to constrain the gluon distribution, provided that theoretical uncertainties from higher-order QCD and electroweak corrections are brought under control, such as by inclusion of anticipated next-to-next-to-leading order QCD corrections.}


\arxivnumber{1304.2424}

\begin{document}

\begin{flushright}
  FERMILAB-PUB-13-106-PPD \\
  ZU-TH 07/13 \\
  28th November 2013
\end{flushright}

\maketitle

\section{Introduction}
The ATLAS and CMS experiments at the Large Hadron Collider (LHC) have now each collected more than 20~fb$^{-1}$ of data at a centre-of-mass energy of 8~TeV.  Searches for new physics beyond the Standard Model (SM) have all returned results that are consistent with the SM and have ruled out a large parameter space of new physics scenarios. While the LHC has now suspended its operation to prepare for the upgrade to 13~TeV, the data from the 8~TeV run is still being analysed.  Indeed, there is a lot of interesting physics that can be done with this data, from testing novel analysis techniques and making precision SM measurements to tuning and improving the Monte Carlo simulations in readiness for the 13~TeV run.  One of the priorities during the LHC shutdown will be to improve our understanding of the SM processes that are backgrounds to new physics searches and hence our sensitivity to new physics.  A key ingredient to making theoretical predictions at hadron colliders is the parton distribution functions (PDFs) of the proton (see ref.~\cite{Forte:2013wc} for a recent review), and knowledge of the PDFs can be improved using LHC data.

At the 7~TeV LHC, measurements have been made of $W/Z$ inclusive (or differential in rapidity) cross sections~\cite{Aad:2011dm,CMS:2011aa,Chatrchyan:2011wt,Aaij:2012vn,Aaij:2012mda}, the inclusive $W$ charge asymmetry~\cite{Aad:2011dm,Chatrchyan:2011jz,Chatrchyan:2012xt,Aaij:2012vn}, the $W/Z$ transverse momentum ($p_T$) distributions~\cite{Aad:2011fp,Aad:2011gj,Chatrchyan:2011wt} and the $W/Z$+jets process~\cite{Aad:2010ab,Chatrchyan:2011ne,Aad:2011qv,Aad:2012en,Aad:2013ysa,Aaij:2013nxa}.  The ratio of the $W$ and $Z$ cross sections with exactly one jet has been measured as a function of the jet transverse momentum threshold~\cite{Aad:2011xn}.  Some preliminary measurements have also been made at the 8~TeV LHC.  However, no measurement has been made so far of the ratio of $W$ and $Z$ (or $W^+$ and $W^-$) cross sections as a function of the boson $p_T$.  The main goal of this paper is to motivate such a measurement as a constraint on the $Z(\to\nu\bar{\nu})$+jets background to new physics searches and also on the PDFs of the proton.

These two applications require somewhat different theoretical tools, and hence the optimal measurement enabling a constraint will also be slightly different.  New physics searches, typically in the region of high transverse momenta, $p_T\sim \mathcal{O}$(0.1--10~TeV), require calculations of $W/Z$ production in association with multiple jets, usually obtained by merging predictions for different hard-parton multiplicities after matching to a parton shower.  Here, there are uncertainties from merging/matching prescriptions and tunable parameters associated with the parton shower, in addition to the usual theoretical uncertainties arising from fixed-order QCD calculations.  Precise measurements can therefore be used to validate these calculations.  For the purposes of constraining the proton PDFs, these additional uncertainties can be avoided by considering only the inclusive boson $p_T$ distributions without explicitly demanding jets.  However, aiming to describe the whole $p_T$ range would introduce undesirable extra uncertainties from the need to include $p_T$-resummation (in the low-$p_T$ region of $p_T\ll M_{W,Z}$) and from matching to the fixed-order calculation (in the intermediate-$p_T$ region of $p_T\lesssim M_{W,Z}$).  It is therefore preferable to restrict only to the high-$p_T$ region of $p_T\gtrsim M_{W,Z}$ where the fixed-order calculations should be reliable without invoking $p_T$-resummation.  Consequently, we will focus on only this high-$p_T$ region.

The content of this paper is as follows.  In section~\ref{sec:Znunu} we review different methods for estimating the $Z(\to\nu\bar{\nu})$+jets background in new physics searches at the LHC, and we explain the utility of a precise measurement of the $W/Z$ ratio at large boson $p_T$.  In section~\ref{sec:flavour} we explore how the $W$ and $Z$ cross sections as a function of boson $p_T$ depend on the flavour of the initial-state partons and how the cross-section ratios depend on the multiplicity of associated jets.  In section~\ref{sec:theory} we carry out a detailed evaluation of theoretical uncertainties, on both the differential cross sections (${\rm d}\sigma/{\rm d}p_T$) and the cross-section ratios, arising from higher-order QCD and electroweak corrections and the choice of PDF set.  In section~\ref{sec:CoM} we compare predictions for the cross-section ratios at two LHC centre-of-mass energies (8~TeV and 13~TeV).  Finally, we conclude in section~\ref{sec:conclusions}.

\section{Estimating the $Z(\to\nu\bar{\nu})$+jets background to new physics} \label{sec:Znunu}

The production of a $Z$ boson in association with jets and its subsequent decay to neutrinos constitutes a major irreducible background in searches for new physics that involve missing transverse energy.  Searches for Supersymmetry (SUSY) where the lightest SUSY particle is neutral and weakly interacting, Large Extra Dimensions where the graviton escapes into the extra dimensions, and the direct production of dark matter candidates such as Weakly Interacting Massive Particles, all give rise to the missing transverse energy signature.  In some searches, the $Z(\to\nu\bar{\nu})$+jets process can make up 70\% or more of the total SM background~\cite{CMS-PAS-EXO-12-048,ATLAS:2012ky,ATLAS-CONF-2012-109,Chatrchyan:2012jx,Chatrchyan:2012lia}.  To reduce uncertainties from higher-order corrections and Monte Carlo modelling, the backgrounds are estimated using techniques that rely on data.

Three data-driven methods~\cite{CMS-PAS-SUS-08-002} have been used to estimate the $Z(\to\nu\bar{\nu})$ background, all of which exploit the similarities in the kinematic characteristics between $Z(\to\nu\bar{\nu})$+jets and $V$+jets events, where $V = Z(\to\ell\ell)$, $\gamma$, or $W(\to\ell\nu)$.  The presence of new physics will contribute to each of these three channels differently and it is therefore important to have a cross-check of the background prediction.  This also enables the predictions from various methods to be combined to achieve greater precision.  Searches for new physics that use at least one of these methods to estimate the background can be found in refs.~\cite{CMS-PAS-EXO-12-048,CMS-PAS-SUS-12-024,Chatrchyan:2012wa,Chatrchyan:2012jx,Chatrchyan:2012lia,Chatrchyan:2011ida,ATLAS-CONF-2013-024,ATLAS-CONF-2012-165,ATLAS:2012ky,ATLAS-CONF-2012-103,Aad:2012fqa,ATLAS-CONF-2012-109}.  The three methods are summarised below:
\begin{enumerate}
\item $Z(\to\ell\ell)$+jets.  The fully reconstructable decay of a $Z$ boson to dileptons is a `standard candle' process for many analyses.  It is conceptually the simplest method used to derive the $Z(\to\nu\bar{\nu})$+jets background.  The only theoretical input is the ratio of branching fractions for $(Z\to\ell\ell)/(Z\to\nu\bar{\nu})$, which is very well known, to within 0.3$\%$~\cite{Beringer:1900zz}.  However, the method has a large statistical uncertainty owing to limited $Z(\to\ell\ell)$+jets statistics, in particular in the regions of phase space in which searches for new physics are conducted.
\item $\gamma$+jets.  The $\gamma$+jets channel has a significantly higher production rate than $Z(\to\ell\ell)$+jets, but it pays a price for gauge boson substitution and relies on the theoretical prediction of the $\gamma/Z$ cross section.  There has been considerable work on estimating and reducing the QCD uncertainty on this theoretical ratio, which currently stands at less than 10\%~\cite{Bern:2011pa,Ask:2011xf,Bern:2012vx}.
\item $W(\to\ell\nu)$+jets.  This channel is again statistically more powerful than $Z(\to\ell\ell)$+jets but with a non-negligible contribution from background processes such as $t\bar{t}$.  It also incurs an additional systematic uncertainty from the substitution of a $Z$ boson with a $W$ boson, which enters in the ratio of the $W/Z$ cross sections in the regions of high transverse momentum that are typical of searches.
\end{enumerate}

In this paper, we concentrate on the last method. The $W/Z$ ratio is a major theoretical input and contributes as one of the largest systematic uncertainties on the determination of the $Z(\to\nu\bar{\nu})$+jets background from $W(\to\ell\nu)$+jet events.  This is shown in ref.~\cite{ATLAS:2012ky} where it is the dominant source of uncertainty on the total background prediction in two of the four search regions.  It is assumed in ref.~\cite{ATLAS:2012ky} that the ratio of the $Z+$jets and $W+$jets cross sections is well modelled in the simulation and this is to an extent supported by the measurement of the $W/Z$ ratio as a function of the jet transverse momentum threshold~\cite{Aad:2011xn}.  The uncertainty on the ratio is evaluated by comparing the background prediction using $Z/W$ distributions from the generators \textsc{alpgen}~\cite{Mangano:2002ea} and \textsc{sherpa}~\cite{Gleisberg:2008ta}.  The detailed study of the theoretical uncertainties on this ratio presented in this paper will already be useful input to the searches for new physics, to be supplemented by a future measurement of the ratio by the LHC experiments.

In addition to $Z(\to\nu\bar{\nu})$+jets events, $W+$jet events in which the $W$ decays leptonically and the charged lepton is not reconstructed, thus mimicking missing transverse energy, are also backgrounds to searches for new physics.  In many searches, this background is estimated together with other processes in which a lepton is not reconstructed, such as $t\bar{t}$, by using a data control sample of single lepton+jet events~\cite{Chatrchyan:2012wa,Chatrchyan:2012jx,ATLAS-CONF-2012-103}.  However, in the ATLAS and CMS monojet analyses~\cite{CMS-PAS-EXO-12-048,ATLAS:2012ky}, the $W\to \mu\nu$ (and $W\to e\nu$) control sample is used to estimate only the background from $W$+jet events where the lepton is not reconstructed.  A complementary method to estimate this important background, which accounts for roughly 30--50\% of monojet events, could be to use a control sample of $Z\to\ell\ell$ events, follow a similar procedure as in refs.~\cite{CMS-PAS-EXO-12-048,ATLAS:2012ky}, and then correct for the difference in the $W$ and $Z$ cross sections using the $W/Z$ ratio.  Hence, it adds a further motivation to measure this ratio.

Searches for Supersymmetry typically define search regions using event variables such as the $\not\!\! H_{\mathrm{T}}$, which is a vector sum of the jets above a certain $p_T$ threshold~\cite{Chatrchyan:2011ida}. The boson $p_T$ is numerically very close to the $\not\!\! H_{\mathrm{T}}$, thus making it a good choice of variable that represents well the overall kinematics of the event.  Hence, a study of the $W/Z$ ratio as a function of the boson $p_T$ should be applicable to a wide range of new physics searches.  The boson $p_T$ is also a good choice of variable for measurements intended to constrain PDFs, due to its close correspondence to the kinematics of the initial-state partons.

\section{Flavour decomposition and dependence of ratios on jet multiplicity} \label{sec:flavour}

In this section we use the leading-order (LO) Monte Carlo event generator \textsc{madgraph}~5~\cite{Alwall:2011uj} to study the flavour decomposition of $W^\pm$ and $Z^0$ production and the dependence of the ratios on the number of jets, for the LHC at a centre-of-mass-energy of $\sqrt{s} = 8$~TeV.  Samples of $W^\pm(\to\ell^\pm\nu) + n$~hard-partons and $Z^0(\to\ell^+\ell^-) + n$~hard-partons are produced using \textsc{madgraph} for each of $n\in\{0,1,2,3,4\}$, with $p_T^{\rm parton}>10$~GeV and $|\eta^{\rm parton}|<5$, matched to \textsc{pythia}~6.42~\cite{Sjostrand:2006za} (tune Z2*~\cite{Chatrchyan:2011id}) using the \textsc{mlm}~\cite{Alwall:2007fs} shower matching prescription with a matching threshold of 10~GeV.  The $Z^0$ process includes the effect of a virtual photon ($\gamma^*$), therefore we restrict the invariant mass of the produced lepton-pair ($\ell^+\ell^-$) to the region of the $Z^0$ mass, $M_{\ell^+\ell^-}\in[60,120]$~GeV.  The CTEQ6L1~\cite{Pumplin:2002vw} PDFs are used and the renormalisation and factorisation scales are set to $\mu_R = \mu_F = \sqrt{M_V^2 +\sum_{\rm partons} (p_{T}^{\rm parton})^2}$, where $M_V$ is the mass of the vector boson.  We then define inclusive $N$-jet multiplicity subsamples, for $N\in\{0,1,2,3,4\}$, each with at least $N$ jets selected according to the anti-$k_T$ algorithm~\cite{Cacciari:2008gp} with a distance parameter of $R=0.5$, $p_T^{\rm jet}>10$~GeV and $|\eta^{\rm jet}|<5$.  The inclusive $N=0$ samples therefore contain exact tree-level $\mathcal{O}(\alpha_s^n)$ contributions from each of $n\in\{0,1,2,3,4\}$ hard-partons, together with additional radiation from the parton shower.  Note that jets can originate either from the \textsc{madgraph} hard-partons or from the \textsc{pythia} parton shower, and we cannot distinguish the two sources of origin.

\subsection{Flavour decomposition of boson $p_T$ distributions}

\begin{figure}
  \centering
  \subfigure[$W^+$]{\includegraphics[width=0.5\textwidth]{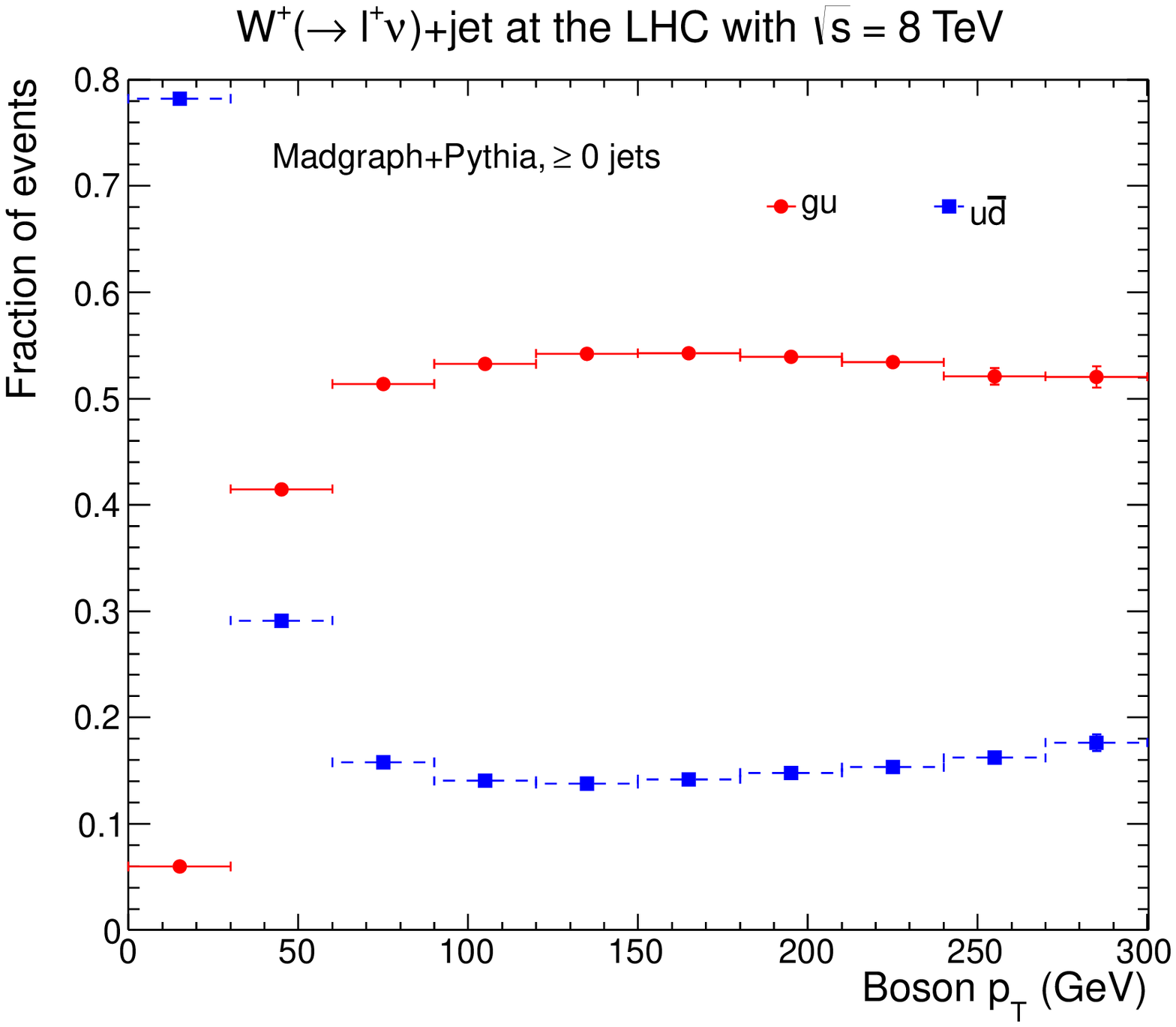}}%
  \subfigure[$W^-$]{\includegraphics[width=0.5\textwidth]{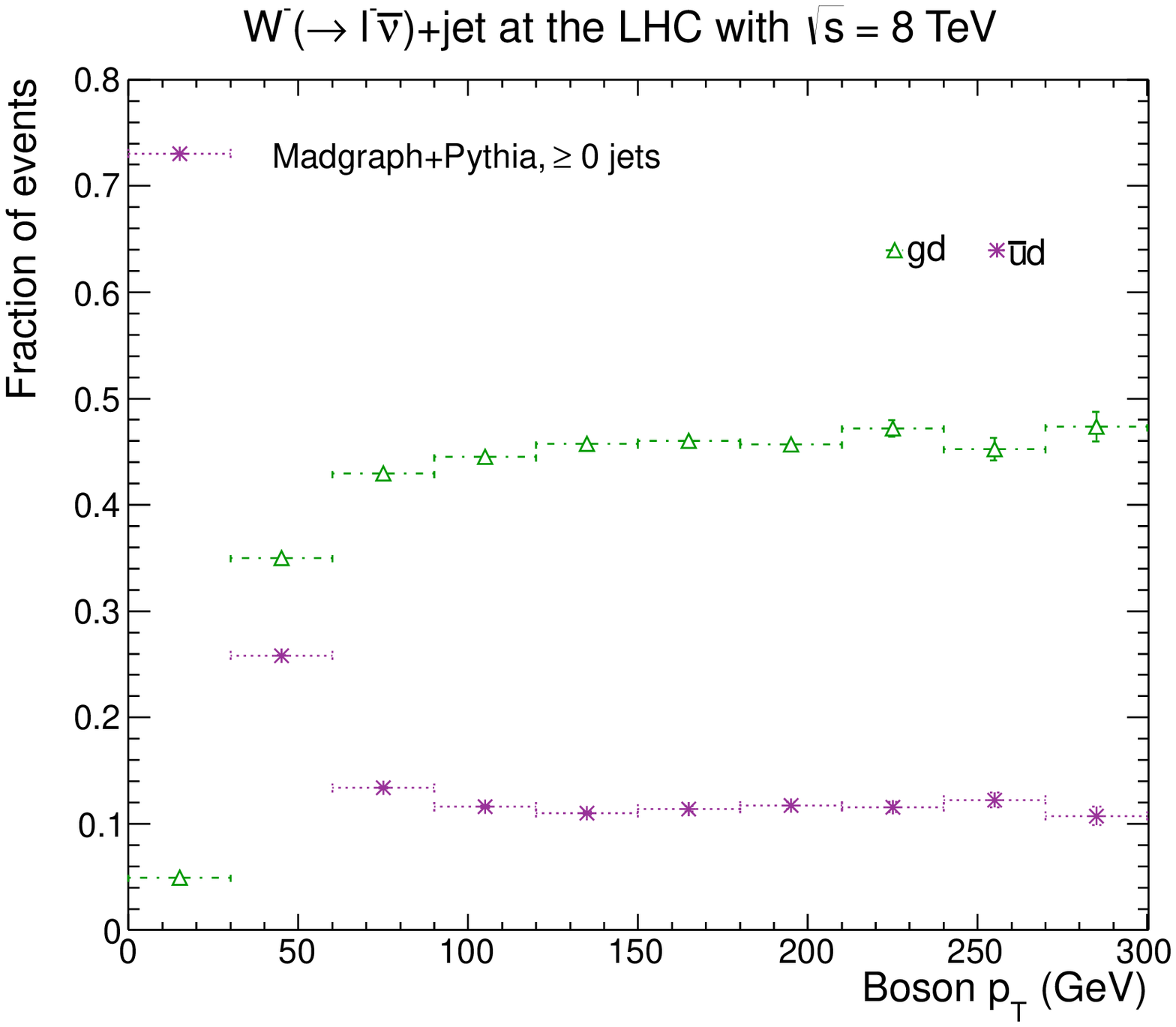}}
  \subfigure[$W^\pm$]{\includegraphics[width=0.5\textwidth]{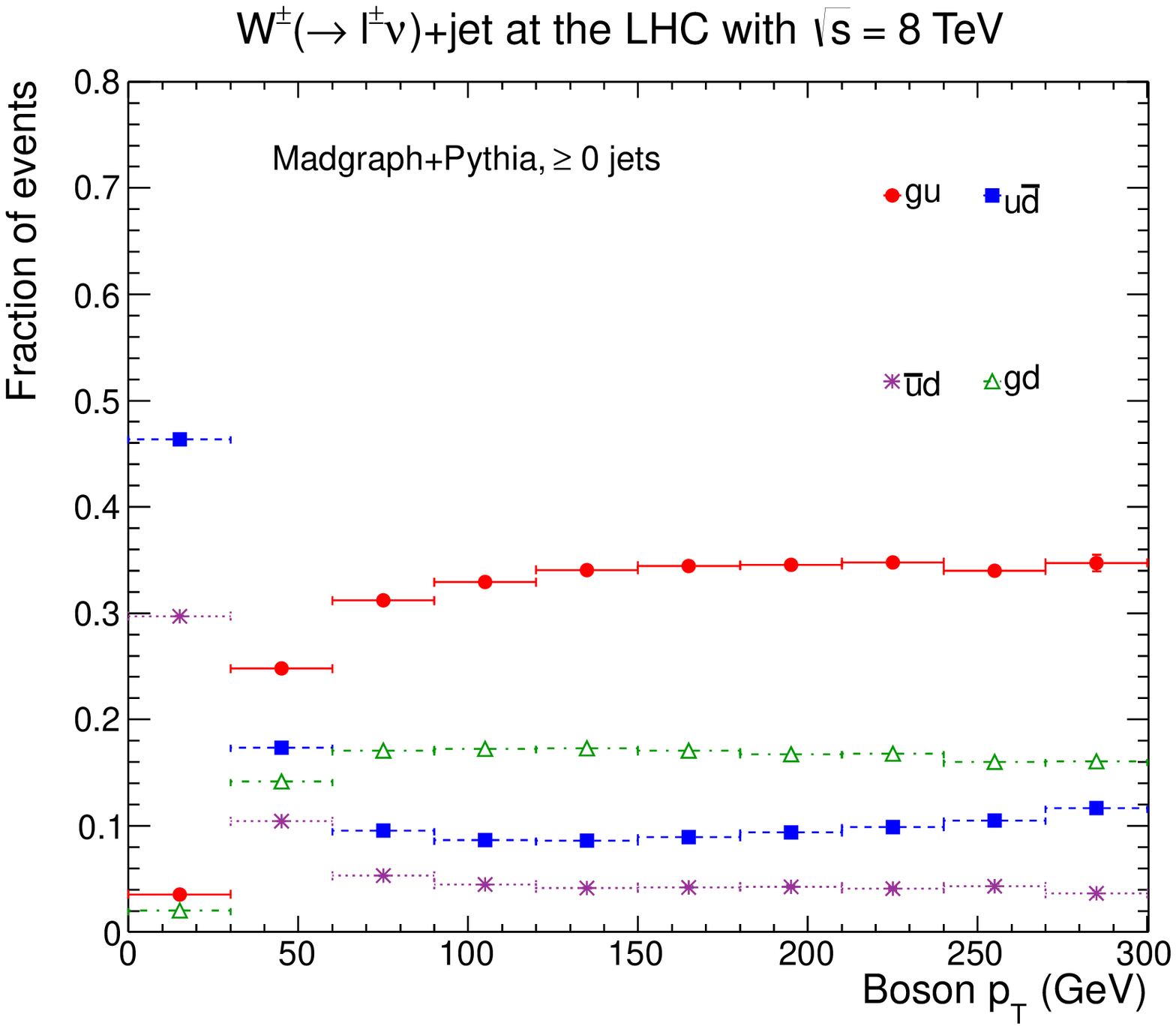}}%
  \subfigure[$Z^0$]{\includegraphics[width=0.5\textwidth]{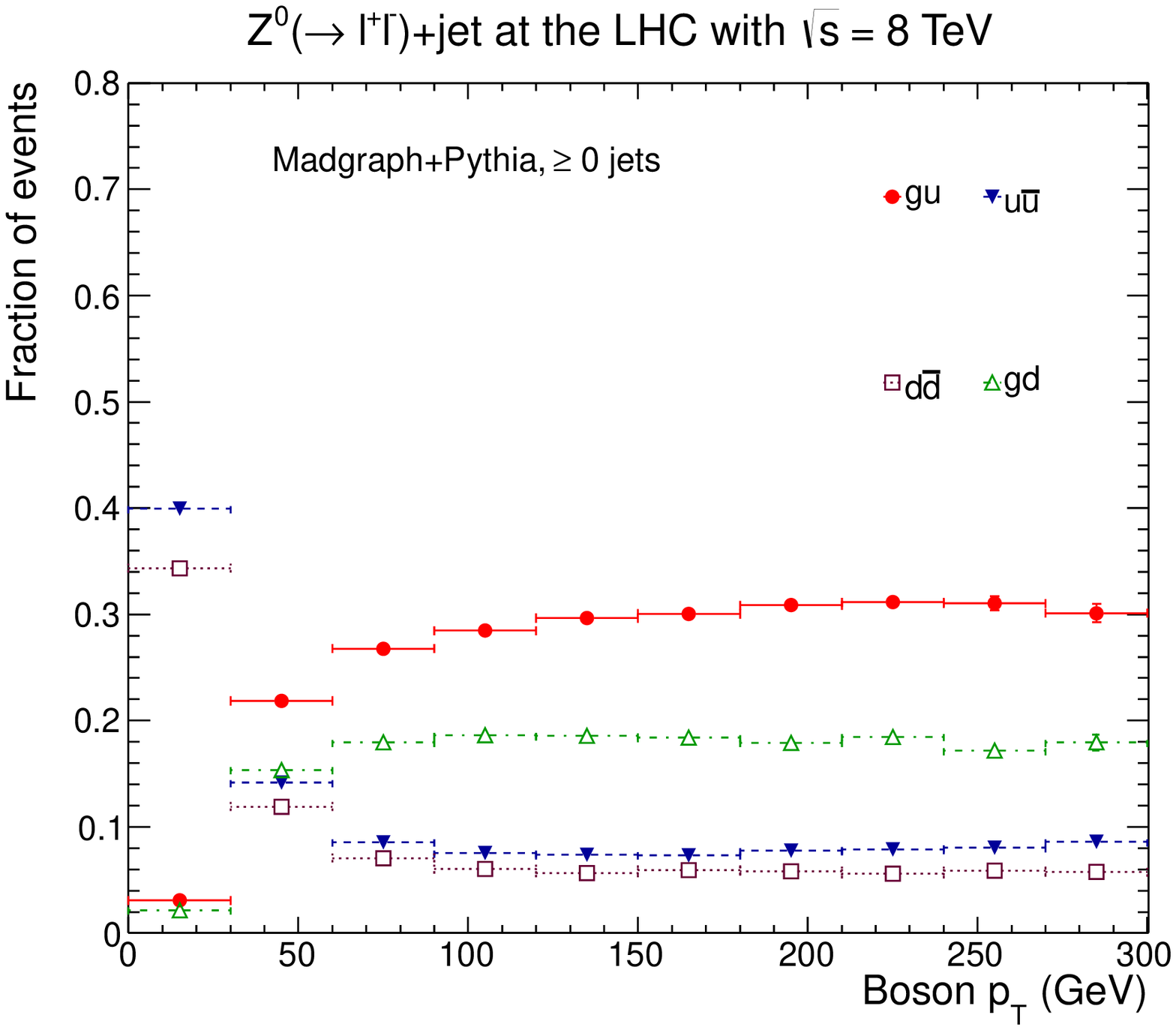}}
  \caption{Decomposition of the dominant initial-state partons contributing to (a)~$W^+$, (b)~$W^-$, (c)~$W^\pm$ and (d)~$Z^0$ production, as a function of the boson $p_T$, as predicted by \textsc{madgraph}+\textsc{pythia}.}
  \label{fig:flavour_WZ}
\end{figure}
In figure~\ref{fig:flavour_WZ} we show the decomposition of the initial-state partons contributing to $W^+$, $W^-$, $W^\pm$ ($\equiv W^++W^-$) and $Z^0$ production as a function of the boson $p_T$, for the inclusive samples with $\ge0$ jets.  For clarity we only show the largest partonic contributions.  The very low $p_T$ region is dominated by the initial state $u\bar{d}$ for $W^+$, $\bar{u}d$ for $W^-$ and $u\bar{u}$ for $Z^0$ production.  As the boson acquires transverse momentum, the dominant initial state becomes $gu$ for $W^+$ and $gd$ for $W^-$, contributing to roughly 50\% of the total $W$ production, and these remain dominant for the entire $p_T$ region studied.  Note from figure~\ref{fig:flavour_WZ}(c,d) that the sum $W^\pm\equiv W^++W^-$ at large $p_T$ has a very similar flavour decomposition, dominated by $gu$ and $gd$ configurations, as $Z^0$.  The precise details of figure~\ref{fig:flavour_WZ} beyond these general features may depend on the nature of the calculation, such as, for example, the inclusion of higher-order corrections and the choice of factorisation scheme/scale.  However, it can be seen in figures 4 and 5 of ref.~\cite{Brandt:2013hoa} that the dominance of the $qg$ channel over the $q\bar{q}$ channel at large $p_T$ is also found in a conventional next-to-leading order (NLO) calculation in the $\overline{\rm MS}$ factorisation scheme with a factorisation scale $\mu_F = \sqrt{M_V^2+p_T^2}$.  We have produced the corresponding plots as in figure~\ref{fig:flavour_WZ} for the other inclusive $N$-jet multiplicities, $N\in\{1,2,3,4\}$, and we find that the main components of the flavour decomposition are very similar for different jet multiplicities.  This observation demonstrates insensitivity to higher-order corrections, given that, at large boson $p_T$, the inclusive $N$-jet multiplicity ($N\ge2$) may be considered as a real $\mathcal{O}(\alpha_S^{N-1})$ correction to the inclusive $1$-jet multiplicity.  In particular, the $gg$ initial state, which first appears only with $n\ge2$ hard-partons, is found to contribute at below the 5\% level even for the higher jet multiplicities.

\subsection{Dependence of ratios on jet multiplicity}

\begin{figure}
  \centering
  \subfigure[$W^+/W^-$]{\includegraphics[width=0.5\textwidth]{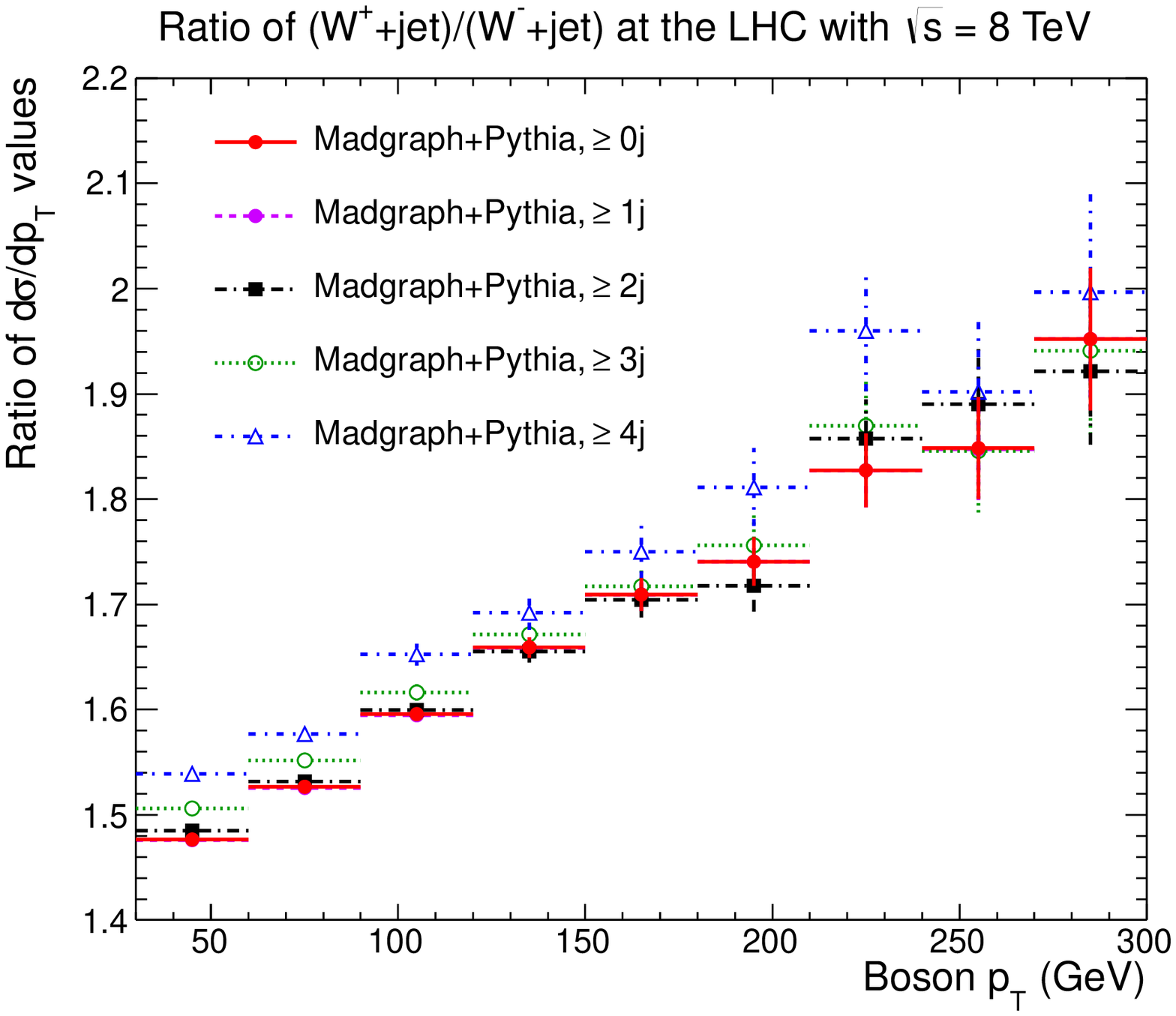}}%
  \subfigure[$W^+/Z^0$]{\includegraphics[width=0.5\textwidth]{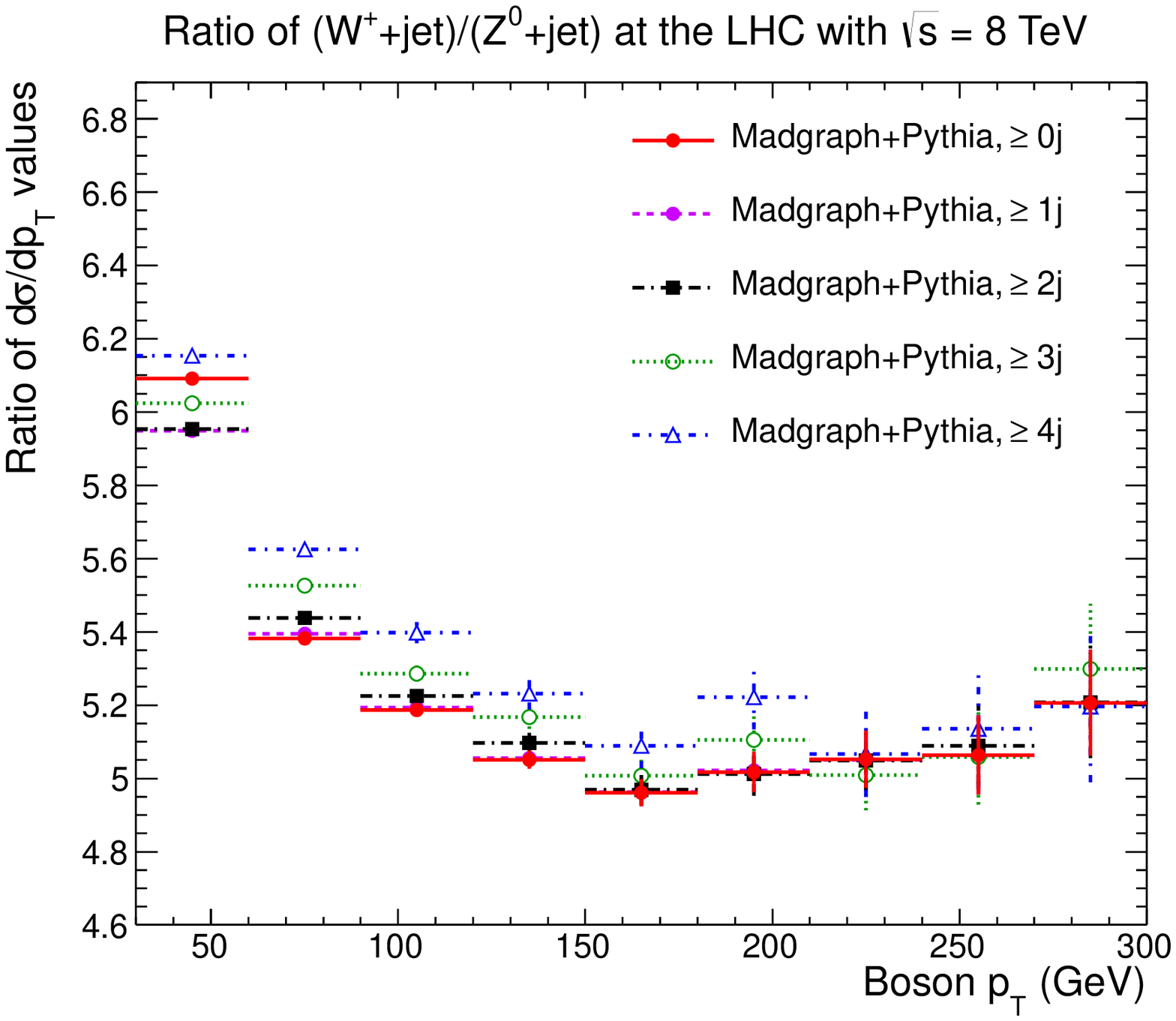}}
  \subfigure[$W^-/Z^0$]{\includegraphics[width=0.5\textwidth]{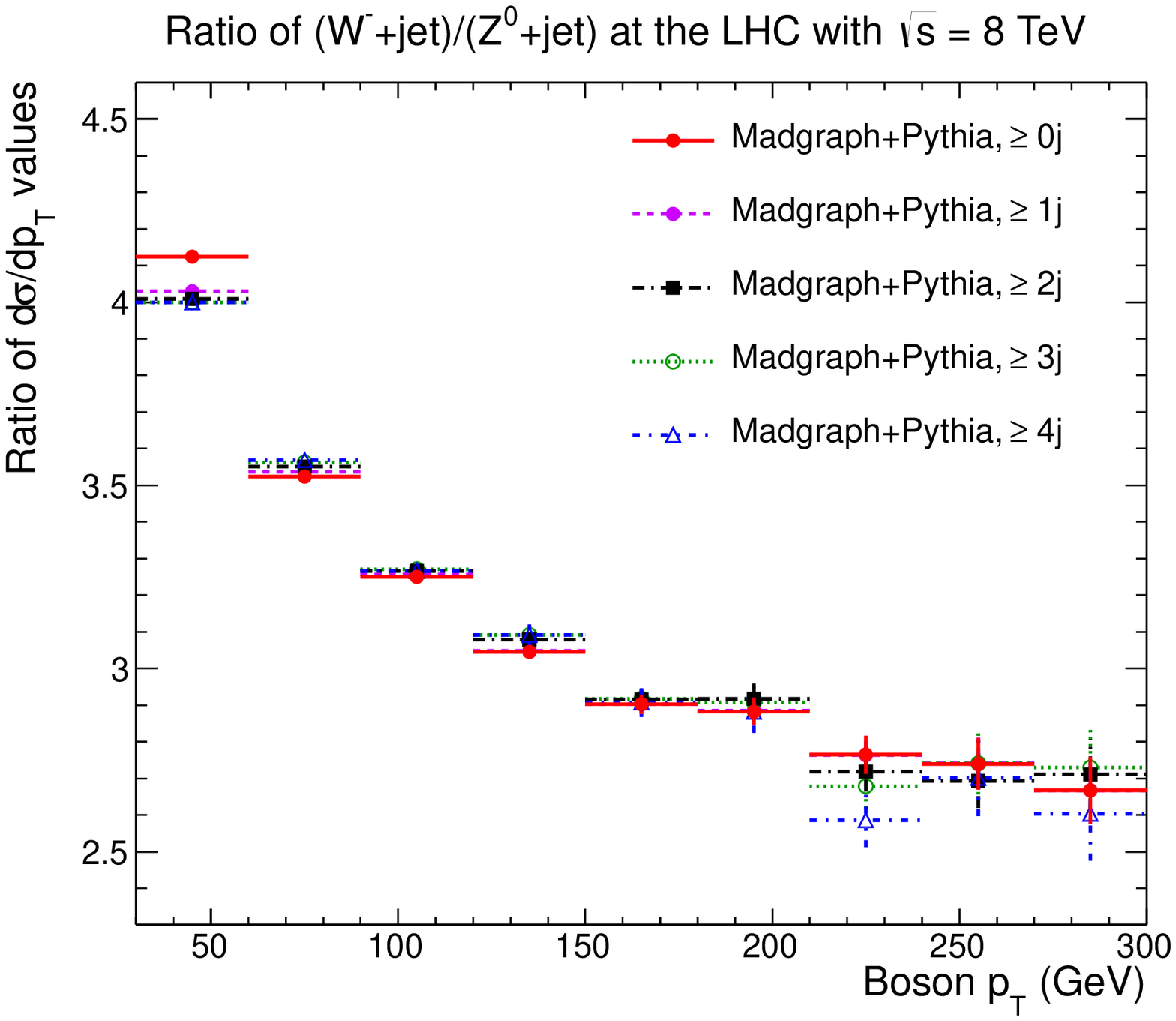}}%
  \subfigure[$W^\pm/Z^0$]{\includegraphics[width=0.5\textwidth]{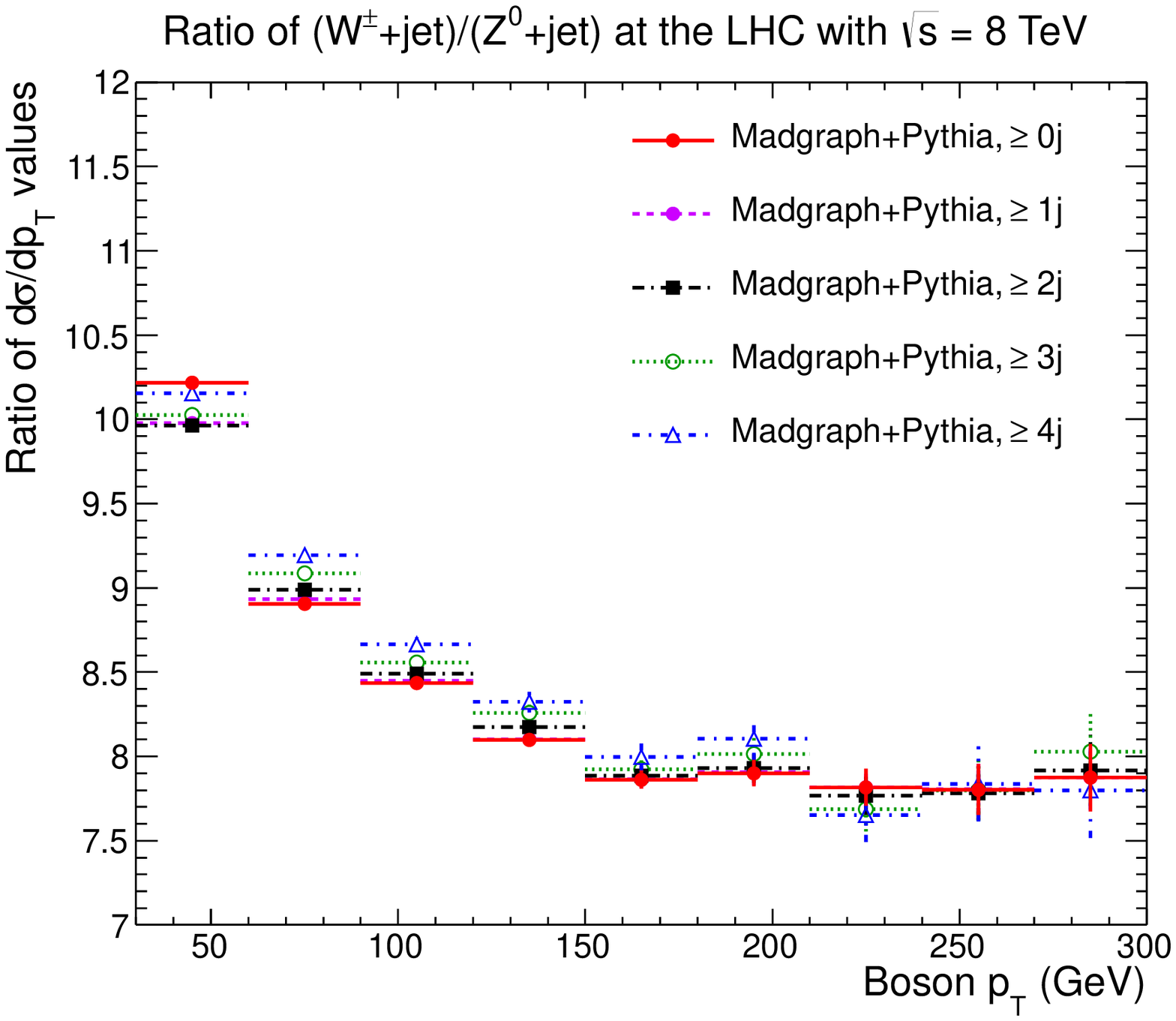}}
  \caption{Ratios of (a)~$W^+/W^-$, (b)~$W^+/Z^0$, (c)~$W^-/Z^0$ and (d)~$W^\pm/Z^0$ for various inclusive jet multiplicities: $\ge0$, $\ge1$, $\ge2$, $\ge3$ and $\ge4$, as predicted by \textsc{madgraph}+\textsc{pythia}.}
  \label{fig:jetmult}
\end{figure}
In figure~\ref{fig:jetmult} we show the ratios of $W^+/W^-$, $W^+/Z^0$, $W^-/Z^0$ and $W^\pm/Z^0$ for the following inclusive jet multiplicities: $\ge0$, $\ge1$, $\ge2$, $\ge3$ and $\ge4$, as predicted by \textsc{madgraph} matched to \textsc{pythia} as described above.  We omit the region of boson $p_T<30$~GeV from the plots, where the result is most influenced by the parton shower and depends on the $p_T^{\rm jet}>10$~GeV selection.  The dependence of the ratios on boson $p_T$ is not strongly dependent on the jet multiplicity and, in particular, the differences between the $\ge0$~jet and $\ge1$~jet samples are very small.  Each of the four ratios has an interesting dependence on boson $p_T$, that can be understood by considering the dominant initial-state parton configurations, namely $gu$ for $W^+$ or $Z^0$ and $gd$ for $W^-$.  Then the $W^+/W^-$ ratio reflects the $u/d$ ratio of PDFs, which increases going to larger boson $p_T$ as higher values of the momentum fraction $x$ are being probed.  Conversely, the $W^-/Z^0$ ratio reflects the $d/u$ ratio, and so decreases with increasing boson $p_T$.  These two ratios ($W^+/W^-$ and $W^-/Z^0$) change by around $30\%$ in going from $p_T\sim 50$~GeV to $p_T\sim 300$~GeV, whereas the $W^+/Z^0$ ratio only changes by around 10\% and $W^\pm/Z^0$ changes by around 20\%.  The behaviour  of the various $W/Z$ ratios at smaller $p_T$ values is driven by kinematic differences between $W$ and $Z$ production due to the different boson masses, $M_W$ and $M_Z$.  These kinematic effects are most important for boson $p_T\lesssim M_{W,Z}$, but then the $\sim$10\% difference between $M_W$ and $M_Z$ becomes irrelevant for $p_T\gg M_{W,Z}$.  Taking appropriate limits to find the dominant behaviour of a simple calculation for ${\rm d}\sigma/{\rm d}p_T$ given in eq.~(28) of ref.~\cite{Watt:2003vf} gives a factor in the $W/Z$ ratios of $(M_Z^2+p_T^2)/(M_W^2+p_T^2)$, which numerically takes a value of 1.2 at $p_T=50$~GeV, 1.1 at $p_T=100$~GeV, and then rapidly approaches 1 for larger $p_T$, in qualitative agreement with the decrease with increasing $p_T$ of the $W/Z$ ratios at smaller $p_T$ values observed in figure~\ref{fig:jetmult}.  The limiting behaviour of the various $W/Z$ ratios at very large $p_T\gg M_{W,Z}$ is then driven by the PDF dependence, to be discussed further in section~\ref{sec:bosonratioPDF}.  The slight increase of the $W^+/W^-$ ratio at fixed $p_T$ with increasing jet multiplicity can be understood by the fact that the typical partonic invariant masses (and hence the $x$ values) increase with the number of jets, and the $u/d$ ratio increases with increasing $x$ values.

As for the flavour decomposition in figure~\ref{fig:flavour_WZ}, the remarkable insensitivity of the ratios in figure~\ref{fig:jetmult} to different jet multiplicities also demonstrates an insensitivity of the ratios to higher-order QCD corrections, as we will examine in more detail in the next section in the context of a fixed-order calculation.  Our study of the theoretical uncertainties below is carried out using \textsc{mcfm}~\cite{Campbell:2010ff} with $V+\ge1$~jet but it is equally applicable to the inclusive $\ge0$ jet sample.  In fact, we encourage the experimental measurement to be carried out in the inclusive channel, where greater precision should be achievable without demanding the presence of associated jets.

\section{Theoretical uncertainties in $V$+jet production} \label{sec:theory}

Inclusive vector boson production including leptonic decay has been calculated at next-to-next-to-leading order (NNLO) in perturbative QCD, that is, $\mathcal{O}(\alpha_S^2)$~\cite{Melnikov:2006kv,Catani:2009sm}.  However, requiring large boson $p_T$ with either $\ge 0$ or $\ge 1$ jets means that at least one hard-parton must be emitted, and so the lowest non-vanishing perturbative order is $\mathcal{O}(\alpha_S)$.  The LO calculation for the boson $p_T$ distribution at large $p_T$ is therefore a $2\to2$ process and the NLO calculation is $\mathcal{O}(\alpha_S^2)$~\cite{Ellis:1981hk,Arnold:1988dp,Gonsalves:1989ar}.  Note that at LO for the $V$+jet process, the boson $p_T = p_T^{\rm jet}$, and so we expect that the $p_T$ distribution at large boson $p_T$ for the $V$+jet process should be very similar to the result for inclusive $V$ production (that is, without explicitly demanding a jet).  Therefore, our findings presented below apply equally to the $p_T$ distributions at large boson $p_T$ for inclusive $V$ production.  We have checked explicitly that varying the minimum $p_T^{\rm jet}$ cut only affects NLO predictions for the boson $p_T$ distributions if the boson $p_T$ is less than or very close to the minimum $p_T^{\rm jet}$ cut.

We use the \textsc{mcfm} (v6.4) code~\cite{Campbell:2010ff} for the $V$+jet process, where $V\in\{W^+,W^-,Z^0\}$, including the appropriate leptonic decay of the vector boson: $W^+\to\ell^+\nu$, $W^-\to\ell^-\bar{\nu}$ or $Z^0\to\ell^+\ell^-$.  Jets are defined according to the anti-$k_T$ algorithm~\cite{Cacciari:2008gp} with a distance parameter of $R=0.5$, $p_T^{\rm jet}>10$~GeV and $|\eta^{\rm jet}|<5$.  As stated before for \textsc{madgraph}, the $Z^0$ process in \textsc{mcfm} includes the effect of a virtual photon ($\gamma^*$), therefore we restrict the invariant mass of the produced lepton-pair ($\ell^+\ell^-$) to the region of the $Z^0$ mass, $M_{\ell^+\ell^-}\in[60,120]$~GeV.  We make a dynamic choice for the central renormalisation and factorisation scales, $\mu_R=\mu_F=\mu_0\equiv \sqrt{M^2+p_T^2}$, where $M\in\{M_{\ell^+\nu},M_{\ell^-\bar{\nu}},M_{\ell^+\ell^-}\}$ and $p_T$ is the boson transverse momentum.  To smooth statistical fluctuations, results are averaged over a large number ($\sim 100$) of independent \textsc{mcfm} runs, each with different seeds for the \textsc{vegas} integration.  We do not investigate theoretical uncertainties due to more realistic event simulation, such as inclusion of the underlying event (multiple interactions) or hadronisation of the parton-level jets, but these effects should not be important at large boson $p_T$ values, and they should largely cancel in cross-section ratios.

\subsection{Higher-order QCD corrections}

\subsubsection{Boson $p_T$ distributions} \label{sec:bosonPTtheory}

\begin{figure}
  \centering
  \subfigure[$W^+$]{\includegraphics[width=0.5\textwidth]{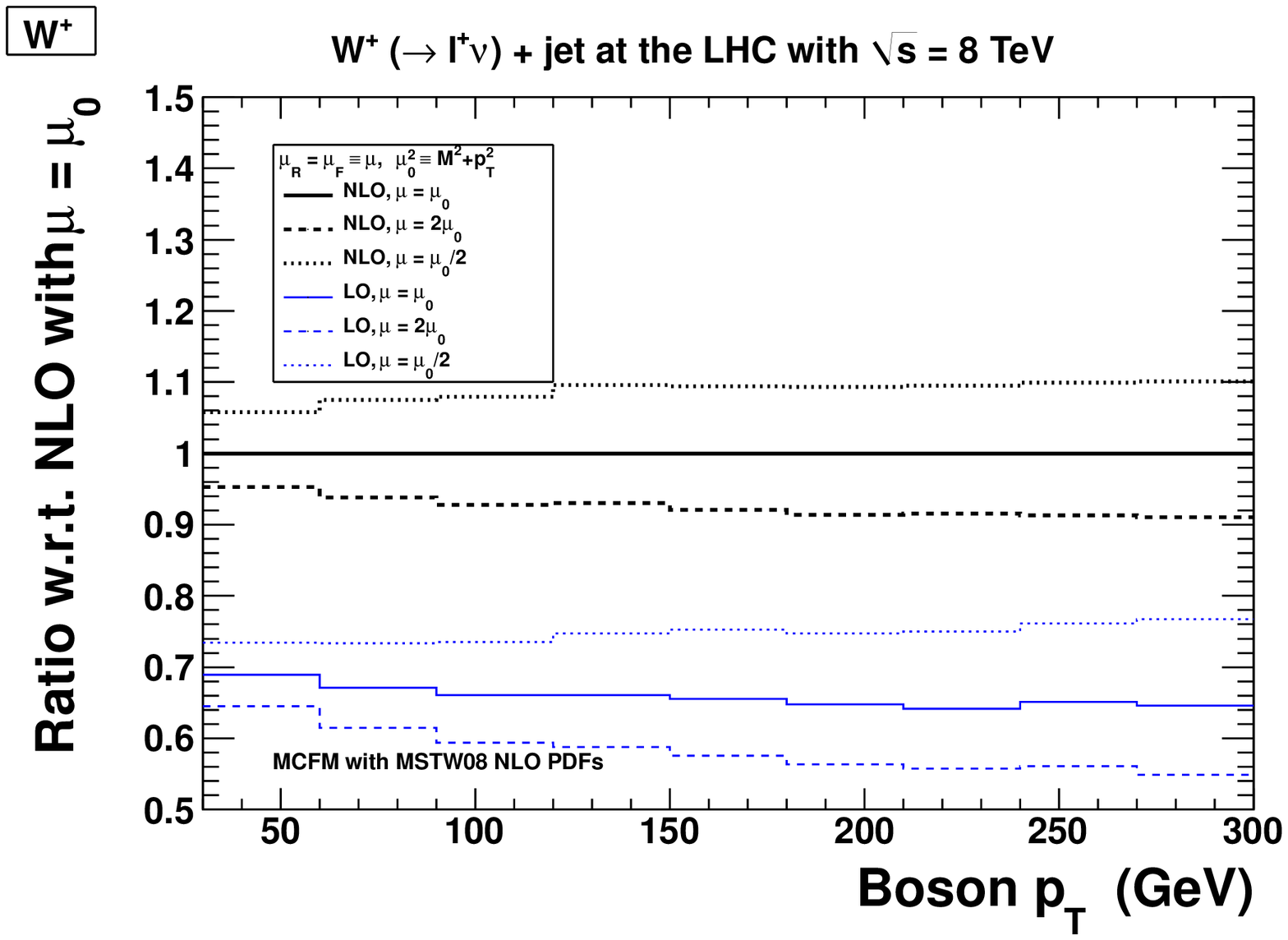}}%
  \subfigure[$W^-$]{\includegraphics[width=0.5\textwidth]{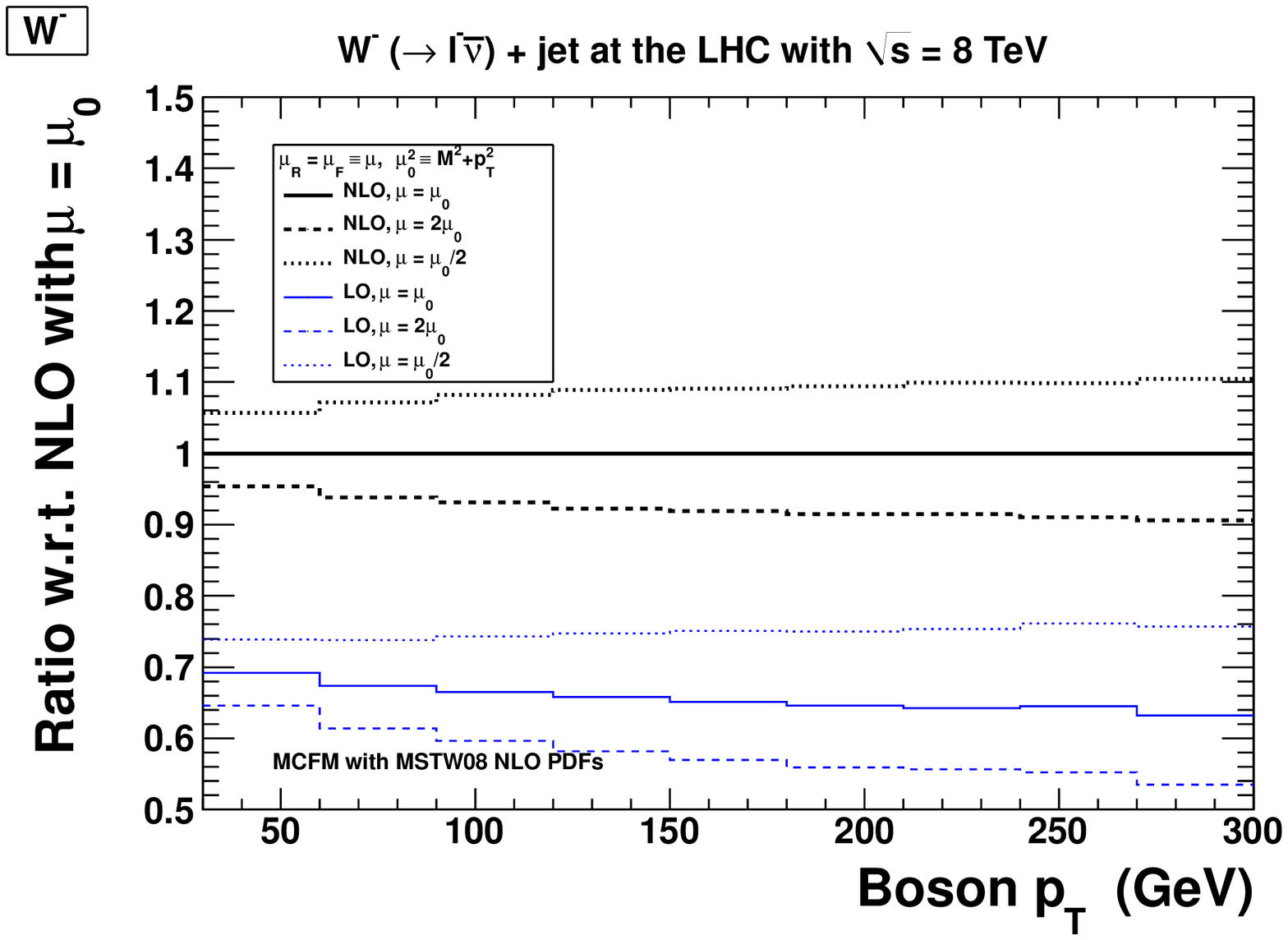}}
  \subfigure[$W^\pm$]{\includegraphics[width=0.5\textwidth]{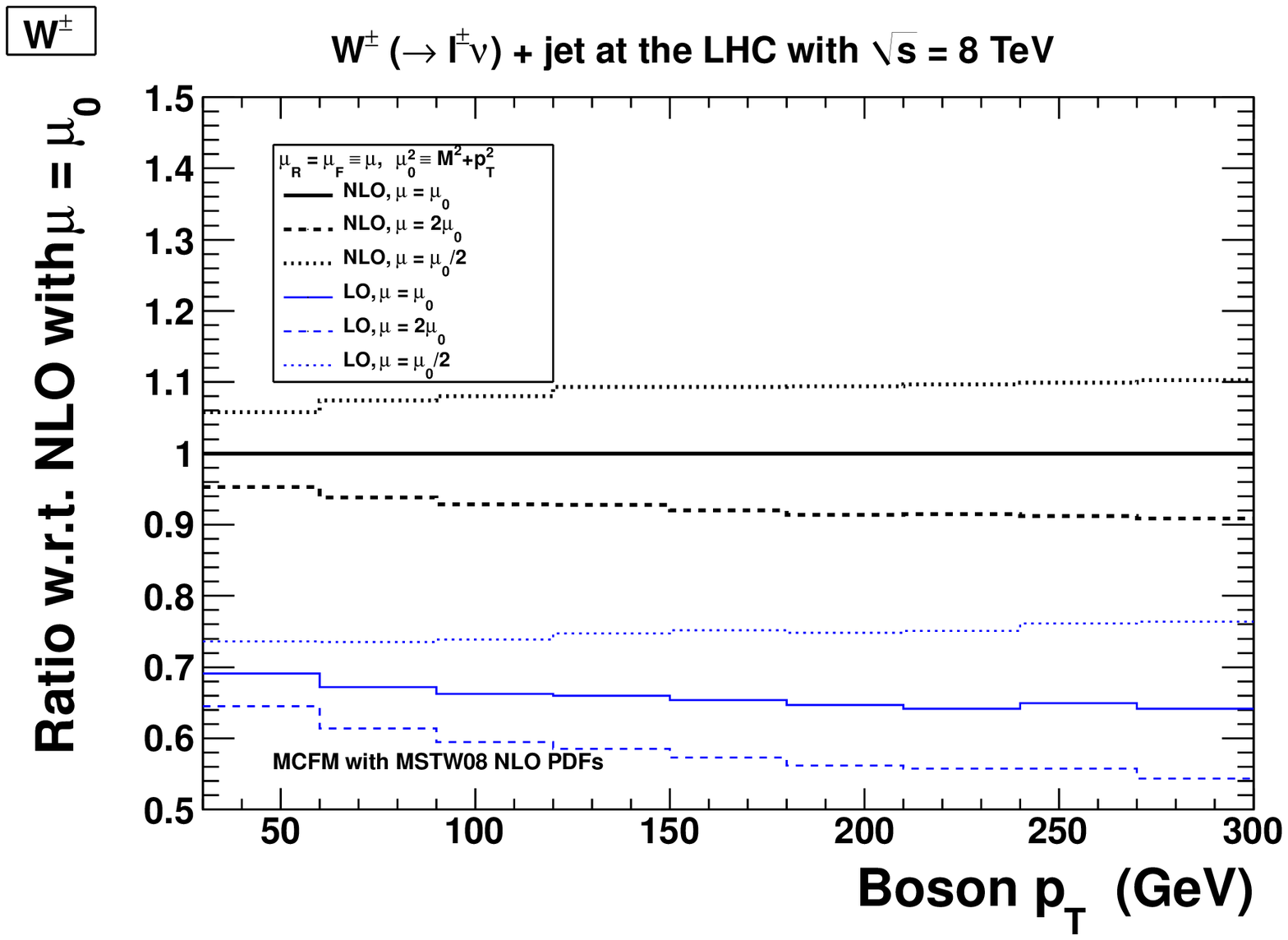}}%
  \subfigure[$Z^0$]{\includegraphics[width=0.5\textwidth]{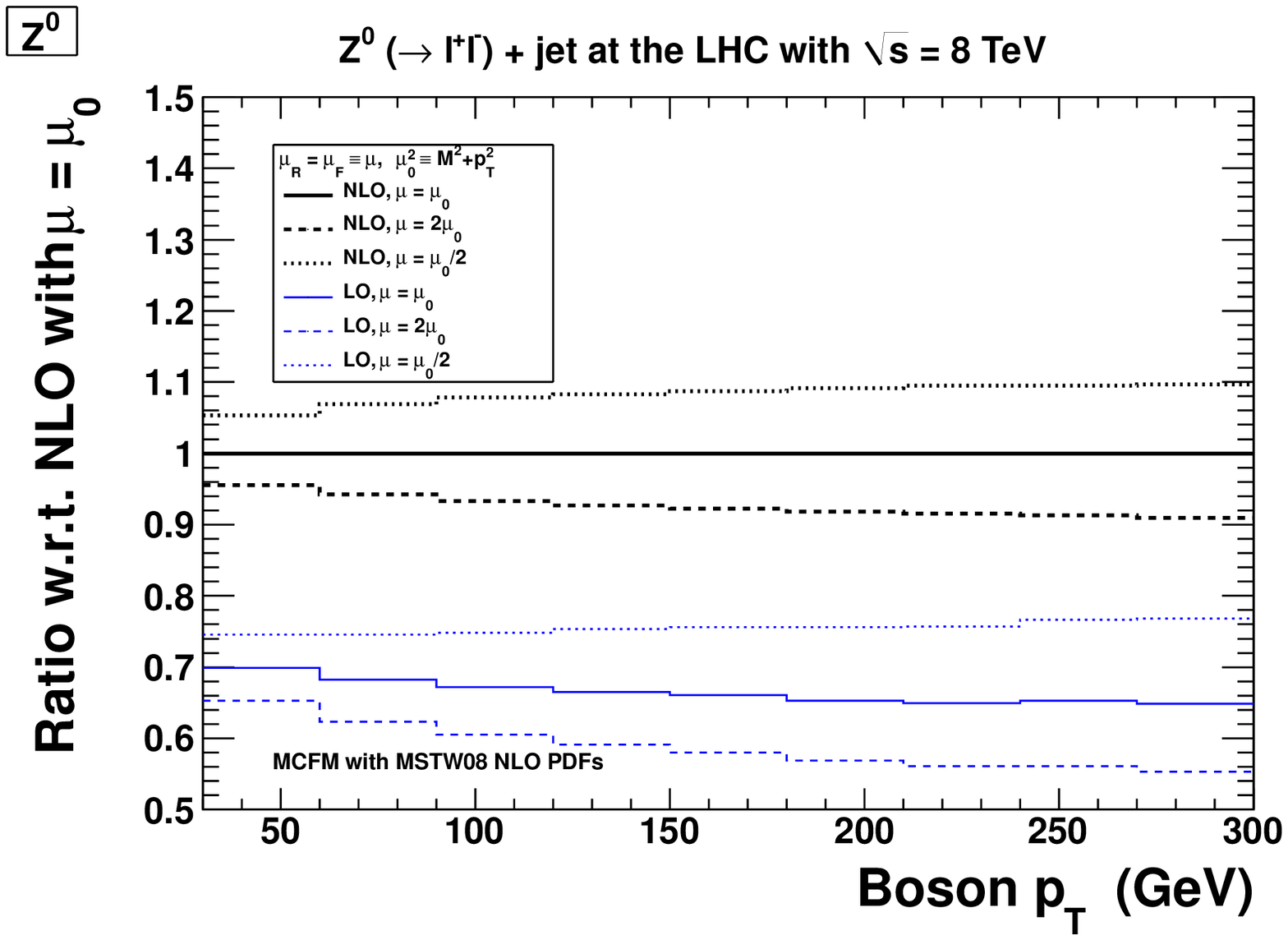}}
  \caption{Differential cross sections, ${\rm d}\sigma/{\rm d}p_T$, for the $V$+jet process as a function of boson $p_T$, taking the ratio to the central NLO prediction, for (a)~$W^+$, (b)~$W^-$, (c)~$W^\pm$ and (d)~$Z^0$.}
  \label{fig:dsdpt}
\end{figure}
In figure~\ref{fig:dsdpt} we show the differential cross sections, ${\rm d}\sigma/{\rm d}p_T$, as a function of the boson $p_T$, normalised to the central NLO prediction, for the LHC at a centre-of-mass-energy of $\sqrt{s} = 8$~TeV.  We show predictions at both LO (thinner lines) and NLO (thicker lines), each for three scale choices $\mu_R=\mu_F\in\{\mu_0/2,\mu_0,2\mu_0\}$.  In all cases we use the best-fit MSTW 2008 NLO PDF set~\cite{Martin:2009iq} with the corresponding value of $\alpha_S(M_Z^2)$.  The four plots in figure~\ref{fig:dsdpt} correspond to (a)~$V=W^+$, (b)~$V=W^-$, (c)~$V=W^\pm$ ($\equiv W^++W^-$) and (d)~$V=Z^0$.  We concentrate on the region of large boson $p_T>30$~GeV to minimise the impact of the $p_T^{\rm jet}>10$~GeV cut and the need to resum large logarithms of $M_V/p_T$ (most important for $p_T\ll M_V$), either analytically or by matching to a parton shower.  We see from figure~\ref{fig:dsdpt} that the scale dependence and the NLO/LO ratio are very similar for all four observables.  The scale uncertainty grows with increasing $p_T$, reaching almost 20\% at LO and around 10\% at NLO at the highest $p_T\sim 300$~GeV.  It is interesting that the LO and NLO scale uncertainty bands do not overlap, with the NLO/LO ratio growing from around a factor 1.4 at $p_T\sim 30$~GeV to a factor 1.6 at $p_T\sim 300$~GeV.  We have also investigated the effect of taking $\mu_R\ne\mu_F$.  We find that independently varying $\mu_R$ and $\mu_F$ by factors of two relative to $\mu_0$ slightly increases the scale uncertainty bands only in the lowest two $p_T$ bins; in all other $p_T$ bins the choice $\mu_R=\mu_F$ provides the largest scale variation.  Finally, taking the central scale choice to be $\mu_R=\mu_F=H_T$, the scalar sum of the transverse momenta of all final state particles, gives very similar scale uncertainty bands as our default choice $\mu_R=\mu_F=\mu_0$.

The relatively large NLO/LO ratio of $\sim 1.5$ is indicative that as-yet-unknown NNLO QCD corrections, that is, $\mathcal{O}(\alpha_S^3)$, to the boson $p_T$ distributions could be substantial, possibly larger than the estimated scale uncertainty at NLO.  The relevant two-loop QCD helicity amplitudes have been computed for $q\bar{q}\to Vg$ and $qg\to Vq$~\cite{Gehrmann:2011ab}, and recently also for $gg\to Zg$~\cite{Gehrmann:2013vga}.  An approximate method for estimating NNLO QCD corrections to the $Z$+jet process is discussed in ref.~\cite{Rubin:2010xp}, but it does not perform well for the $Z$ $p_T$ distribution.  Resummation of threshold logarithms has been performed at next-to-next-to-leading logarithmic accuracy~\cite{Becher:2011fc,Kidonakis:2012sy,Becher:2012xr}, and recently even at next-to-next-to-next-to-leading logarithmic accuracy~\cite{Becher:2013vva}, giving results consistent with the NLO predictions, but generally with smaller theoretical uncertainties.

It has long been recognised (see, for example, refs.~\cite{Dokshitzer:1978dr,Collins:1984kg}) that, in the low-$p_T$ region where $p_T\ll M_V$, the convergence of a fixed-order perturbative expansion in powers of $\alpha_S$ is spoiled by the presence of large logarithmic terms of the form $\ln^n(M_V^2/p_T^2)$, which must be resummed to all orders in $\alpha_S$.  The extent to which the inclusion of such resummed terms changes the fixed-order prediction at $p_T>30$~GeV depends on the details of the resummation formalism and, in particular, on the recipe used to join the two well-defined regions of $p_T\ll M_V$ (dominated by the resummed terms) and $p_T\gtrsim M_V$ (dominated by the fixed-order terms).  The popular \textsc{resbos} code~\cite{Balazs:1997xd} does not converge exactly to the fixed-order prediction for very large $p_T\gg M_V$, where it displays unphysical behaviour~\cite{Ellis:1997ii}, and a switch must be made by hand between the \textsc{resbos} prediction and the fixed-order prediction, for example, at the crossing point typically between $p_T=M_V/2$ and $p_T=M_V$.  A resummed calculation matched more carefully to the fixed-order result can be seen in figure~3 of ref.~\cite{Brandt:2013hoa}, where the total prediction deviates substantially from the fixed-order result only for very low $p_T<15$~GeV, while the deviation is small for $p_T\sim 30$--$50$~GeV and completely negligible for higher $p_T$ values.  Similar results can be seen in figure~1 of ref.~\cite{Bozzi:2010xn}, where it is observed that the difference between the total prediction (including resummation) and the fixed-order prediction reduces when increasing the perturbative order, and is again small for $p_T\sim 30$--$50$~GeV.  We note that the introduction of resummation complicates the initial-state flavour decomposition, in that only the sum of $q\bar{q}$ and $qg$ contributions is well-defined, and generally results in the PDFs being evaluated at multiple factorisation scales.  Clearly, in order to provide a meaningful PDF constraint using fixed-order calculations, it is necessary to restrict to the $p_T\gtrsim M_V$ region where $p_T$-resummation should not play any r\^ole.  A cut of $p_T>30$~GeV is likely to be sufficient, but to be more conservative a stronger cut could easily be imposed, such as $p_T>M_V/2$ or even $p_T>M_V$.

\subsubsection{Ratios of boson $p_T$ distributions}

\begin{figure}
  \centering
  \subfigure[$W^+/W^-$]{\includegraphics[width=0.5\textwidth]{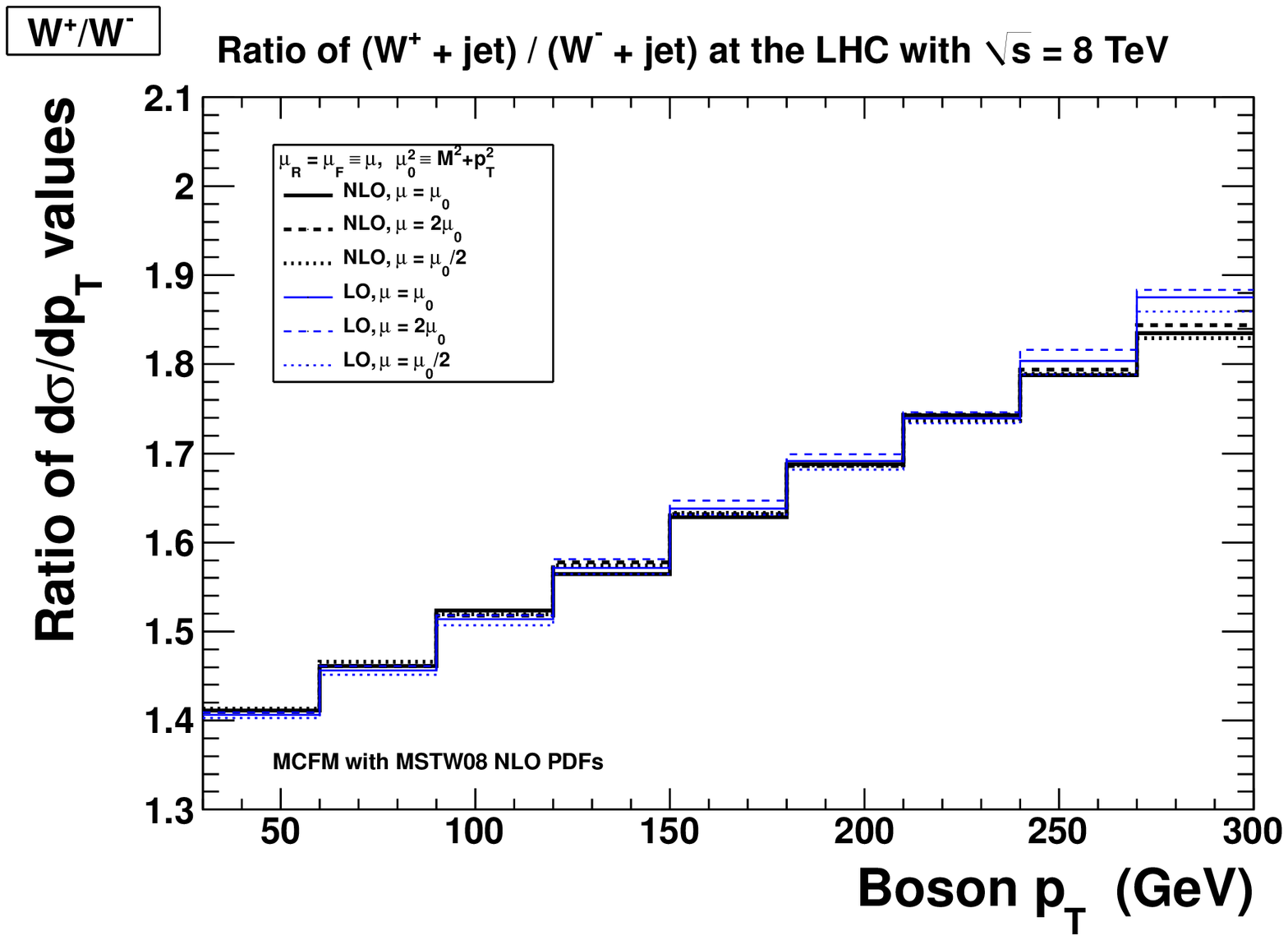}}%
  \subfigure[$W^+/Z^0$]{\includegraphics[width=0.5\textwidth]{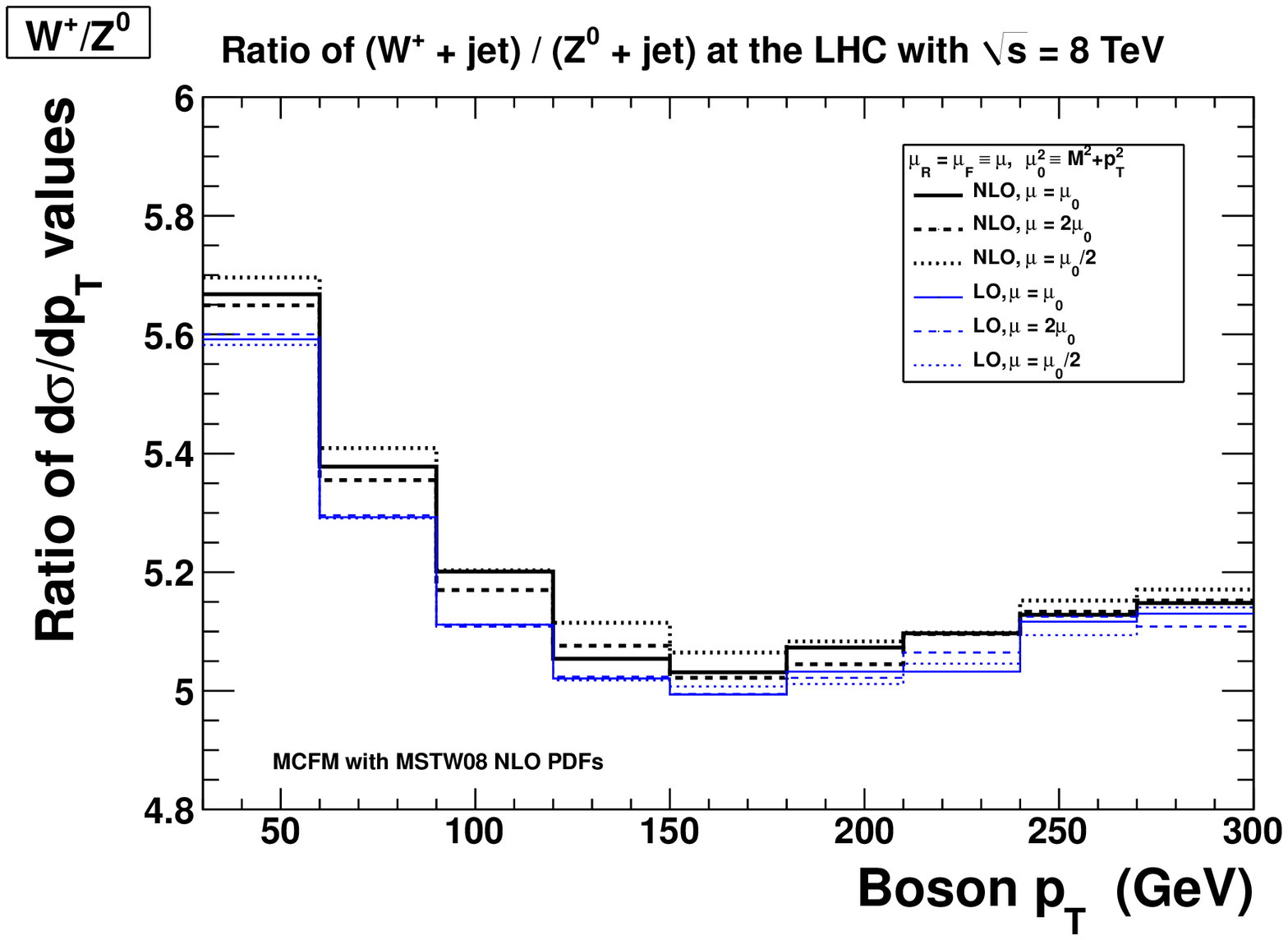}}
  \subfigure[$W^-/Z^0$]{\includegraphics[width=0.5\textwidth]{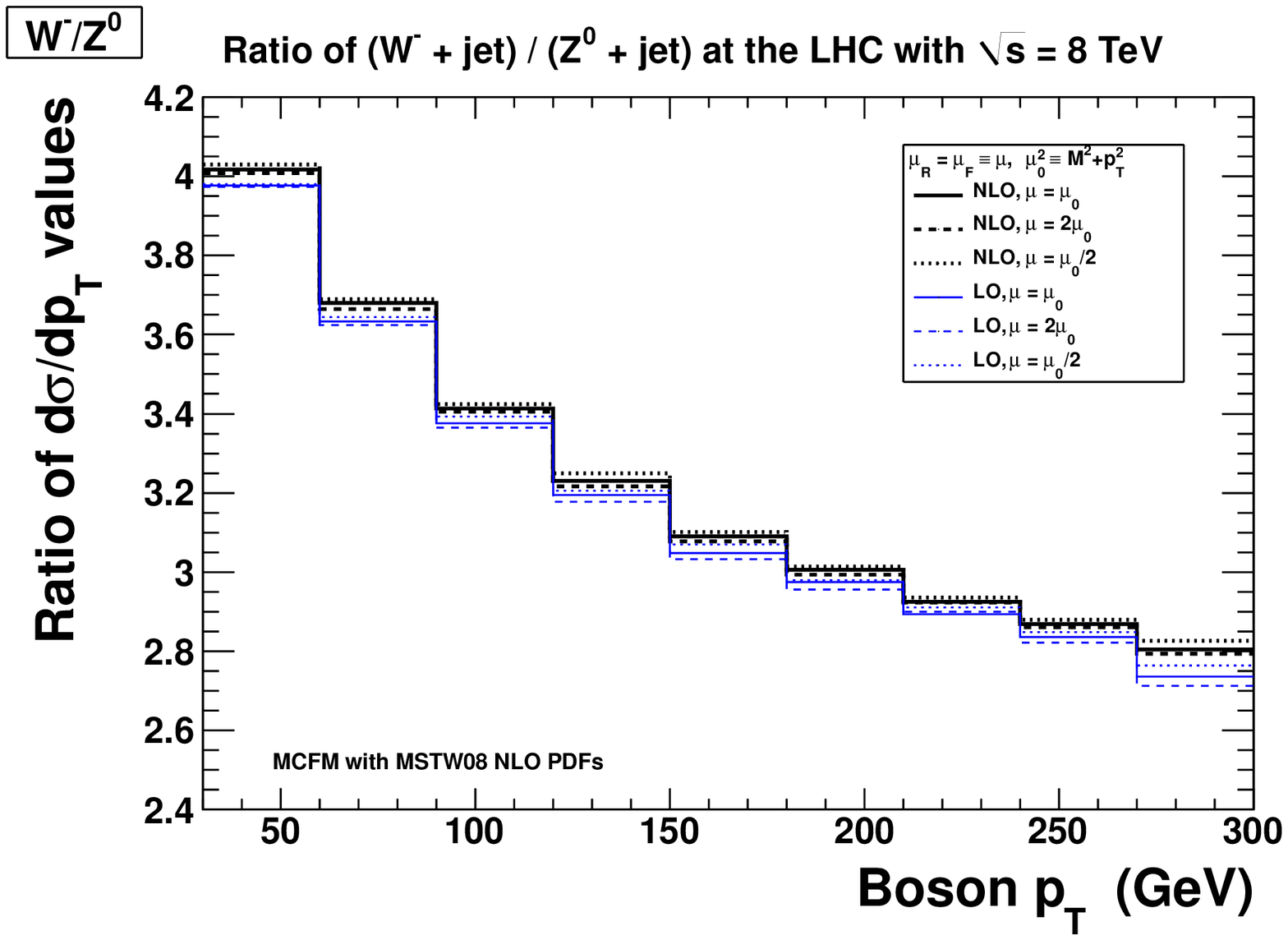}}%
  \subfigure[$W^\pm/Z^0$]{\includegraphics[width=0.5\textwidth]{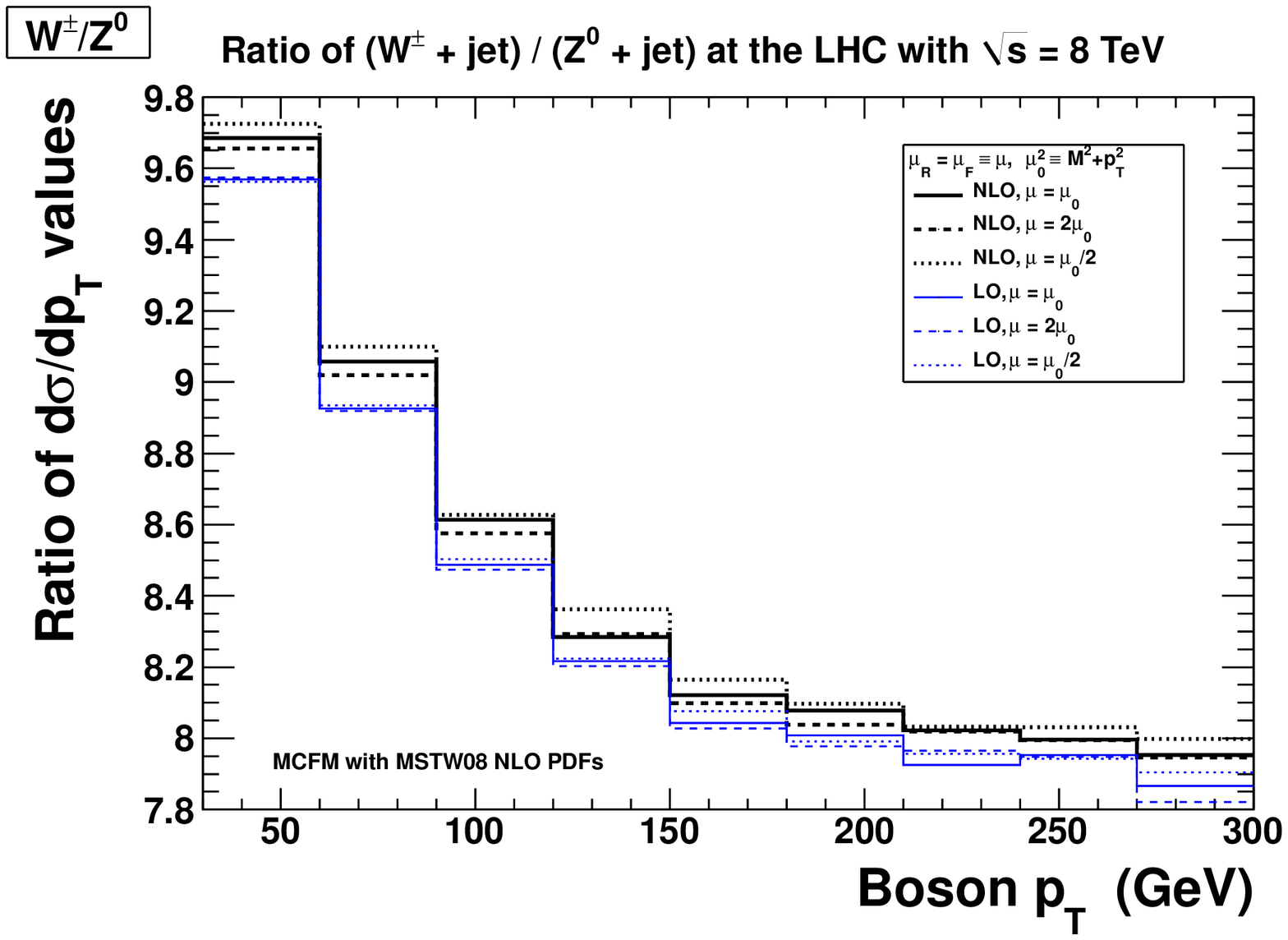}}
  \caption{Ratios of differential cross sections for the $V$+jet process as a function of boson $p_T$, for (a)~$W^+/W^-$, (b)~$W^+/Z^0$, (c)~$W^-/Z^0$ and (d)~$W^\pm/Z^0$, at LO and NLO for different scales.}
  \label{fig:Rdsdpt}
\end{figure}

In figure~\ref{fig:Rdsdpt} we show the various cross-section ratios at LO and NLO with different scale choices, exactly as in figure~\ref{fig:dsdpt}.  The four plots correspond to (a)~$W^+/W^-$, (b)~$W^+/Z^0$, (c)~$W^-/Z^0$ and (d)~$W^\pm/Z^0$ where $W^\pm\equiv W^++W^-$.  Here, we are making the reasonable assumption that scale variations are fully correlated between numerator and denominator in the cross-section ratios.  This assumption is easily justified from the similarity of the four plots in figure~\ref{fig:dsdpt}, for example, in going from $\mu_0$ to $2\mu_0$ (or to $\mu_0/2$) the four independent cross sections decrease (or increase) by a very similar amount.  More quantitatively, considering pairs of cross sections, $(A,B)$, for the three scale choices, $\mu_R=\mu_F\in\{\mu_0/2,\mu_0,2\mu_0\}$, separately at LO and NLO, we find that the correlation coefficient is essentially always greater than $+0.999$ (indicating a very strong correlation) for each of the four pairs of cross sections, $(A,B)\in\{(W^+,W^-),(W^+,Z^0),(W^-,Z^0),(W^\pm,Z^0)\}$.  Both the scale dependence and the difference betweeen LO and NLO cancels almost completely in the $W^+/W^-$ ratio, with residual differences being smaller than the statistical fluctuations.  The cancellation is not quite as complete for the $W^+/Z^0$, $W^-/Z^0$ and $W^\pm/Z^0$ ratios, where the NLO prediction generally lies about 1\% above the LO prediction, but this should be compared with a much larger NLO correction of $\sim$50\% for the separate ${\rm d}\sigma/{\rm d}p_T$ distributions (see figure~\ref{fig:dsdpt}).

Of course, other central scale choices are possible apart from our default choice of $\mu_R=\mu_F=\mu_0\equiv \sqrt{M^2+p_T^2}$, and we have investigated two alternative choices.  Taking $\mu_R=\mu_F=H_T$ gives results close to $\mu_R=\mu_F=2\mu_0$, so that $\mu_R=\mu_F=H_T/2$ would give results very similar to $\mu_R=\mu_F=\mu_0$.  Taking $\mu_R=\mu_F=M$ is not sensible at large boson $p_T$ values where it does not represent a typical hard scale of the process, and leads to results for ${\rm d}\sigma/{\rm d}p_T$ larger at NLO by around 20\% at $p_T=300$~GeV compared to $\mu_R=\mu_F=\mu_0$.  However, even with this unreasonable scale choice, the $W^+/W^-$ ratio is completely consistent with the default prediction within the statistical fluctuations, while the $W^+/Z^0$, $W^-/Z^0$ and $W^\pm/Z^0$ ratios are only 2\% larger at $p_T = 300$~GeV.  The insensitivity of the ratios to different scale choices is further evidence that the ratios are very stable with respect to higher-order QCD corrections.  Note that at very large boson $p_T$ values, $p_T\gg M_V$, then the boson mass $M_V$ becomes irrelevant and $p_T$ becomes the only sensible scale choice.  The boson $p_T$ distribution at very large $p_T$ is therefore effectively a single-scale observable analogous to inclusive jet production, which is already used in current global PDF fits, but with a cleaner final state from the leptonic decay of the vector boson rather than the hadronic jets.

\begin{figure}
  \centering
  \subfigure[$W^+/W^-$]{\includegraphics[width=0.5\textwidth]{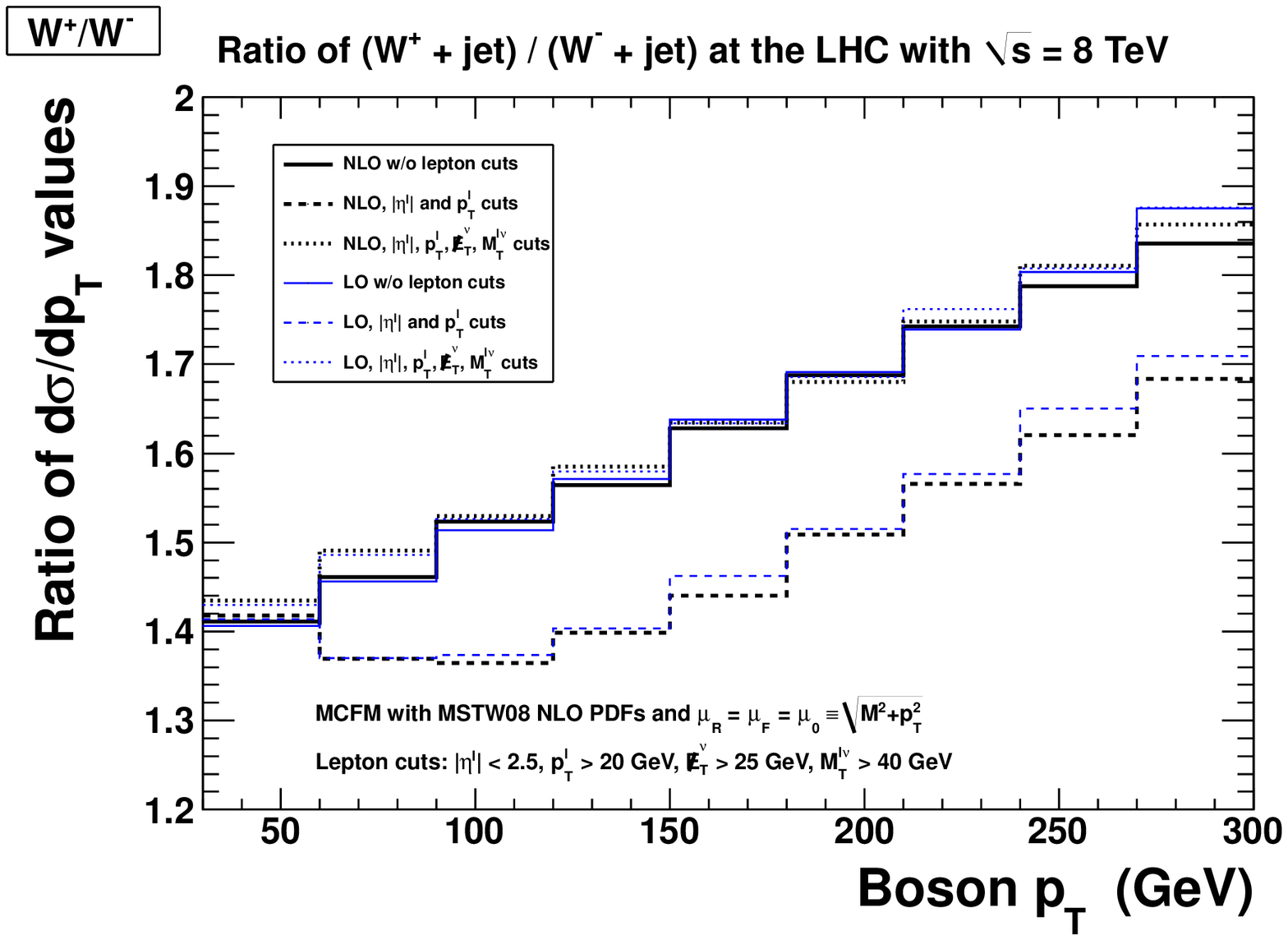}}%
  \subfigure[$W^+/Z^0$]{\includegraphics[width=0.5\textwidth]{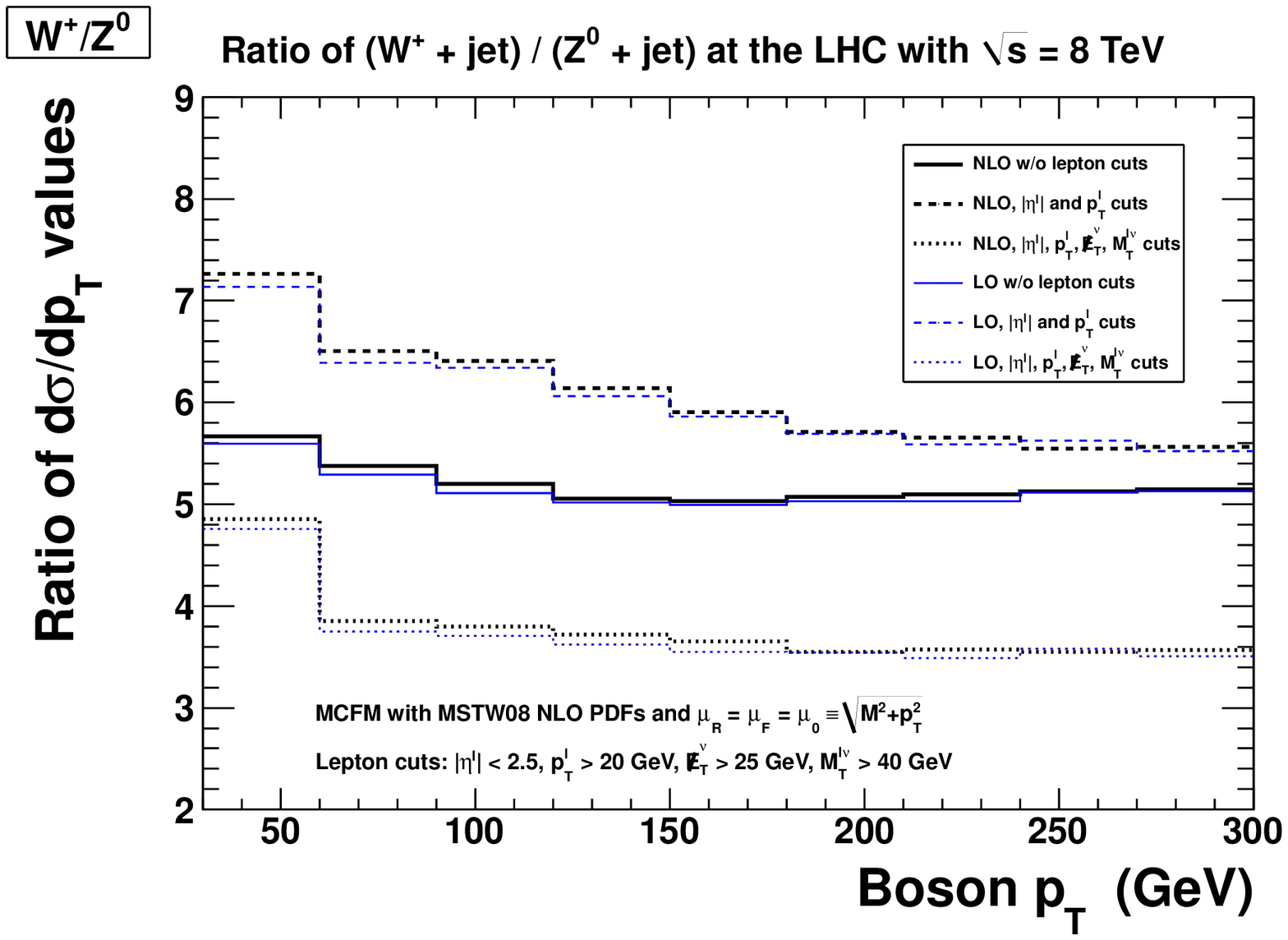}}
  \subfigure[$W^-/Z^0$]{\includegraphics[width=0.5\textwidth]{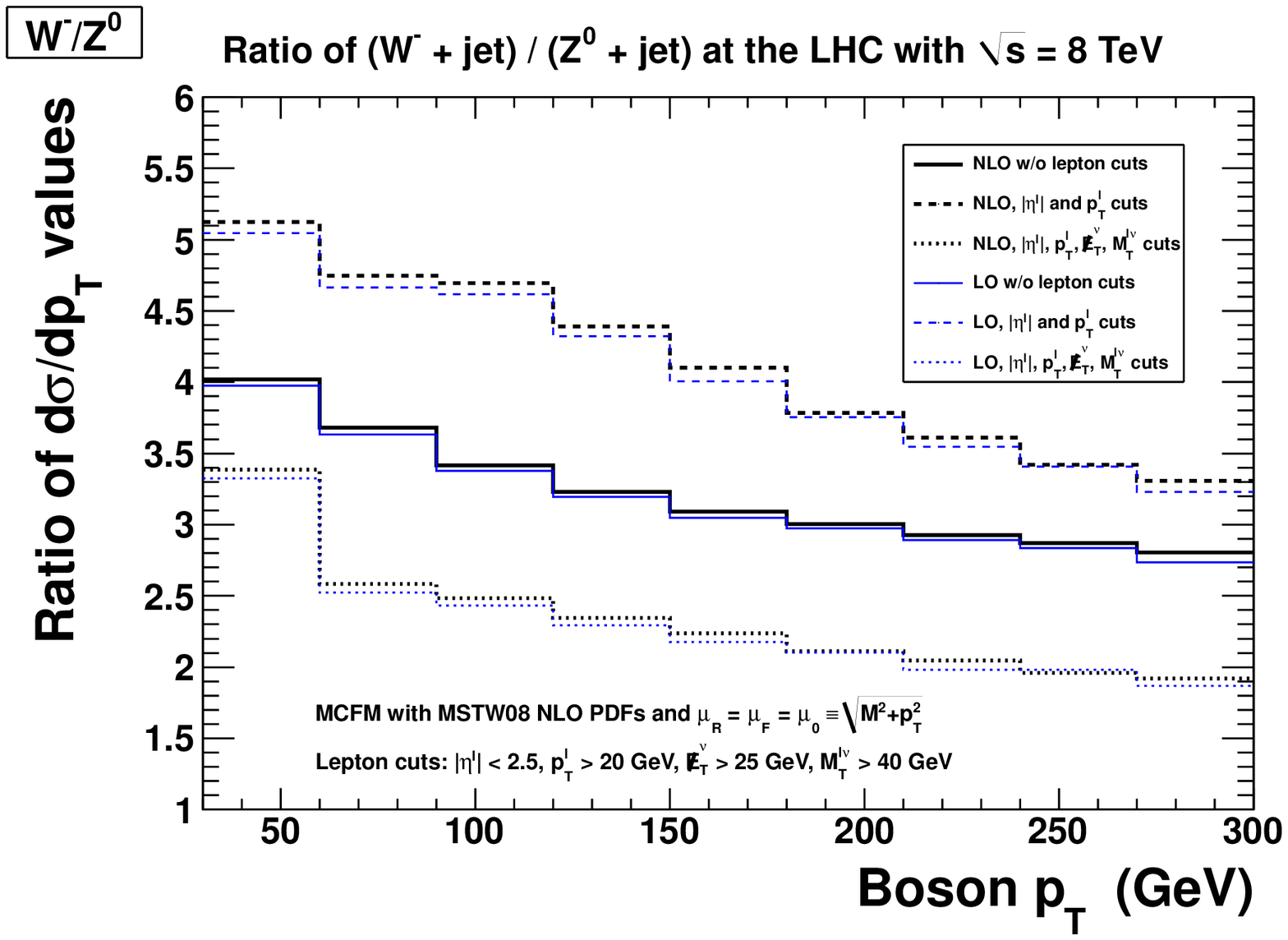}}%
  \subfigure[$W^\pm/Z^0$]{\includegraphics[width=0.5\textwidth]{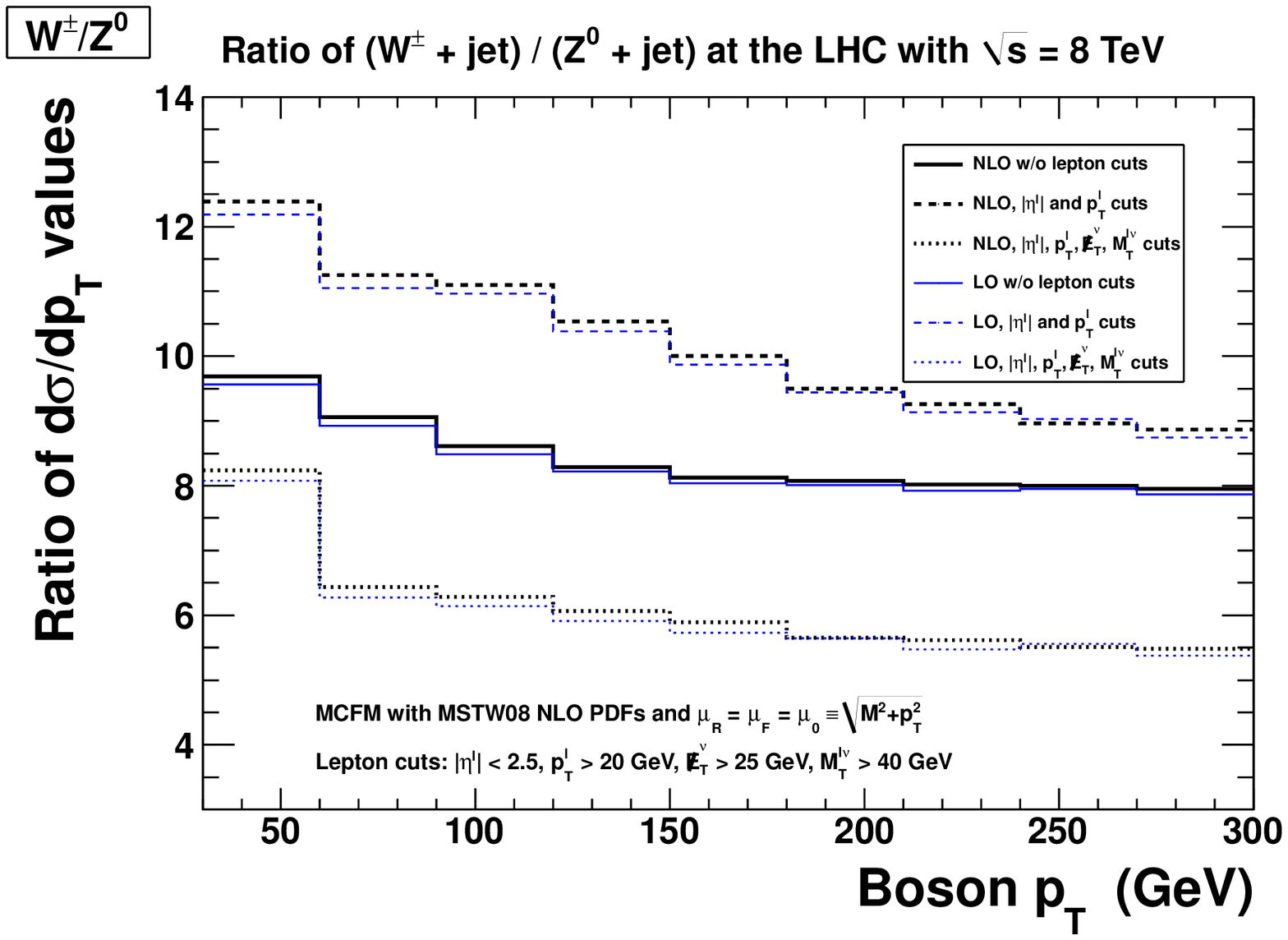}}
  \caption{Ratios of differential cross sections for the $V$+jet process as a function of boson $p_T$, for (a)~$W^+/W^-$, (b)~$W^+/Z^0$, (c)~$W^-/Z^0$ and (d)~$W^\pm/Z^0$, at LO and NLO with typical kinematic cuts imposed on the leptonic decay products.}
  \label{fig:RdsdptLepCut}
\end{figure}
We have not attempted so far to impose realistic cuts on the leptonic decay products.  However, to avoid additional theoretical uncertainties arising from extrapolating the experimental data to the full phase space, it would be better to compare data and theory within a common fiducial phase space.  In figure~\ref{fig:RdsdptLepCut} we show the effect of imposing typical acceptance cuts on the charged-lepton pseudorapidity and transverse momentum of $|\eta^\ell| < 2.5$ and $p_T^\ell>20$~GeV.  These cuts reduce the $Z^0\to\ell^+\ell^-$ cross sections (with two charged leptons) more than the $W^\pm\to\ell^\pm\nu$ cross sections (with only one charged lepton), with somewhat more effect on $W^+$ than $W^-$, hence the $W^+/W^-$ ratio decreases while the $W^+/Z^0$, $W^-/Z^0$ and $W^\pm/Z^0$ ratios all increase after these cuts are imposed.  In figure~\ref{fig:RdsdptLepCut} we also show the effect of imposing additional cuts for the $W\to\ell\nu$ decay, used by ATLAS~\cite{Aad:2011xn}, on the missing transverse energy $\not\mathrel{E}_T^\nu>25$~GeV and transverse mass $M_T^{\ell\nu}=\sqrt{2p_T^\ell\not\mathrel{E}_T^\nu(1-\cos\Delta\phi_{\ell\nu})}>40$~GeV, where $\Delta\phi_{\ell\nu}$ is the azimuthal separation between the directions of the charged lepton and neutrino.  These additional cuts further reduce the $W$ cross sections, now with somewhat more effect on $W^-$ than $W^+$.  The net effect is that the $W^+/W^-$ ratio is now very close to the result with no lepton cuts, while the $W^+/Z^0$, $W^-/Z^0$ and $W^\pm/Z^0$ ratios are now reduced compared to the result with no lepton cuts (as can also be seen by comparing the two plots in figure~4 of ref.~\cite{Aad:2011xn}).  However, although the numerical value of the cross-section ratios appears to be quite sensitive to the precise lepton cuts imposed, the trend between LO and NLO is very similar, as is the qualitative dependence on the boson $p_T$.  In the remainder of this paper, for simplicity we do not impose cuts on the leptonic decay products, which will generally be different for ATLAS and CMS, and for electrons and muons, hence the calculations will need to be repeated with the precise cuts after the measurements are made.  However, we expect our findings regarding theoretical uncertainties not to be substantially altered when restricted acceptance cuts are imposed (again, as can also be seen by comparing the two plots in figure~4 of ref.~\cite{Aad:2011xn}).

Similar observations regarding perturbative stability have been made for the $Z^0/W^+$ and $Z^0/W^-$ ratios computed at LO and NLO and plotted versus jet $p_T$ in the $Z^0$ + 4 jets process~\cite{Ita:2011wn}.  Note that the $gg$ channel is absent at LO for $V$ + 1 jet, but present for higher jet multiplicities.  Predictions have been made at both LO and NLO for the $Z^0/W^+$ total cross-section ratio with up to 4 jets~\cite{Ita:2011wn}, for the $W^+/W^-$ total cross-section ratio with up to 4 jets~\cite{Berger:2010zx}, and recently also for the $W^+/W^-$ total cross-section ratio with up to 5 jets~\cite{Bern:2013gka}.  The difference between LO and NLO predictions for these ratios has some dependence on the jet multiplicity, but is never more than a few percent.  However, the results in refs.~\cite{Ita:2011wn,Berger:2010zx,Bern:2013gka} use LO PDFs (and $\alpha_S$) with the LO calculation and NLO PDFs (and $\alpha_S$) with the NLO calculation.  It is therefore difficult to isolate the genuine impact of NLO corrections to the hard-scattering process from the impact of using different PDFs (and $\alpha_S$) in the LO and NLO calculations.  Recall that in figure~\ref{fig:Rdsdpt} we use the same NLO PDFs (and $\alpha_S$) in both the LO and NLO calculations.

The assumption that renormalisation and factorisation scale variations are fully correlated between numerator and denominator in a ratio of cross sections is almost always made in the literature when considering two independent, but similar, processes.  For example, this assumption is made when considering the ratio of double-Higgs to single-Higgs production~\cite{Djouadi:2012rh,Goertz:2013kp} or when extracting the strong coupling $\alpha_S$ from the ratio of the inclusive 3-jet cross section to the inclusive 2-jet cross section~\cite{Chatrchyan:2013txa}.  Of course, assuming uncorrelated scale variations would result in the ratio having a larger theoretical uncertainty than the individual cross sections, hence removing a prime motivation for taking the ratio.  In reality, the degree of correlation will be somewhere in between these two extremes.  However, all available evidence, such as the similarity of the four plots in figure~\ref{fig:dsdpt}, the correlation coefficients $\gtrsim0.999$, the stability of the ratios at LO and NLO for various scale choices, and the very mild dependence of the ratios on jet multiplicity (see figure~\ref{fig:jetmult}), indicates that higher-order QCD corrections to $W^+$, $W^-$ and $Z^0$ production at large boson $p_T$ are very similar, such that the assumption of fully-correlated scale variation is justified.  Stability of the ratios at NNLO (when known) would further justify this assumption.  Despite the different initial-state flavours and electroweak couplings to quarks involved in $W^+$, $W^-$ and $Z^0$ production, the theory of QCD is flavour-blind meaning that gluons couple with equal strength to all quark flavours, while the small kinematic difference due to the different $W$ and $Z$ masses becomes irrelevant for $p_T\gg M_{W,Z}$, all together explaining the similarity of the higher-order QCD corrections.

\begin{figure}
  \centering
  \subfigure[$W^+/W^-$]{\includegraphics[width=0.5\textwidth]{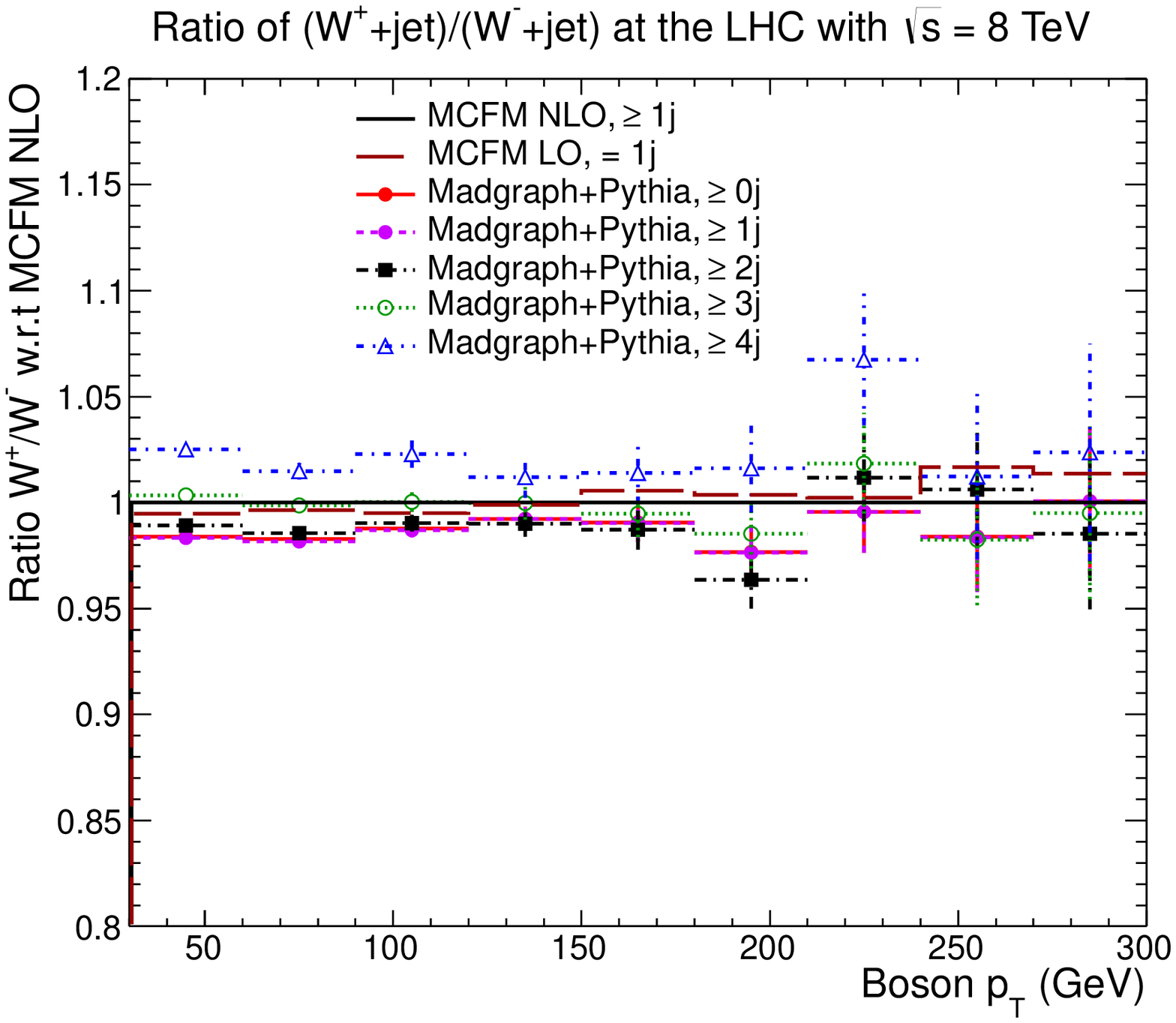}}%
  \subfigure[$W^+/Z^0$]{\includegraphics[width=0.5\textwidth]{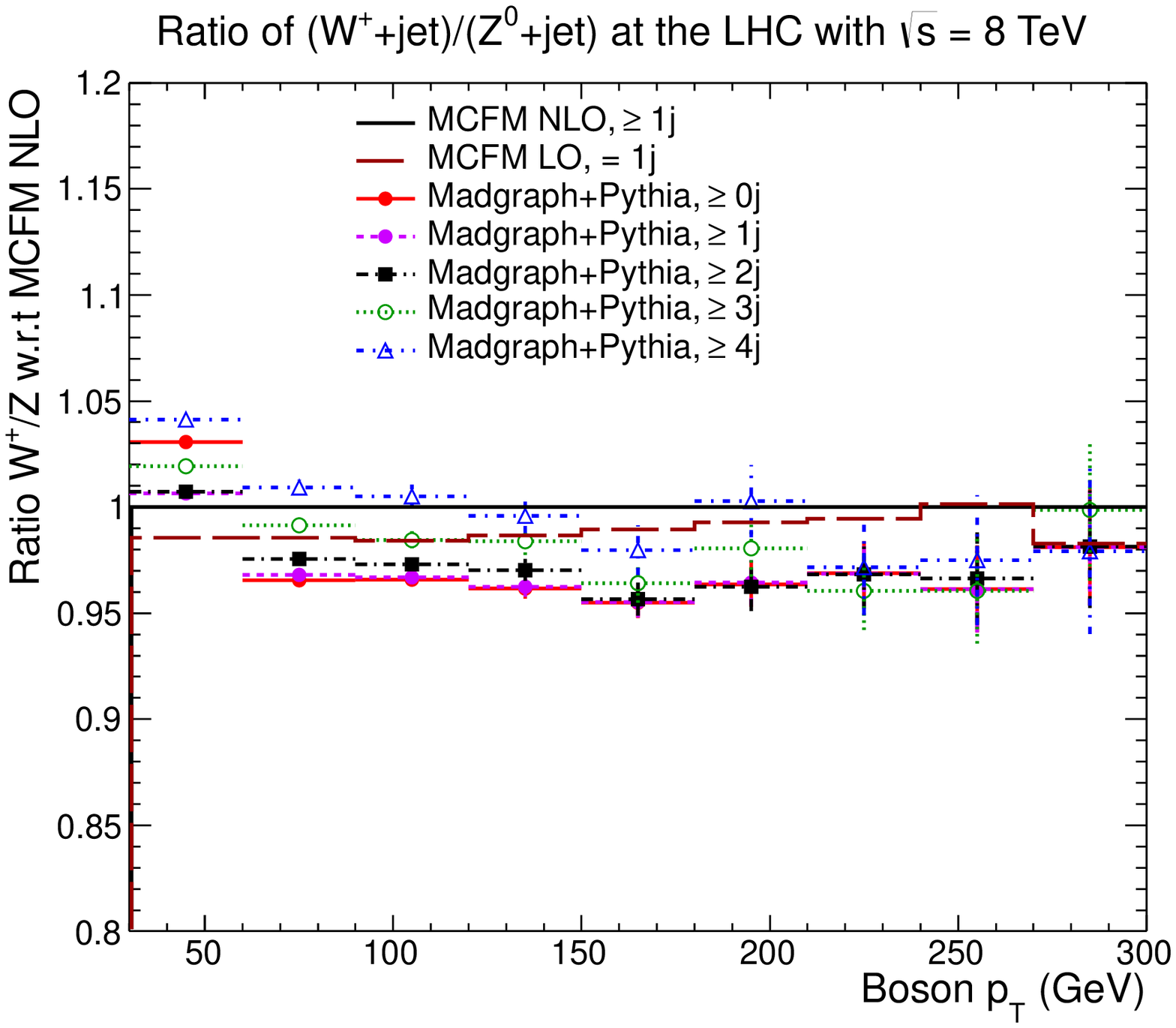}}
  \subfigure[$W^-/Z^0$]{\includegraphics[width=0.5\textwidth]{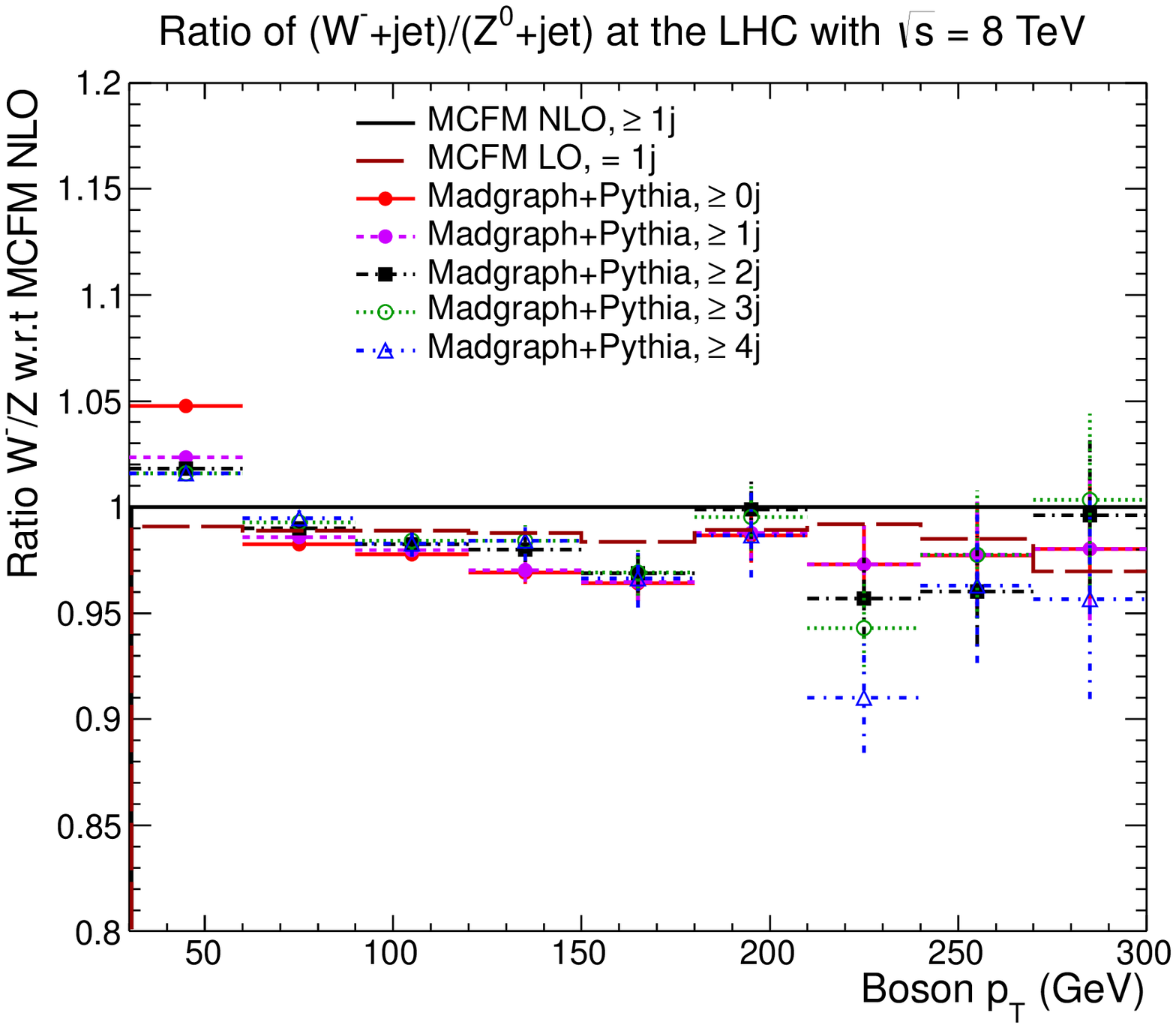}}%
  \subfigure[$W^\pm/Z^0$]{\includegraphics[width=0.5\textwidth]{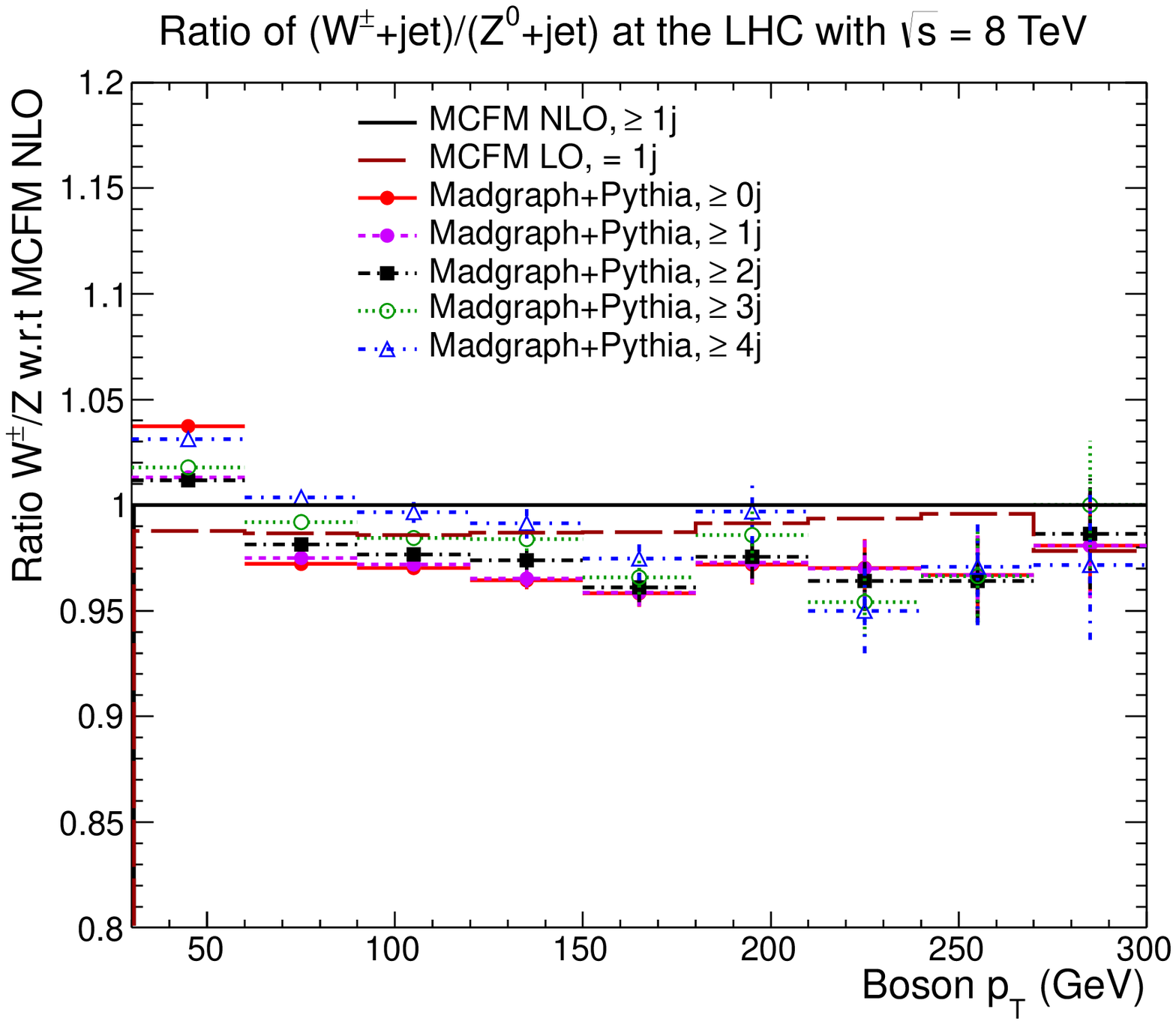}}
  \caption{Comparison of ratios of (a)~$W^+/W^-$, (b)~$W^+/Z^0$, (c)~$W^-/Z^0$ and (d)~$W^\pm/Z^0$, as predicted by \textsc{madgraph}+\textsc{pythia} and \textsc{mcfm} both using CTEQ6L1 PDFs.}
  \label{fig:compare}
\end{figure}
An almost complete cancellation of correlated scale uncertainties in the ($Z$+$N$-jet)/$(\gamma$+$N$-jet) ratios calculated at NLO, where $N\in\{2,3\}$, was observed in refs.~\cite{Bern:2011pa,Bern:2012vx}, where the theoretical QCD uncertainty was instead estimated by taking the difference between the ratios given by either the NLO fixed-order calculation or a LO matrix-element-matched-to-parton-shower calculation (\textsc{sherpa}~\cite{Gleisberg:2008ta}).  In principle, by comparing the \textsc{madgraph}+\textsc{pythia} predictions in figure~\ref{fig:jetmult} with the \textsc{mcfm} predictions in figure~\ref{fig:Rdsdpt}, we can investigate the impact of including higher hard-parton multiplicities and matching to a parton shower on the cross-section ratios.  However, again this issue is complicated by a different choice of PDFs (and the associated $\alpha_S$ value), namely CTEQ6L1 PDFs~\cite{Pumplin:2002vw} in figure~\ref{fig:jetmult} and MSTW08 NLO PDFs~\cite{Martin:2009iq} in figure~\ref{fig:Rdsdpt}.  We therefore repeated the \textsc{mcfm} calculations using the CTEQ6L1 PDFs~\cite{Pumplin:2002vw}.  A consistent comparison is then shown in figure~\ref{fig:compare}.  A remaining complication is that the \textsc{mcfm} predictions include exactly one jet at LO and either one or two jets at NLO, whereas the \textsc{madgraph} sample was generated with up to four hard-partons leading to jets, but the dependence of the cross-section ratios on jet multiplicity is anyway modest (see figure~\ref{fig:jetmult}).  We can then see that matching to a parton shower has almost no impact on the $W^+/W^-$ ratio, while it has a sizeable impact on the $W^+/Z^0$, $W^-/Z^0$ and $W^\pm/Z^0$ ratios only for low boson $p_T\lesssim 50$~GeV, but with still a few percent difference at large boson $p_T$ values.  Predictions from the two codes (\textsc{madgraph}+\textsc{pythia} and \textsc{mcfm} at NLO) for the normalised ${\rm d}\sigma/{\rm d}p_T$ distributions can differ by up to 15\%, but the difference is similar for $W$ and $Z$ bosons and hence almost cancels in the ratio.  We have checked that varying the \textsc{mlm} matching threshold in \textsc{madgraph}+\textsc{pythia} by factors of two from the default value of 10~GeV can have a significant effect of $\mathcal{O}(10\%)$ on the boson $p_T$ distributions (see also figure~20 of ref.~\cite{Alwall:2007fs}), but again this sensitivity largely cancels in the cross-section ratios under the assumption that it is correlated between numerator and denominator.  We have also checked the dependence on the tunable parameters associated with the \textsc{pythia} parton shower by switching to a so-called ``power shower''~\cite{Plehn:2005cq}.  This option allows the \textsc{pythia} parton shower to populate the full phase space by taking the starting scale of the $p_T$-ordered shower to be $\sqrt{s}$ rather than the default $\mu_F$ used in the so-called ``wimpy shower''~\cite{Plehn:2005cq}, leading to harder boson $p_T$ distributions in the absence of matching.  However, in the presence of \textsc{mlm} matching, we find that the $p_T$ distributions at large boson $p_T$ are almost unaffected by the different parton shower parameters.  Similar findings were previously observed in a study of gluino-pair and top-pair production; see figure~3 of ref.~\cite{Alwall:2008qv}.  Note that our use of \textsc{madgraph}+\textsc{pythia} is motivated by the fact that it is a common tool used by the LHC experiments (in particular, by CMS) to estimate the $V$+jets background in many analyses, and it is interesting to show that it gives similar predictions for the cross-section ratios as a fixed-order calculation (\textsc{mcfm}).  However, clearly for precision applications such as PDF fitting, the use of a fixed-order calculation is more appropriate, and hence we concentrate on \textsc{mcfm} predictions for the remainder of this paper.

\subsection{PDF dependence}

To examine the dependence of the boson $p_T$ distributions, and the various cross-section ratios, on the particular PDF choice, we run \textsc{mcfm}~\cite{Campbell:2010ff} at NLO with the central scale choice $\mu_R=\mu_F=\mu_0$ and using the \textsc{lhapdf} (v5.8.8) interface~\cite{Whalley:2005nh} for four modern NLO PDF sets: MSTW08~\cite{Martin:2009iq}, CT10~\cite{Lai:2010vv}, NNPDF2.3~\cite{Ball:2012cx} and ABM11~\cite{Alekhin:2012ig}.  In each case we store the histograms corresponding to the boson $p_T$ distributions for all members of a given PDF set during a single \textsc{mcfm} run, allowing PDF uncertainties to be calculated accurately without suffering from statistical fluctuations.  However, sufficient statistics are needed to distinguish the separate predictions from the four PDF sets, therefore again we average results over a large number ($\sim 100$) of independent \textsc{mcfm} runs, each with different seeds for the \textsc{vegas} integration.

\begin{figure}
  \centering
  \subfigure[Gluon distribution]{\includegraphics[width=0.5\textwidth]{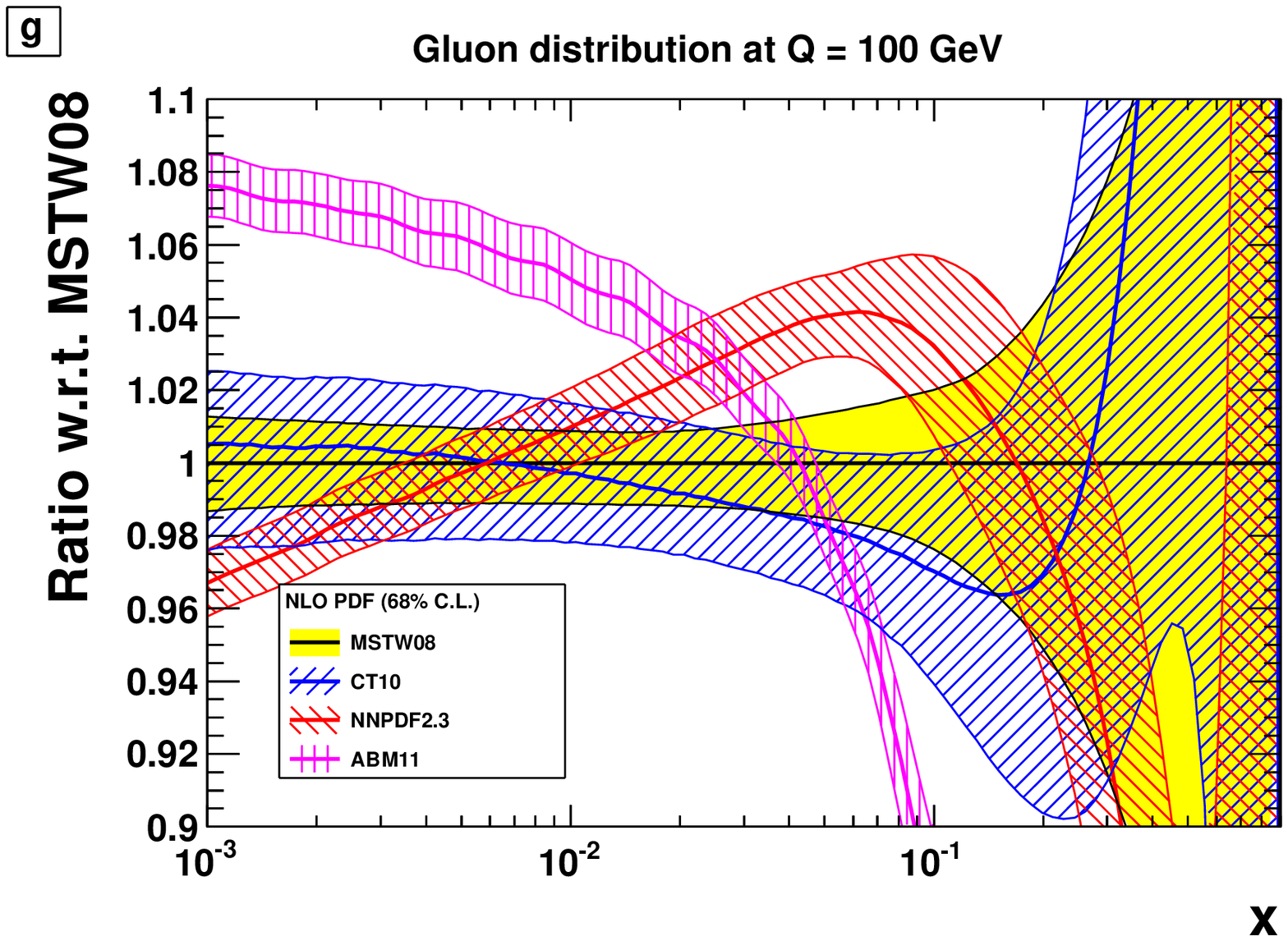}}%
  \subfigure[Ratio of up/down quark distributions]{\includegraphics[width=0.5\textwidth]{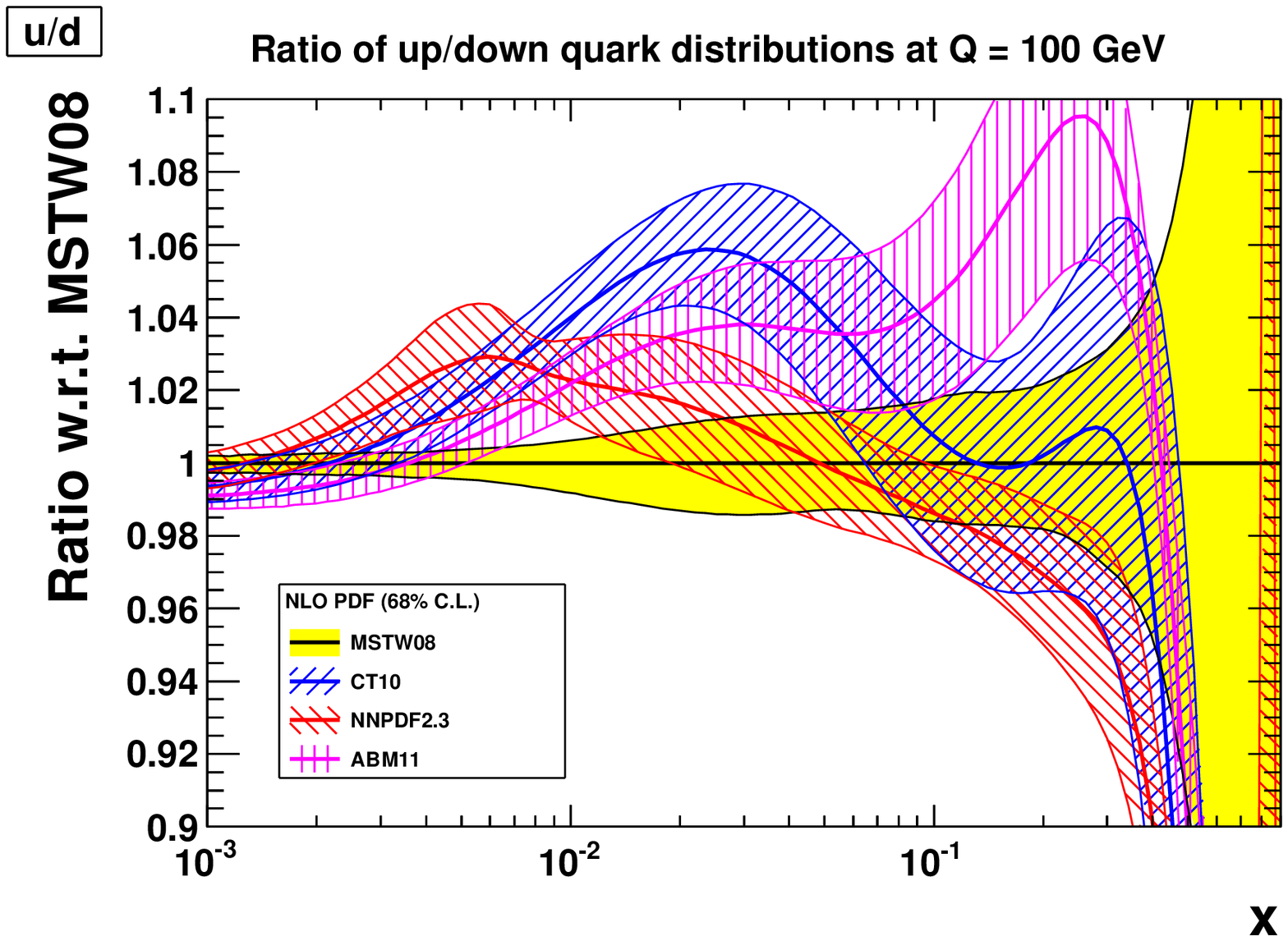}}
  \subfigure[Down-quark distribution]{\includegraphics[width=0.5\textwidth]{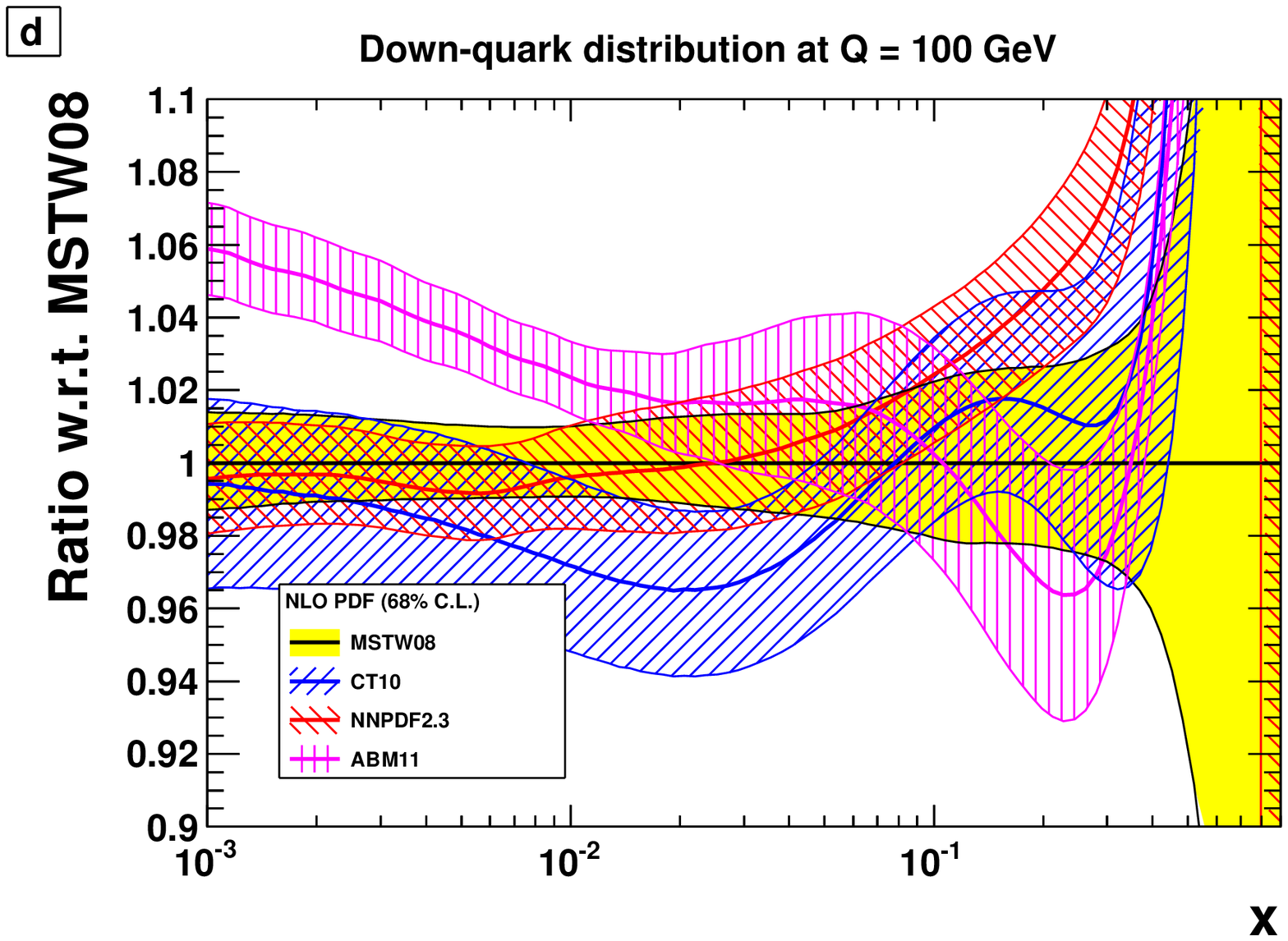}}%
  \subfigure[Up-quark distribution]{\includegraphics[width=0.5\textwidth]{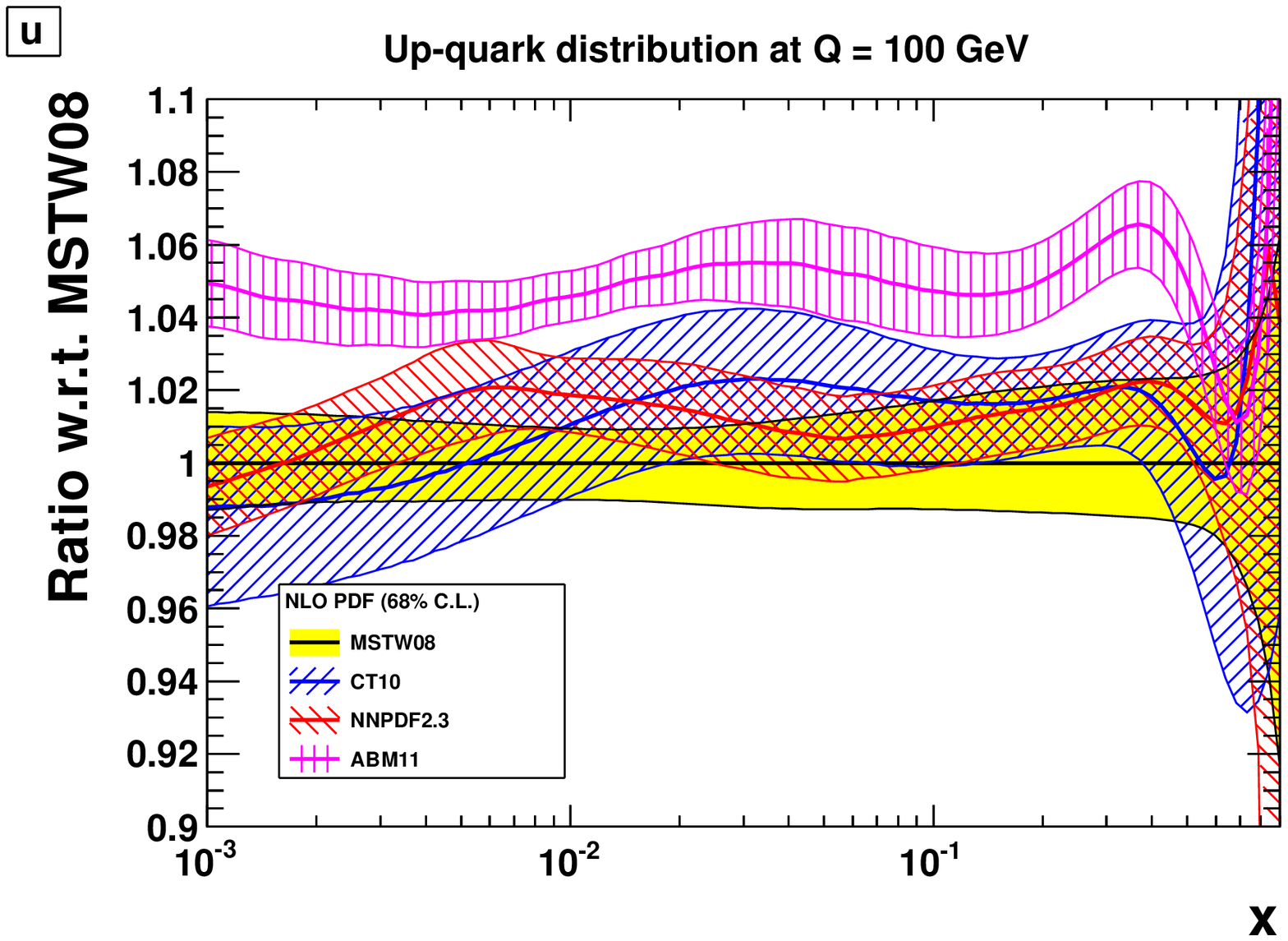}}
  \caption{Different PDF flavours at a scale $Q=100$~GeV versus $x$ for four choices of PDF set, taking the ratio to the MSTW08 value, for (a)~the gluon distribution, (b)~the ratio of the up-quark to the down-quark distributions, (c)~the down-quark distribution, and (d)~the up-quark distribution.}
  \label{fig:pdfsvsx}
\end{figure}
It is instructive to first look at some different PDF flavours plotted versus $x$ for the four choices of NLO PDF set, taking the ratio to the MSTW08 value, shown in figure~\ref{fig:pdfsvsx}.  We calculate PDF uncertainties at an estimated 68\% confidence-level (C.L.) according to the appropriate prescription of each group (see, for example, ref.~\cite{Watt:2011kp}).  The corresponding values of the strong coupling associated with each PDF set are $\alpha_S(M_Z^2) = \{0.1202, 0.1180, 0.1190, 0.1180\}$ for MSTW08, CT10, NNPDF2.3 and ABM11, respectively.  The envelope of the MSTW08, CT10 and NNPDF2.3 predictions therefore implicitly includes an $\alpha_S$ uncertainty of $\alpha_S(M_Z^2)\approx0.119\pm0.001$.  The PDF uncertainties for ABM11 implicitly include an $\alpha_S$ uncertainty of $\alpha_S(M_Z^2)=0.1180\pm0.0012$~\cite{Alekhin:2012ig}.

Reasons for differences between different PDF sets are complex and often not completely understood (see, for example, refs.~\cite{Forte:2013wc,Ball:2012wy}).  The most obvious feature of figure~\ref{fig:pdfsvsx} is that the ABM11 gluon distribution is very different from the others for practically all values of $x$.  This feature is mainly due to the ABM11 fit not including Tevatron data on jet production~\cite{Thorne:2011kq} and also due to the different treatment of heavy-quark contributions to structure functions in deep-inelastic scattering~\cite{Thorne:2012az,Ball:2013gsa}.  The larger ABM11 gluon distribution at low $x$ feeds through to the up- and down-quark distributions at low $x$ via $g\to q\bar{q}$ splitting in the DGLAP evolution, but the difference mostly cancels in the up/down ratio.  The NNPDF2.3 gluon distribution is larger than that from MSTW08 and CT10 for $x\sim 0.01$--$0.1$, but the separate $u$ and $d$ quark distributions agree reasonably well for the three groups.  However, some slight differences are amplified in the $u/d$ ratio shown in figure~\ref{fig:pdfsvsx}(b).  In particular, the MSTW08 $u/d$ ratio is smaller than the others at $x\sim 0.01$ and ABM11 is much larger than the others at $x\sim 0.1$--$0.4$.  We will see shortly that these features directly influence predictions for the $W^\pm$ charge asymmetry at the LHC.

\subsubsection{Boson $p_T$ distributions}

\begin{figure}
  \centering
  \subfigure[$W^+$]{\includegraphics[width=0.5\textwidth]{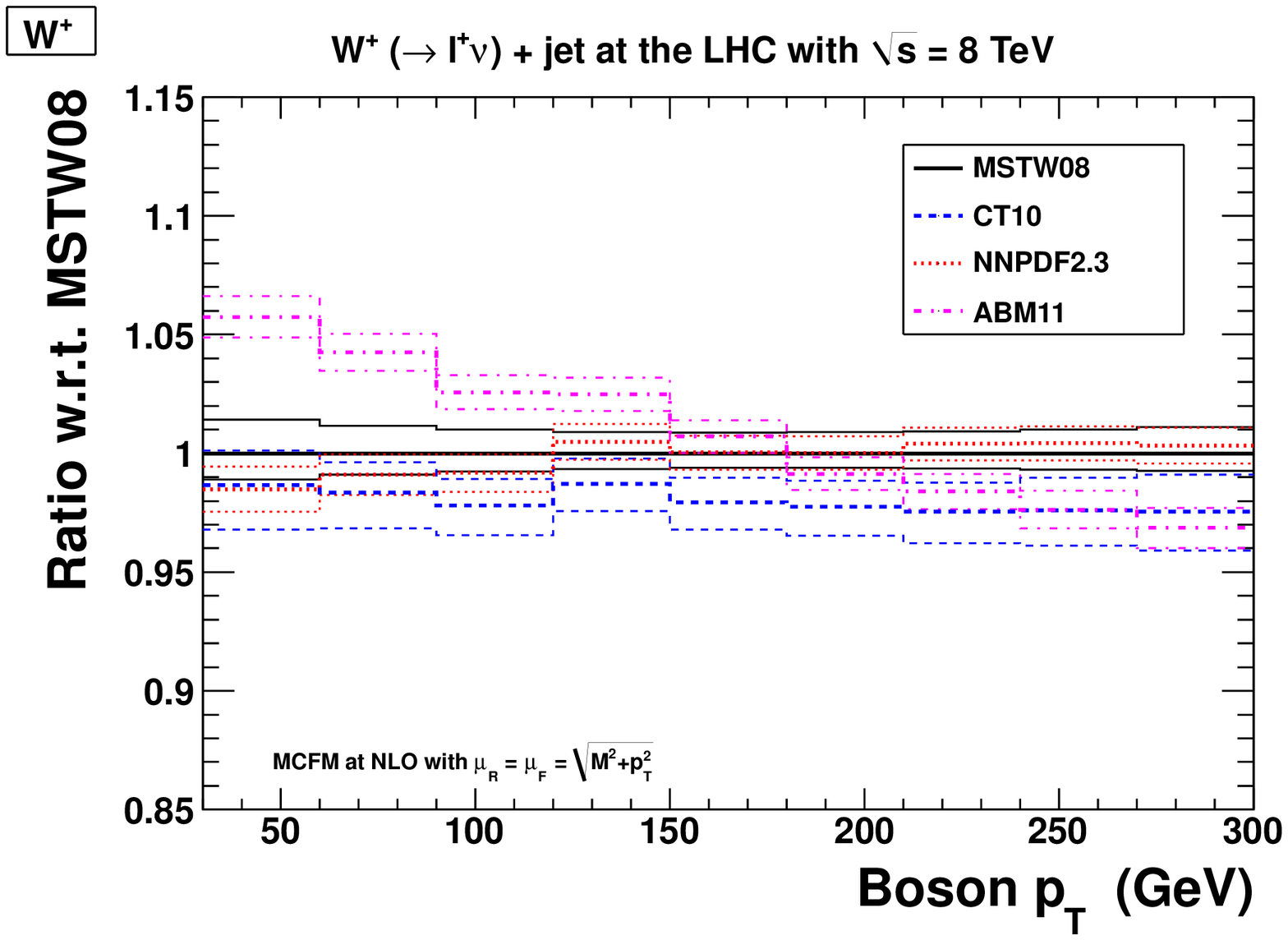}}%
  \subfigure[$W^-$]{\includegraphics[width=0.5\textwidth]{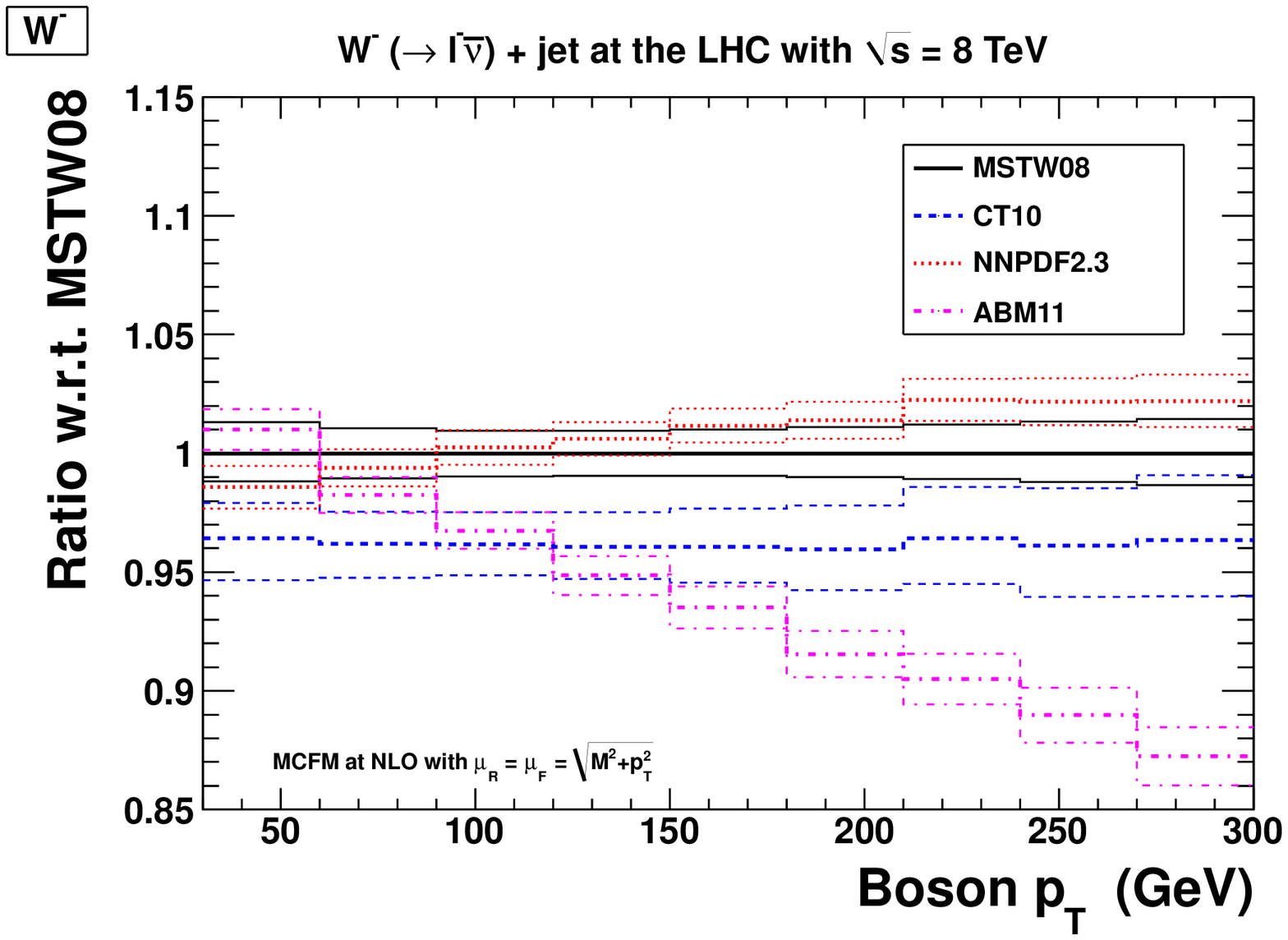}}
  \subfigure[$W^\pm$]{\includegraphics[width=0.5\textwidth]{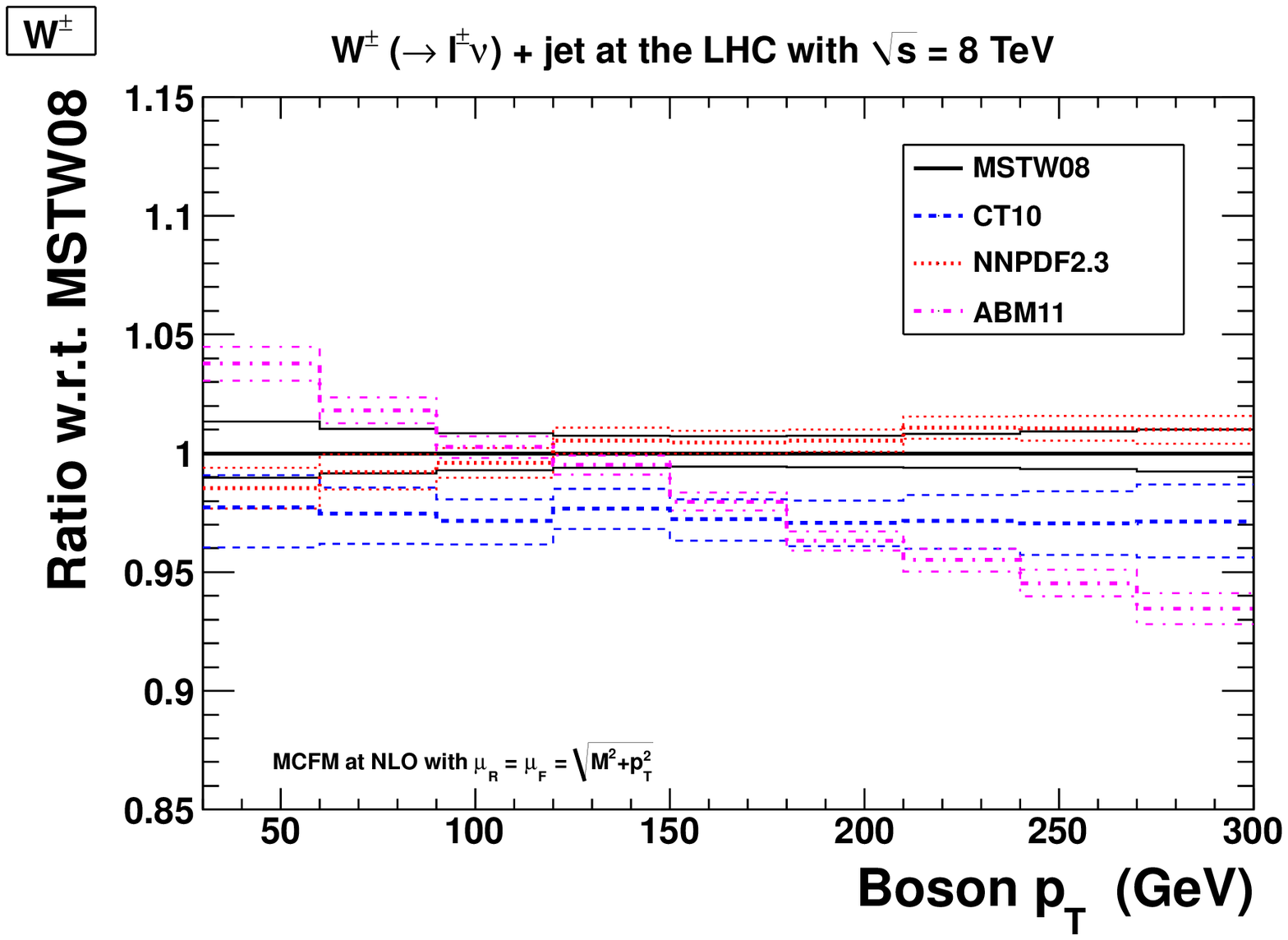}}%
  \subfigure[$Z^0$]{\includegraphics[width=0.5\textwidth]{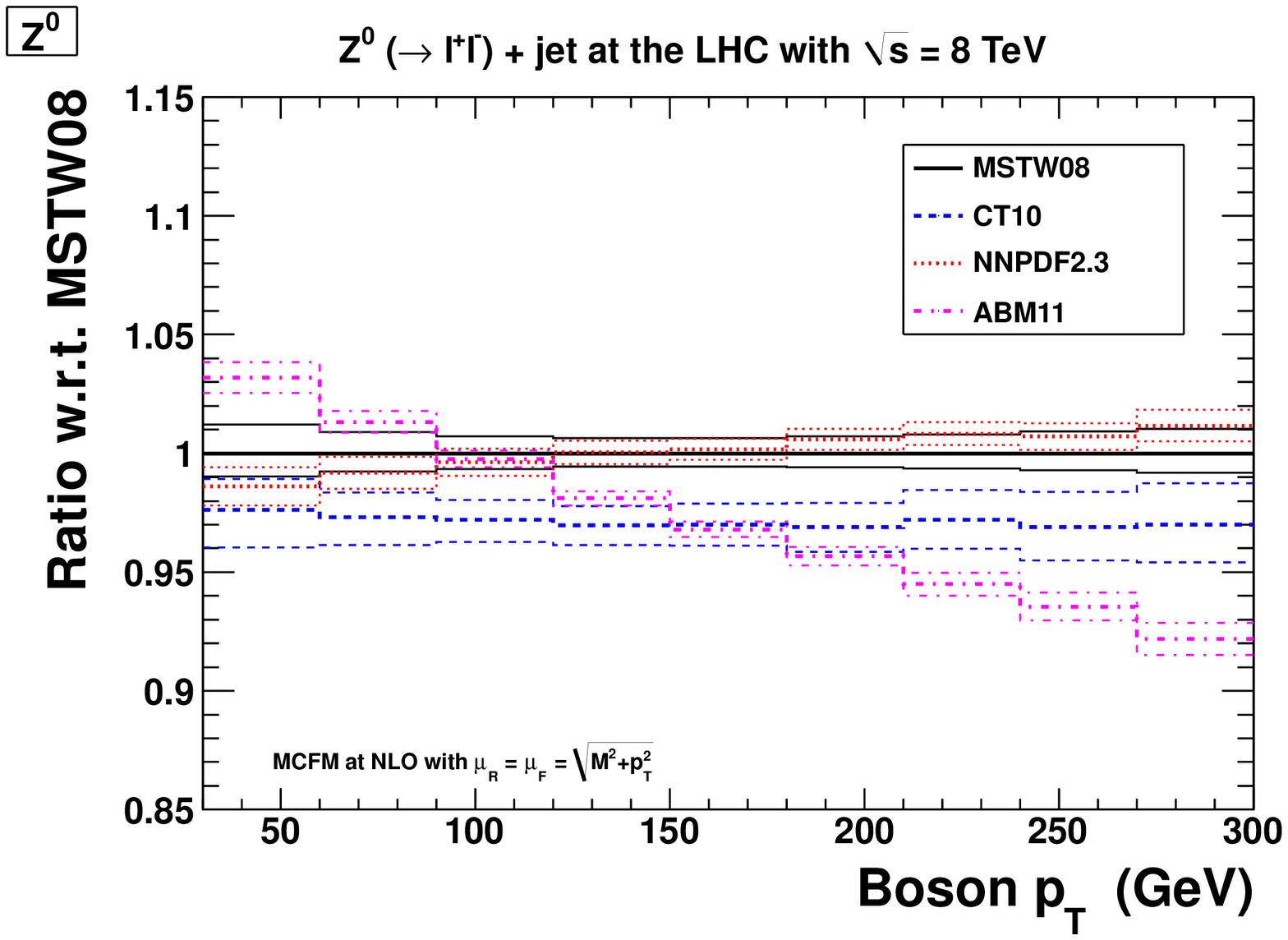}}
  \caption{Differential cross sections, ${\rm d}\sigma/{\rm d}p_T$, for the $V$+jet process as a function of boson $p_T$, taking the ratio to the MSTW08 prediction.  The thinner horizontal lines represent the PDF uncertainties for four choices of PDF set: MSTW08, CT10, NNPDF2.3 and ABM11.}
  \label{fig:PDFdsdpt}
\end{figure}
In figure~\ref{fig:PDFdsdpt} we show the ratio of differential cross sections, ${\rm d}\sigma/{\rm d}p_T$, with respect to the MSTW08 prediction for the different PDF sets, for (a)~$V=W^+$, (b)~$V=W^-$, (c)~$V=W^\pm$ ($\equiv W^++W^-$) and (d)~$V=Z^0$.  The thinner horizontal lines on either side of the central prediction in each $p_T$ bin represent the PDF uncertainties for each of the four choices of PDF set.  The different trends between the $W^+$ and $W^-$ predictions reflect the different dominant parton configurations, $gu$ and $gd$, respectively; see figure~\ref{fig:flavour_WZ}(a,b).  The similarity of the trends between the $W^\pm$ and $Z^0$ predictions reflects the similarity of the initial-state flavour decomposition; see figure~\ref{fig:flavour_WZ}(c,d).  The very different ABM11 prediction compared to the other PDF sets directly reflects the very different gluon distribution; see figure~\ref{fig:pdfsvsx}(a).  Precise measurements of the differential cross sections, ${\rm d}\sigma/{\rm d}p_T$, could therefore potentially constrain the gluon distribution, provided that experimental uncertainties are sufficiently small.  The current problem of large theoretical uncertainties, as discussed in section~\ref{sec:bosonPTtheory}, may be brought under control by the future availability of a NNLO calculation for $V$+jet production.

\subsubsection{Ratios of boson $p_T$ distributions} \label{sec:bosonratioPDF}

\begin{figure}
  \centering
  \subfigure[$W^+/W^-$]{\includegraphics[width=0.5\textwidth]{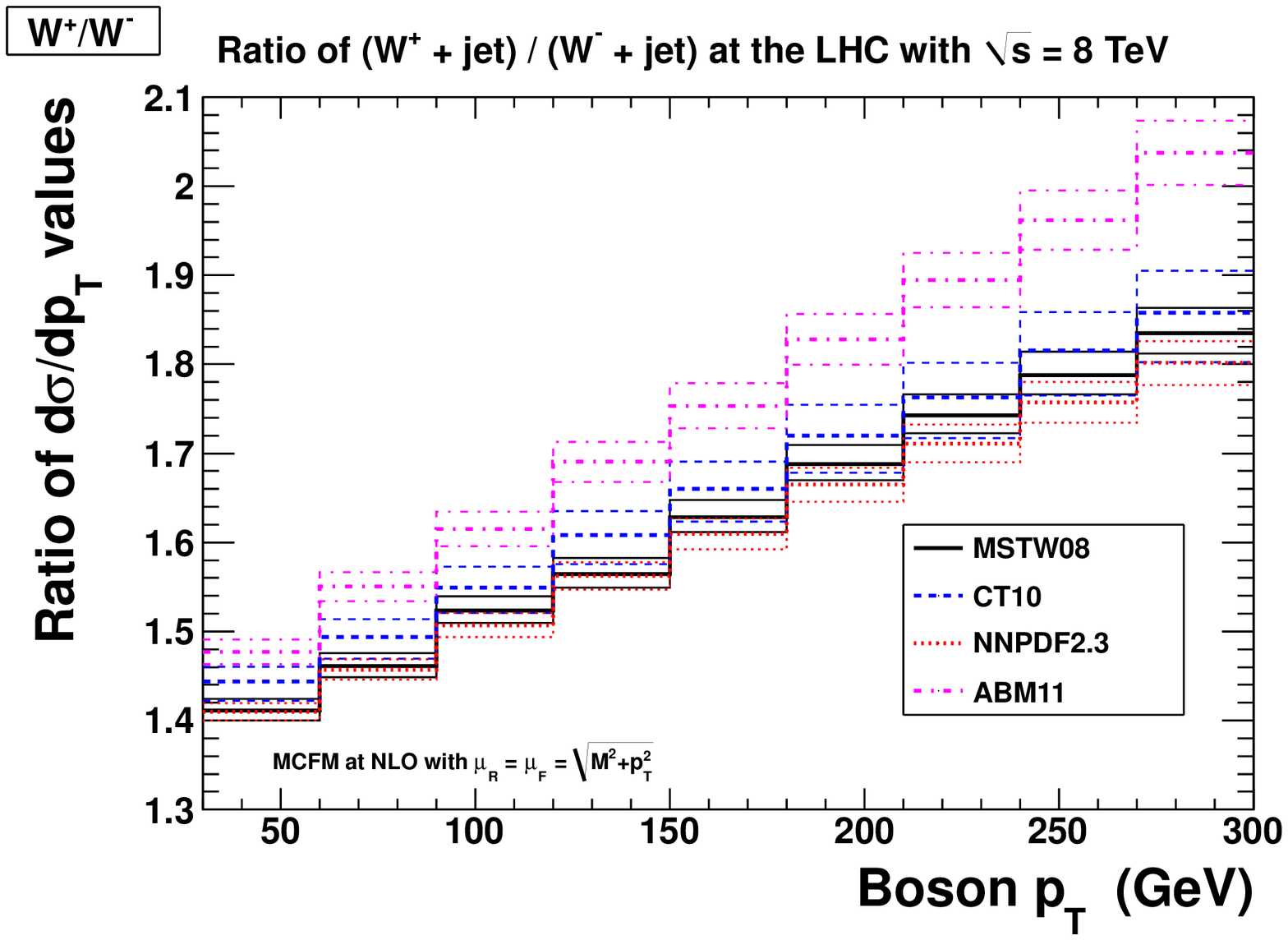}}%
  \subfigure[$W^+/Z^0$]{\includegraphics[width=0.5\textwidth]{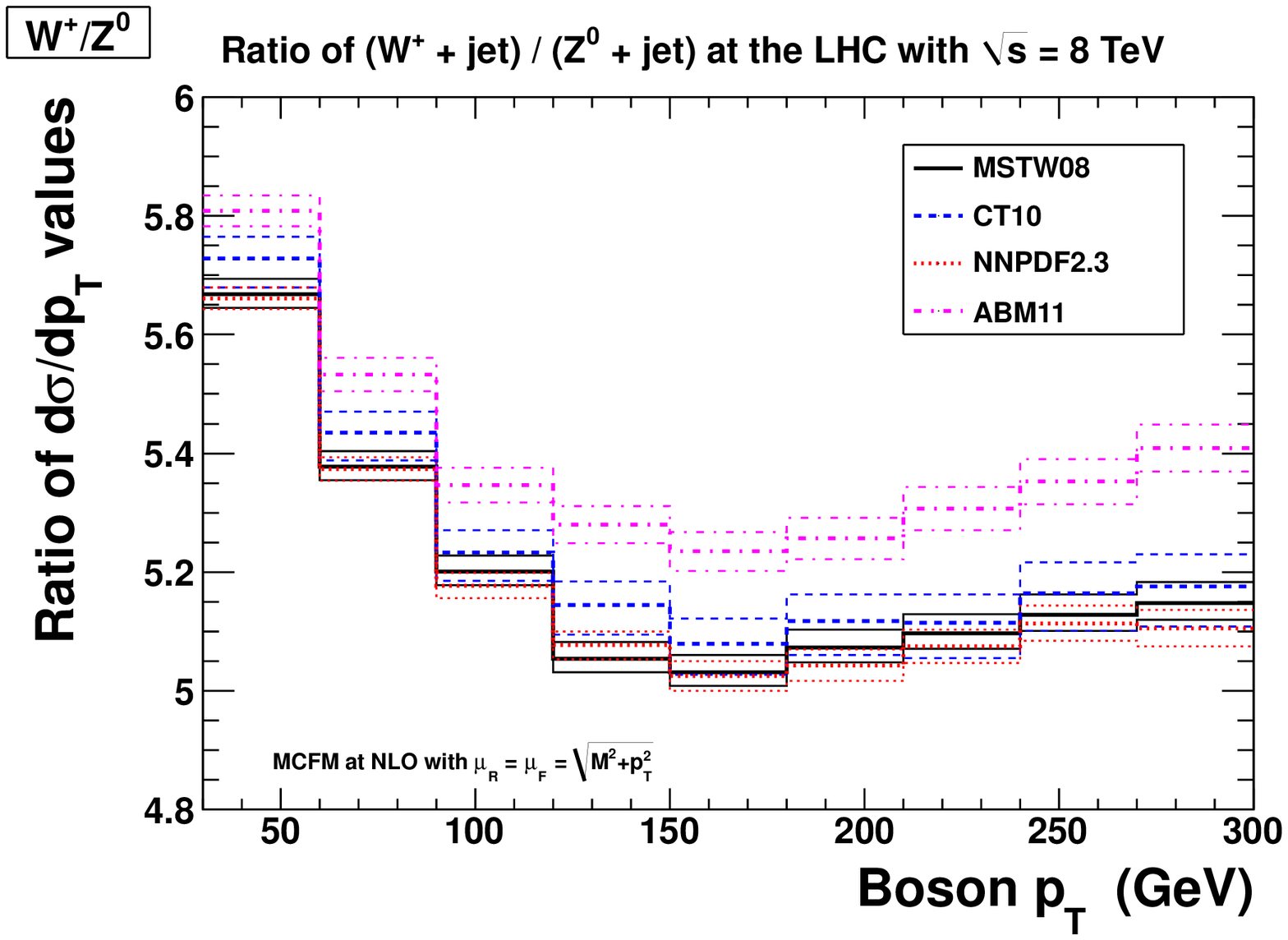}}
  \subfigure[$W^-/Z^0$]{\includegraphics[width=0.5\textwidth]{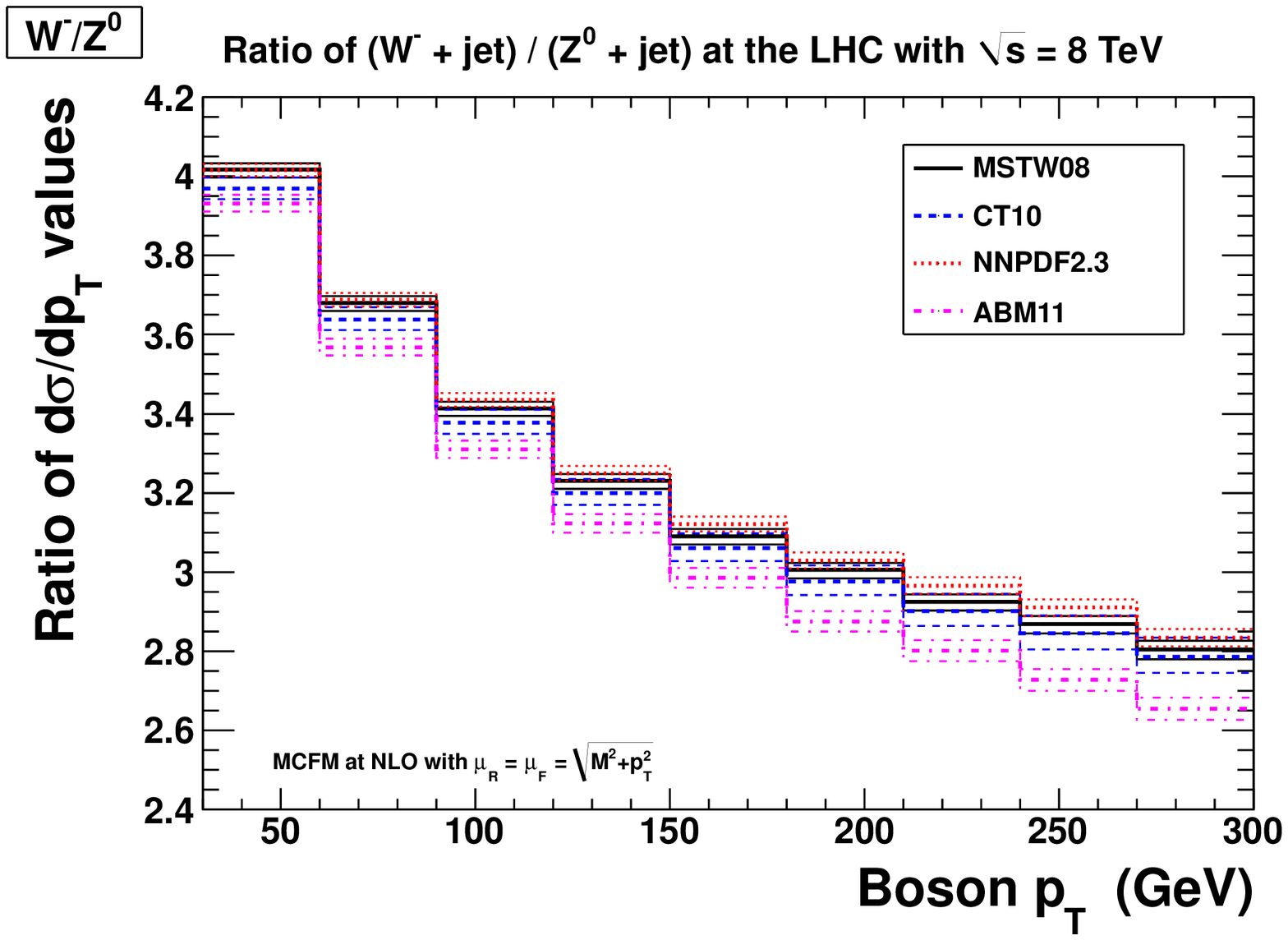}}%
  \subfigure[$W^\pm/Z^0$]{\includegraphics[width=0.5\textwidth]{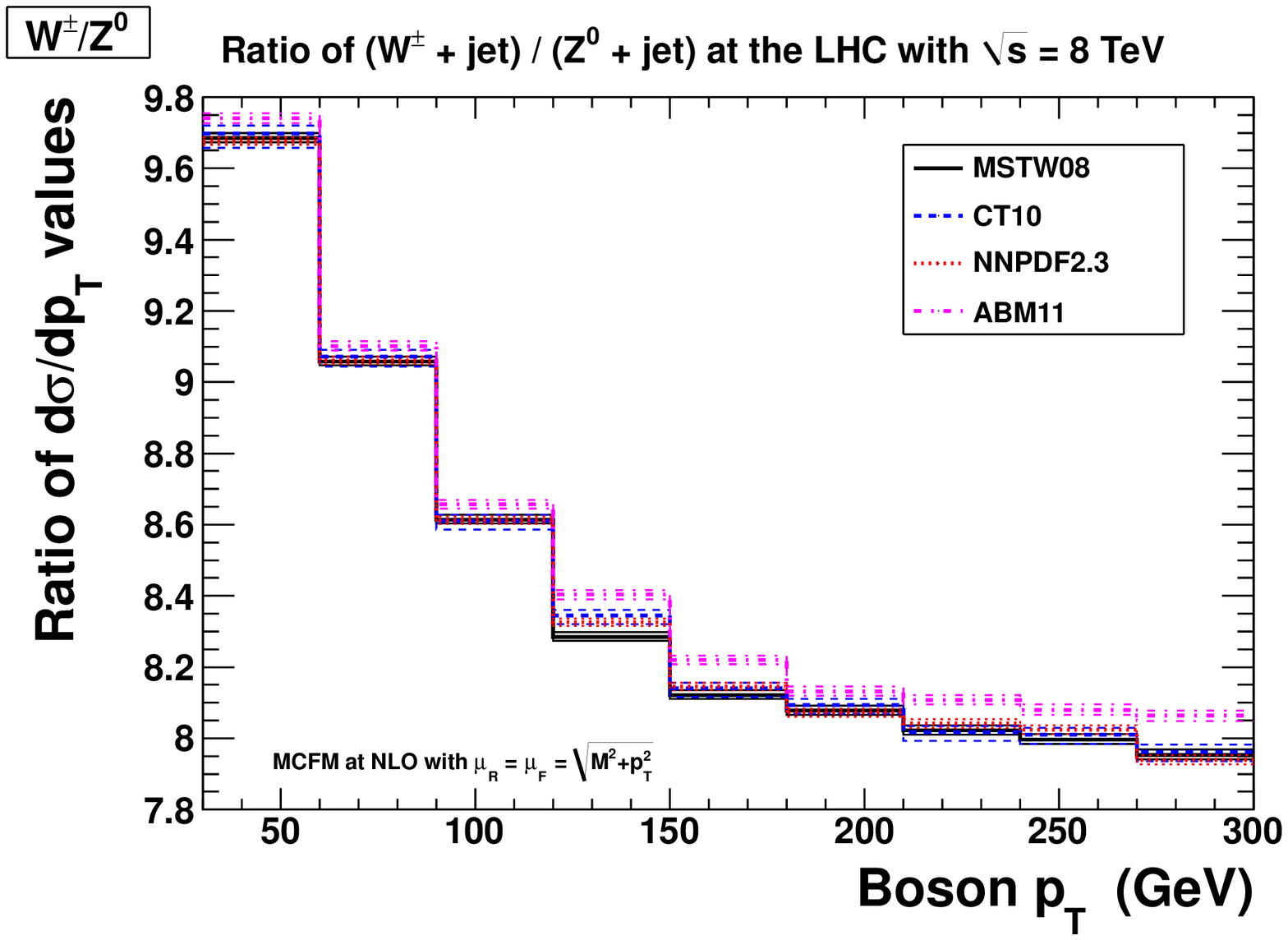}}
  \caption{Ratios of differential cross sections for the $V$+jet process as a function of boson $p_T$.  The thinner horizontal lines represent the PDF uncertainties for four choices of PDF set.}
  \label{fig:PDFRdsdpt}
\end{figure}
\begin{figure}
  \centering
  \subfigure[$W^+/W^-$]{\includegraphics[width=0.5\textwidth]{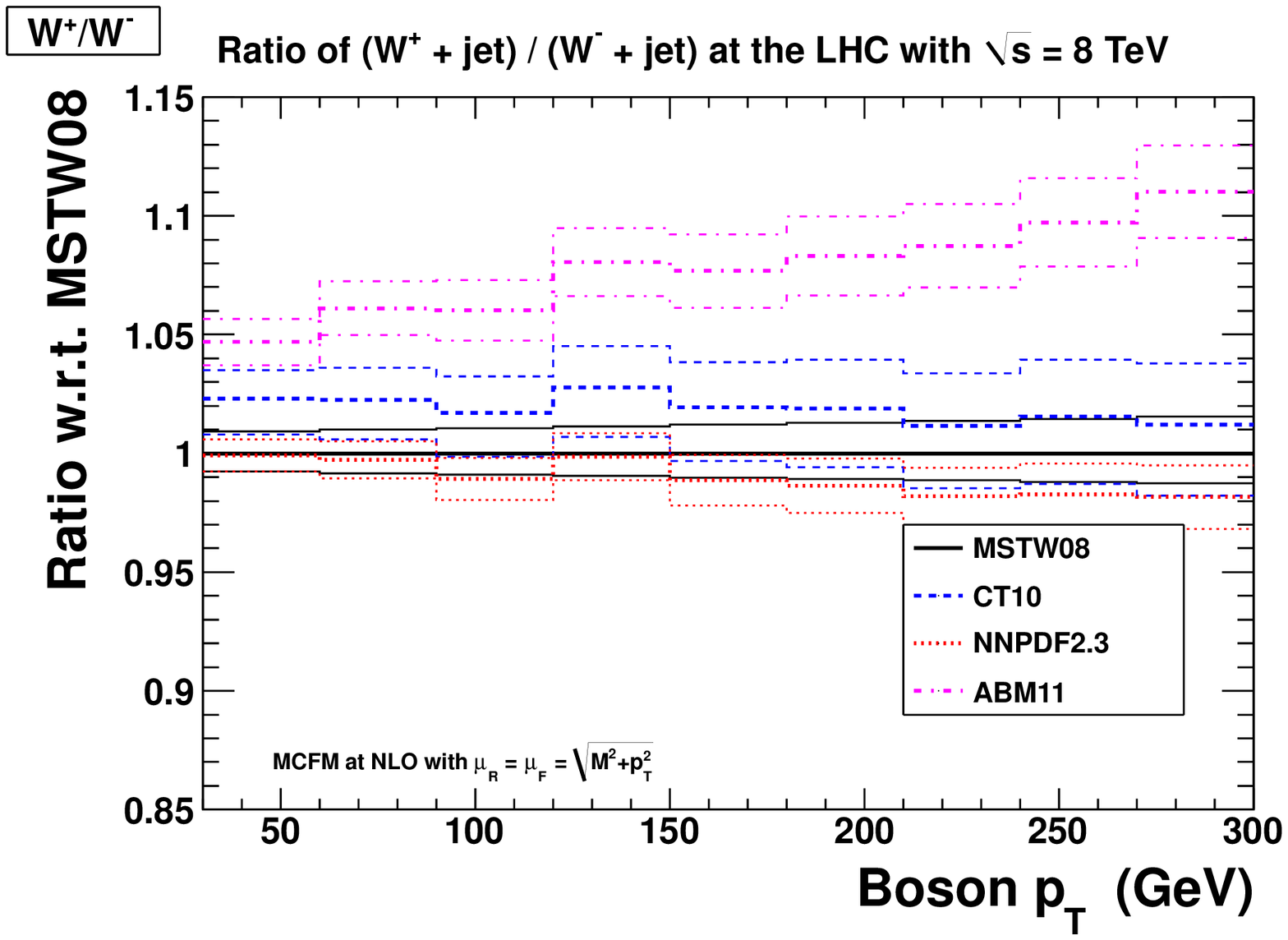}}%
  \subfigure[$W^+/Z^0$]{\includegraphics[width=0.5\textwidth]{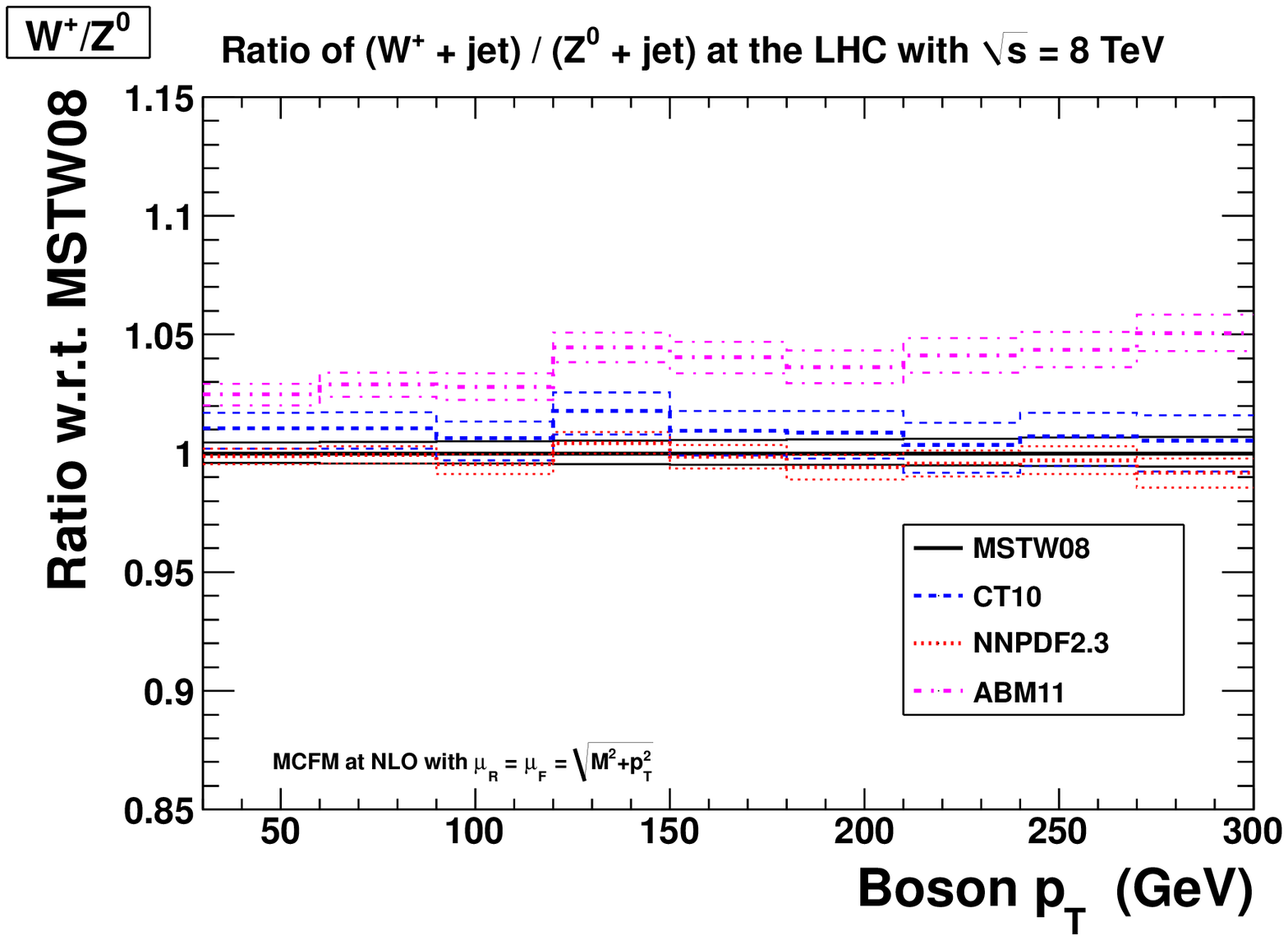}}
  \subfigure[$W^-/Z^0$]{\includegraphics[width=0.5\textwidth]{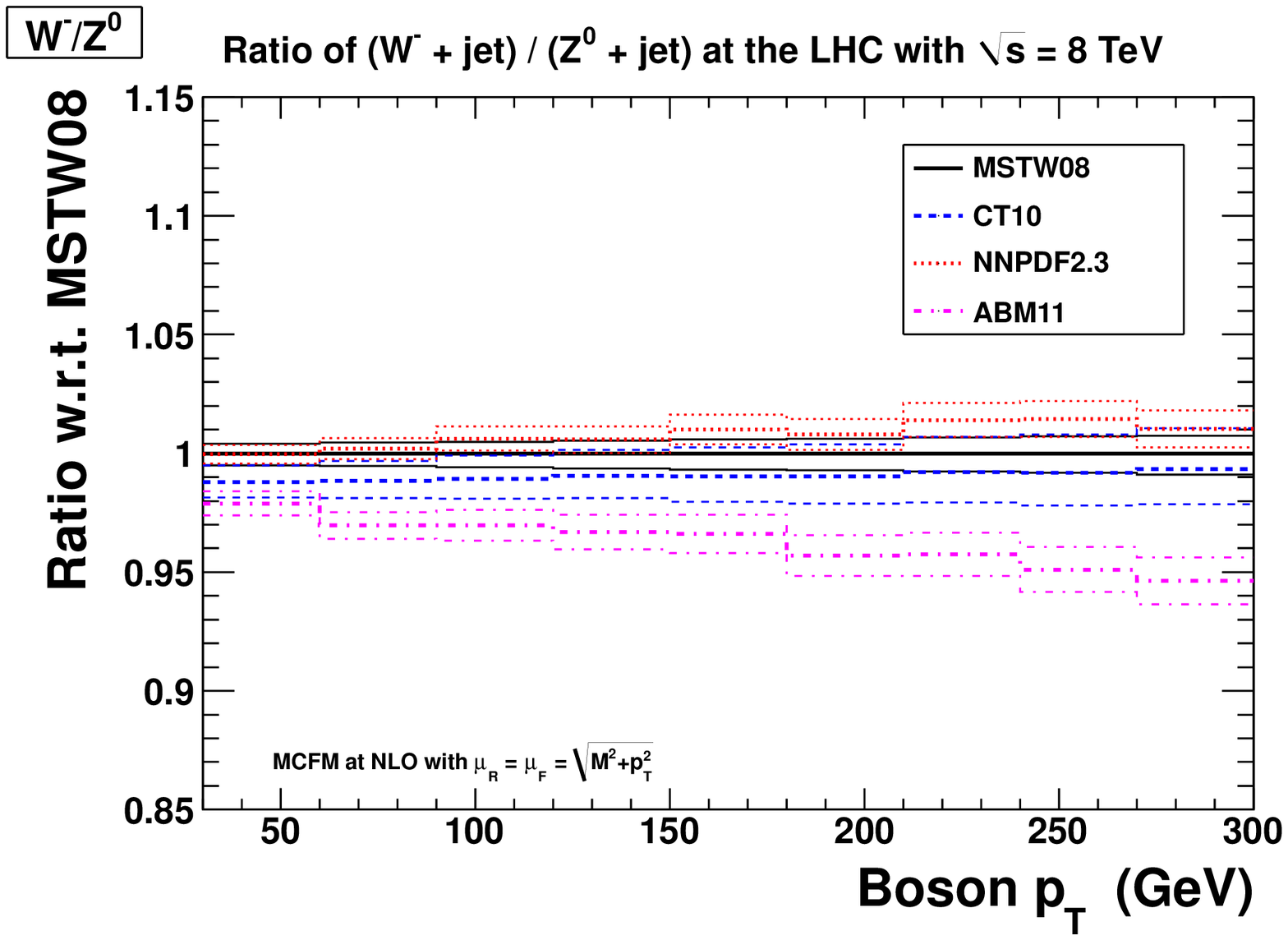}}%
  \subfigure[$W^\pm/Z^0$]{\includegraphics[width=0.5\textwidth]{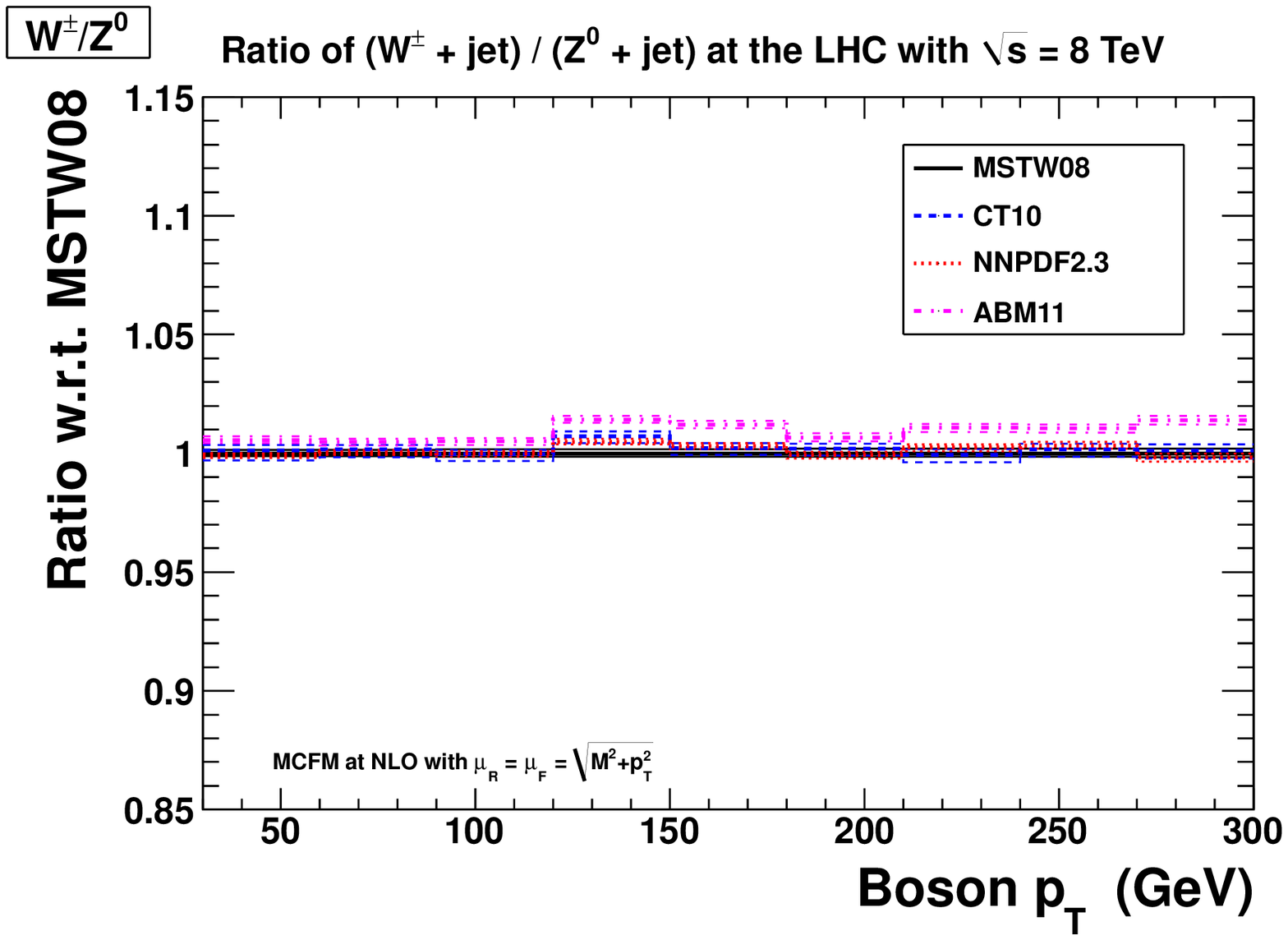}}
  \caption{Ratios of differential cross sections for the $V$+jet process as a function of boson $p_T$, normalised to the MSTW08 predictions.  The thinner horizontal lines represent the PDF uncertainties.}
  \label{fig:divPDFRdsdpt}
\end{figure}
In figure~\ref{fig:PDFRdsdpt} we show the cross-section ratios for different PDFs, and in figure~\ref{fig:divPDFRdsdpt} we show the same results normalised to the MSTW08 predictions.  It is clear that the $W^+/W^-$ ratio is the most sensitive to PDFs, with CT10 and NNPDF2.3 differing from the MSTW08 prediction by up to a couple of percent.  The difference between ABM11 and MSTW08 grows with increasing $p_T$, from 5\% to more than a 10\% difference in the considered $p_T$ range.
\begin{figure}
  \includegraphics[width=\textwidth]{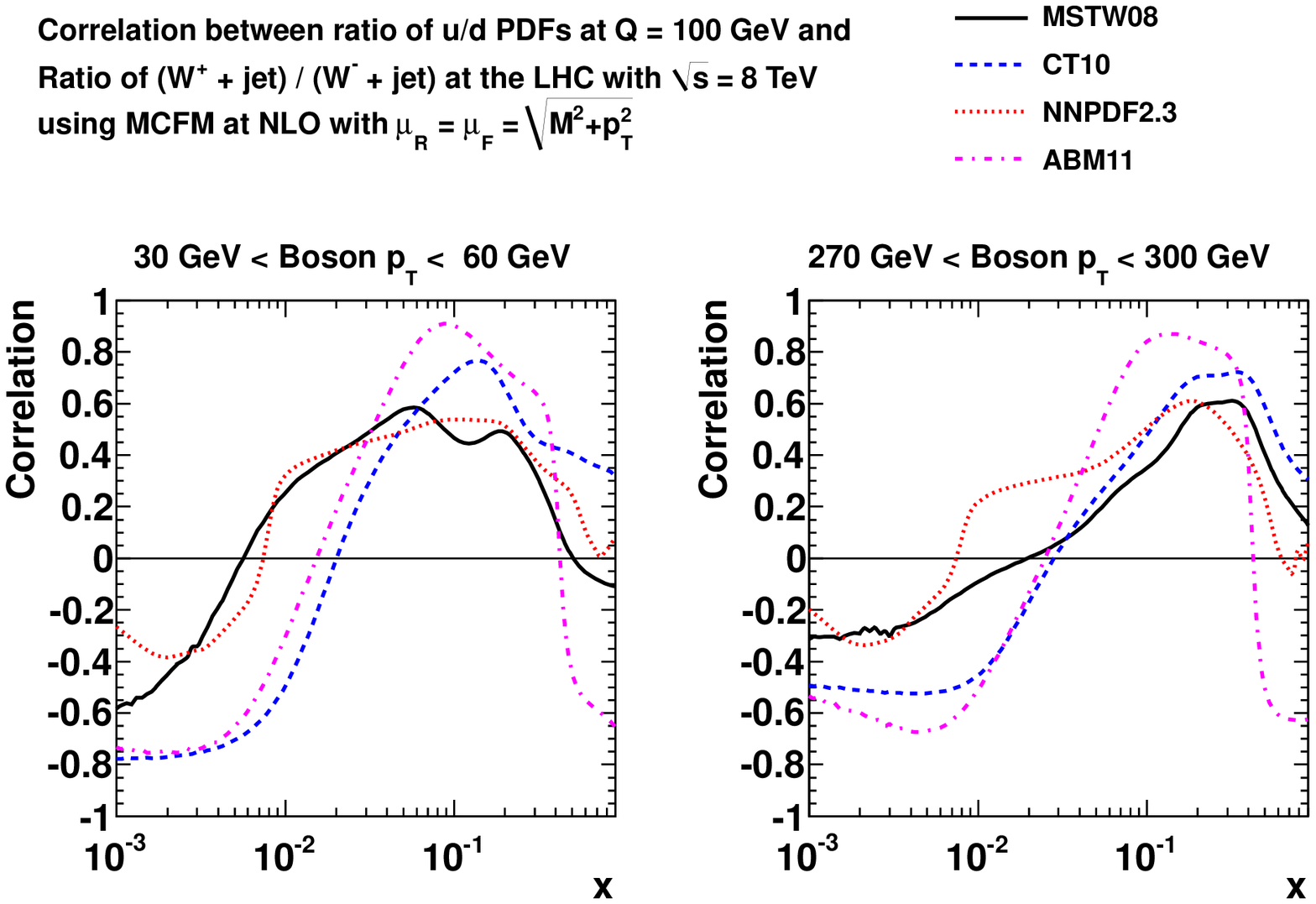}
  \caption{Correlation between ratio of $u/d$ PDFs and ratio of ($W^+$+jet)/($W^-$+jet) production.}
  \label{fig:correlation}
\end{figure}
In an attempt to understand the relevant $x$ values being probed, in figure~\ref{fig:correlation} we show the correlation (see, for example, ref.~\cite{Watt:2011kp}) for each of the four PDF sets between the $u/d$ ratio and the $W^+/W^-$ ratio for the smallest and largest boson $p_T$ bins.  Values close to $\{+1,0,-1\}$ mean that the two quantities are \{correlated, uncorrelated, anticorrelated\}, respectively.  The $x$ range corresponding to a strong correlation becomes slightly narrower and shifts to higher $x$ values as the boson $p_T$ is increased, with the maximum correlation being around $x\sim 0.1$ in the lower $p_T$ bin and around $x\sim 0.2$--$0.3$ in the higher $p_T$ bin, with some dependence on the particular PDF set due to different choices made in the various PDF fits (such as the rigidity of the input parameterisation and the data sets included).  Then we see that the trend between the different PDF sets for the $W^+/W^-$ ratio in figures~\ref{fig:PDFRdsdpt}(a) and \ref{fig:divPDFRdsdpt}(a) is a direct reflection of the $u/d$ ratio in the corresponding $x$ region shown in figure~\ref{fig:pdfsvsx}(b).

A crude simplified argument helps to understand the behaviour of the PDF dependence of the other cross-section ratios in figures~\ref{fig:PDFRdsdpt} and \ref{fig:divPDFRdsdpt}.  In terms of the dominant partonic configurations, we can write:
\begin{equation}
\sigma(W^++{\rm jet}) \sim g\,u,\quad \sigma(W^-+{\rm jet}) \sim g\,d, \quad \sigma(Z^0+{\rm jet}) \sim 0.29\,g\,u+0.37\,g\,d,
\end{equation}
where the numerical values in the last expression are the appropriate sums of the squares of the vector and axial-vector couplings of the $Z^0$ boson to quarks.  Then whereas the $W^+/W^-$ ratio probes the $u/d$ ratio, the $W^\pm/Z^0$ ratio behaves as:
\begin{equation} \label{eq:wzratio}
\frac{\sigma(W^++{\rm jet})+\sigma(W^-+{\rm jet})}{\sigma(Z^0+{\rm jet})}\sim \frac{u+d}{0.29\,u+0.37\,d},
\end{equation}
after cancelling the common factor of the gluon distribution in the numerator and denominator.  The combination of $u$- and $d$-quark contributions is therefore very similar for $W^\pm+{\rm jet}$ and $Z^0+{\rm jet}$, as already seen in figure~\ref{fig:flavour_WZ}(c,d), and so the PDF dependence almost cancels entirely in the $W^\pm/Z^0$ ratio.  Indeed, taking the envelope of the MSTW08, CT10 and NNPDF2.3 predictions in figure~\ref{fig:divPDFRdsdpt}(d) gives a spread of less than 0.5\%, while also including ABM11 would give a spread of about 1\%.  The separate $W^+/Z^0$ and $W^-/Z^0$ ratios retain some sensitivity to the $u/d$ ratio of PDFs, but not as much as the $W^+/W^-$ ratio; see figures~\ref{fig:PDFRdsdpt} and \ref{fig:divPDFRdsdpt}.  Similar arguments have been made to explain the PDF dependence of the inclusive $W^\pm$ and $Z^0$ cross sections in refs.~\cite{Watt:2011kp,Forte:2013wc}.  Note that by writing the $W^\pm/Z^0$ ratio (and the separate $W^+/Z^0$ and $W^-/Z^0$ ratios) given in eq.~\eqref{eq:wzratio} as a function of the $u/d$ ratio, which increases with increasing $x$ and hence with increasing $p_T$, it is possible to infer the limiting behaviour at very large $p_T\gg M_{W,Z}$.  Then we find that the $W^\pm/Z^0$ ratio will very slowly increase (although it is almost constant) with increasing $p_T$, while the $W^+/Z^0$ ratio will increase a little more rapidly, and the $W^-/Z^0$ ratio will slightly decrease with increasing $p_T$.

\subsubsection{Potential for PDF constraints}

\begin{figure}
  \includegraphics[width=\textwidth]{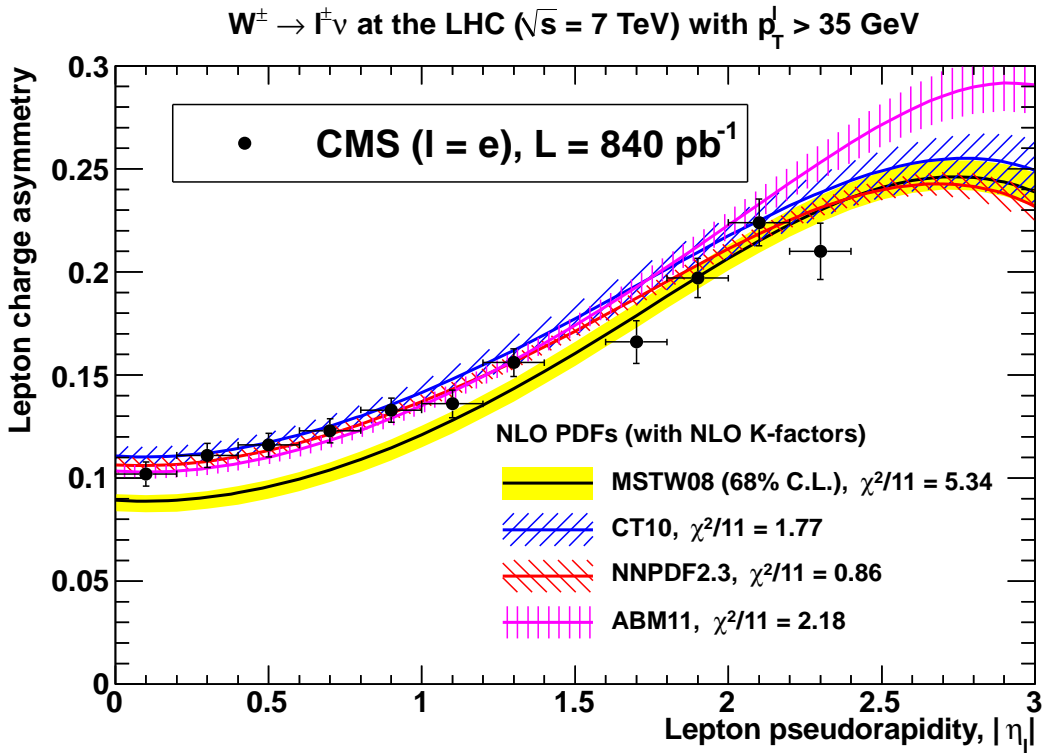}
  \caption{Description of the CMS $W\to e\nu_e$ charge asymmetry data~\cite{Chatrchyan:2012xt} for four PDF sets.}
  \label{fig:cmsasymmetry}
\end{figure}
Perhaps the most discriminating data set to emerge so far from the LHC with respect to providing a PDF constraint is the $W^\pm(\to\ell^\pm\nu)$ charge asymmetry measured as a function of the charged-lepton pseudorapidity ($\eta_\ell$), defined as:
\begin{equation}
  A_\ell(\eta_\ell) = \frac{{\rm d}\sigma(\ell^+)/{\rm d}\eta_\ell-{\rm d}\sigma(\ell^-)/{\rm d}\eta_\ell}{{\rm d}\sigma(\ell^+)/{\rm d}\eta_\ell+{\rm d}\sigma(\ell^-)/{\rm d}\eta_\ell}.
\end{equation}
In figure~\ref{fig:cmsasymmetry} we show the CMS electron charge asymmetry data~\cite{Chatrchyan:2012xt} compared to the predictions of the four NLO PDF sets.  The NLO $K$-factors for ${\rm d}\sigma/{\rm d}\eta_\ell$ are derived using the \textsc{dynnlo} code~\cite{Catani:2009sm}, as discussed in ref.~\cite{Watt:2012tq} and used previously in ref.~\cite{Martin:2012da}.  The goodness-of-fit for the central prediction of each PDF set, $\chi^2$, is calculated simply by adding all experimental uncertainties in quadrature and is indicated in the legend of figure~\ref{fig:cmsasymmetry}.  The worst description of the CMS data is given by the MSTW08 PDF set, particularly at central pseudorapidity values, where the $u/d$ ratio, or more precisely the $u_v-d_v$ difference of valence-quark distributions, is being probed at $x\sim M_W/\sqrt{s}\sim 0.01$.  Indeed, we already saw from figure~\ref{fig:pdfsvsx}(b) that the $u/d$ ratio from MSTW08 at $x\sim 0.01$ lies well below the values predicted by the other PDF groups.  This discrepancy has been resolved by allowing an extended Chebyshev parameterisation form for the fitted input PDFs and more flexible deuteron corrections in a variant of the MSTW08 fit~\cite{Martin:2012da}.  However, we note from figure~\ref{fig:cmsasymmetry} that the ABM11 prediction is much higher than the other PDF sets for $|\eta_\ell|\gtrsim2.5$, beyond the limit of the CMS data (although larger $|\eta_\ell|$ values can be measured by LHCb~\cite{Aaij:2012vn}).  This region is probing PDFs at large $x$ that could instead be accessed by measuring the $W^\pm(\to\ell^\pm\nu)$ charge asymmetry at large boson $p_T$, as we have shown in section~\ref{sec:bosonratioPDF}.  Therefore, measuring the $W^+/W^-$ ratio as a function of boson $p_T$ provides complementary information on the $u/d$ ratio to measuring as a function of $\eta_\ell$.  Another way to access higher $x$ values for the $u/d$ ratio might be to measure the charge asymmetry of high-mass virtual $W^\pm(\to\ell^\pm\nu)$ production, that is, in the region of $M_{\ell\nu}>M_W$.

Note that the necessity to measure the $W^\pm(\to\ell^\pm\nu)$ charge asymmetry as a function of the $\eta_\ell$ variable rather than the preferable $W$ rapidity, which has a closer connection to the initial parton kinematics, arises because the $W$ rapidity cannot be unambiguously reconstructed experimentally due to the unknown longitudinal momentum of the decay neutrino.  However, this problem does not arise when reconstructing the $W$ $p_T$.

In principle, measuring $W^+$, $W^-$ and $Z^0$ differential cross sections, ${\rm d}\sigma/{\rm d}p_T$, and presenting all information on correlated systematic uncertainties, would implicitly include information on cross-section ratios, as done by ATLAS for inclusive $V$ production~\cite{Aad:2011dm}.  However, if directly including the ${\rm d}\sigma/{\rm d}p_T$ observables in a PDF fit, the issue of how to consistently account for possibly large theoretical uncertainties due to electroweak and missing NNLO QCD corrections would need to be addressed.  Therefore, at the present time it is better to measure cross-section ratios explicitly, taking account of all correlations between systematic uncertainties during the experimental measurement.

The $u/d$ ratio at larger values of $x$ can also be probed via the inclusive $W^\pm(\to\ell^\pm\nu)$ asymmetry at the lower centre-of-mass energy ($\sqrt{s} = 1.96$~TeV) of the Tevatron $p\bar{p}$ collider.  However, there have been some problems with the consistency of the existing Tevatron data, particularly when the data are split into bins of the charged-lepton transverse momentum, $p_T^\ell$ (see, for example, refs.~\cite{Watt:2010qt,Thorne:2010kj}).

Using the $W^+/W^-$ ratio at large boson $p_T$ to probe the $u/d$ ratio at large $x$ has the advantage that it is independent of deuteron corrections currently needed for structure functions measured in deep-inelastic scattering from old fixed-target experiments~\cite{Martin:2012da,Ball:2013gsa}.  The $W^+/W^-$ ratio measured as a function of boson $p_T$ could therefore be an important ingredient in a future PDF fit using only `collider' data, or only HERA and LHC data, excluding data from the older fixed-target experiments.

Although the inclusive $W^\pm/Z^0$ ratio is insensitive to PDF uncertainties arising from up- and down-quark distributions, in a similar way to the $(W^\pm+{\rm jet})/(Z^0+{\rm jet})$ ratio, a sensitivity to the strange-quark PDF has been observed~\cite{Aad:2012sb}.  This sensitivity arises from the different combinations $\bar{s}\,c\to W^+$ and $s\,\bar{c}\to W^-$ compared to $s\,\bar{s}\to Z^0$ and $c\,\bar{c}\to Z^0$, and hence also dependence on the charm-quark PDF can be probed~\cite{Halzen:2013bqa}.  But for $V$+jet production, the dependence on the strange-quark (and charm-quark) PDF cancels more completely in the $W^\pm/Z^0$ ratio, because the dominant initial-state combinations are the same, namely $g\,\bar{s}\to W^+\,\bar{c}$, $g\,c\to W^+\,s$, $g\,s\to W^-\,c$ and $g\,\bar{c}\to W^-\,\bar{s}$, compared to $g\,\bar{s}\to Z^0\,\bar{s}$, $g\,c\to Z^0\,c$, $g\,s\to Z^0\,s$ and $g\,\bar{c}\to Z^0\,\bar{c}$.  Moreover, configurations involving initial-state strange and charm quarks are a much smaller contribution to the total for $V$+jet production compared to inclusive $V$ production.  To directly probe the strange (and charm) contents of the proton, the $V$+charm process can be studied~\cite{Stirling:2012vh,ATLAS-CONF-2013-045,Chatrchyan:2013uja}, where the dominant LO processes are $g\,\bar{s}\to W^+\,\bar{c}$, $g\,s\to W^-\,c$, $g\,c\to Z^0\,c$ and $g\,\bar{c}\to Z^0\,\bar{c}$.  Then the same cross-section ratios can be measured as in the present study ($W^+/W^-$, $W^+/Z^0$, $W^-/Z^0$ and $W^\pm/Z^0$), but in the presence of a charm-tagged jet, and also ratios like $(V+{\rm charm})/(V+{\rm jet})$.  Results are available from ATLAS~\cite{ATLAS-CONF-2013-045} and CMS~\cite{Chatrchyan:2013uja} for the $W+{\rm charm}$ cross section and the $(W^++\bar{c})/(W^-+c)$ ratio measured as a function of the pseudorapidity $\eta_\ell$ of the charged-lepton from the $W$ decay.  It would be interesting to also measure these ratios as a function of the boson $p_T$ to probe the PDFs at larger $x$ values.

\subsection{Higher-order electroweak corrections}

Higher-order electroweak corrections to the boson $p_T$ distributions have been calculated for on-shell $Z^0$~\cite{Maina:2004rb,Kuhn:2004em,Kuhn:2005az} and $W^\pm$~\cite{Kuhn:2007qc,Hollik:2007sq,Kuhn:2007cv} bosons, and also for off-shell $W^\pm$~\cite{Denner:2009gj} and $Z^0$~\cite{Denner:2011vu} bosons, in the latter case taking into account the leptonic decays and finite-width effects.  These corrections can reach up to a few tens of percent at very large jet/boson $p_T$, due to large virtual electroweak Sudakov logarithms of $\hat{s}/M_V^2$, where $\sqrt{\hat{s}}$ is the partonic centre-of-mass energy; see, for example, figure~9 (middle-left) of ref.~\cite{Denner:2009gj} for $W^+(\to\ell^+\nu)$+jet production or figure~6 (middle-left) of ref.~\cite{Denner:2011vu} for $Z^0(\to\ell^+\ell^-)$+jet production.  The electroweak corrections are similar for $W^\pm$ and $Z^0$ production, and hence will largely cancel in the cross-section ratios.  While the electroweak corrections cancel almost completely in the $W^+/W^-$ ratio (see figure~10 of ref.~\cite{Kuhn:2007cv}), the effect of electroweak corrections can still decrease the $W^+/Z^0$ and $W^-/Z^0$ ratios (and hence the $W^\pm/Z^0$ ratio) by 4\% at boson $p_T=1$~TeV and by 7\% at boson $p_T=2$~TeV at the 14~TeV LHC (see figure~11 of ref.~\cite{Kuhn:2007cv}), although the shift will be smaller for the lower $p_T$ values in the present study.  Moreover, these shifts are still smaller than the electroweak corrections to the $\gamma/Z^0$ ratio, which is increased by 13\% at boson $p_T=1$~TeV and by 22\% at boson $p_T=2$~TeV at the 14~TeV LHC (see figure~7 of ref.~\cite{Kuhn:2005gv}), and this is a sizeable contribution to the total theoretical uncertainty when the $\gamma$+jets process is used to estimate the $Z^0(\to\nu\bar{\nu})$+jets background~\cite{Bern:2012vx}.  Note that for sufficiently inclusive measurements, the real emission of soft electroweak bosons may partially cancel the effect of the virtual electroweak Sudakov logarithms~\cite{Baur:2006sn,Stirling:2012ak}.  The extent of this cancellation would need to be studied for realistic experimental cuts appropriate to the measurement, and taking into account whether diboson production is considered to be a background to the measurement.  In ref.~\cite{Bern:2012vx} it was found that after imposing typical cuts used in new physics searches, the real electroweak corrections from emission of an extra $W$ or $Z$ boson had a negligible effect (1\% or less) on the $\gamma/Z^0$ ratio.  Similar findings might be expected for the effect of real electroweak corrections on the $W^\pm/Z^0$ ratio with typical cuts used in new physics searches.

In addition to the dominant electroweak corrections to the $V$+jet process arising from the Sudakov logarithms mentioned above, there are of course corrections needed for electroweak radiation off the final-state decay leptons, which will clearly be different for $W^\pm\to\ell^\pm\nu$ (with only one charged lepton) compared to $Z^0\to\ell^+\ell^-$ (with two charged leptons), and hence will modify the $W^\pm/Z^0$ ratio.  In existing measurements of the boson $p_T$ distributions~\cite{Aad:2011fp,Aad:2011gj,Chatrchyan:2011wt}, the data are generally corrected for QED final-state radiation, with a systematic uncertainty assigned to the correction procedure.  However, the ATLAS measurements~\cite{Aad:2011fp,Aad:2011gj} are also presented without correcting for QED final-state radiation, then it would be possible to compare to calculations which include these effects explicitly~\cite{Denner:2009gj,Denner:2011vu}.  The numerical size of these QED corrections to the normalised differential cross sections is typically only a few percent at large boson $p_T$ values~\cite{Aad:2011fp,Aad:2011gj}.

\section{Dependence of ratios on centre-of-mass energy} \label{sec:CoM}

\begin{figure}
  \centering
  \subfigure[$W^+/W^-$]{\includegraphics[width=0.5\textwidth]{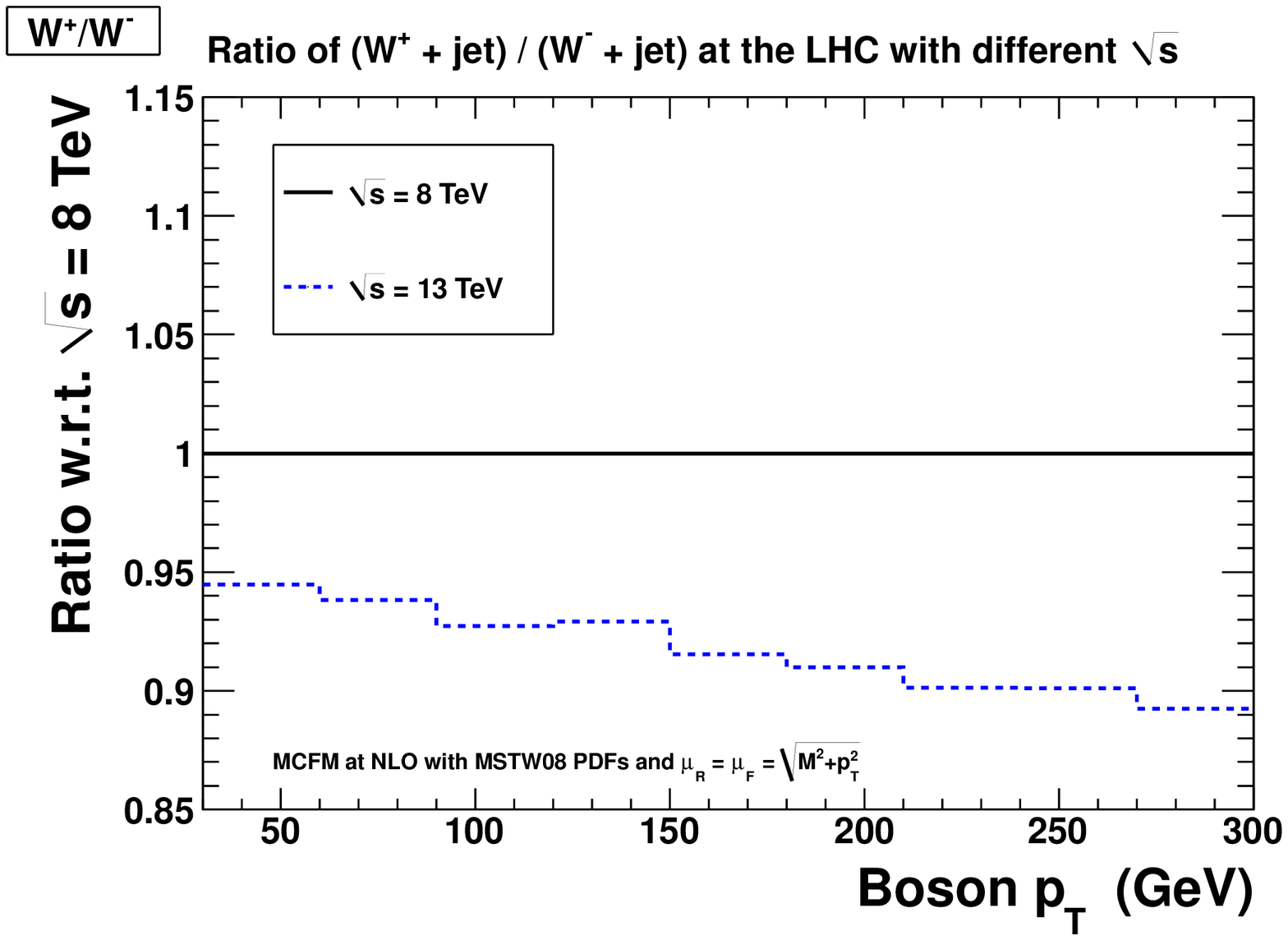}}%
  \subfigure[$W^+/Z^0$]{\includegraphics[width=0.5\textwidth]{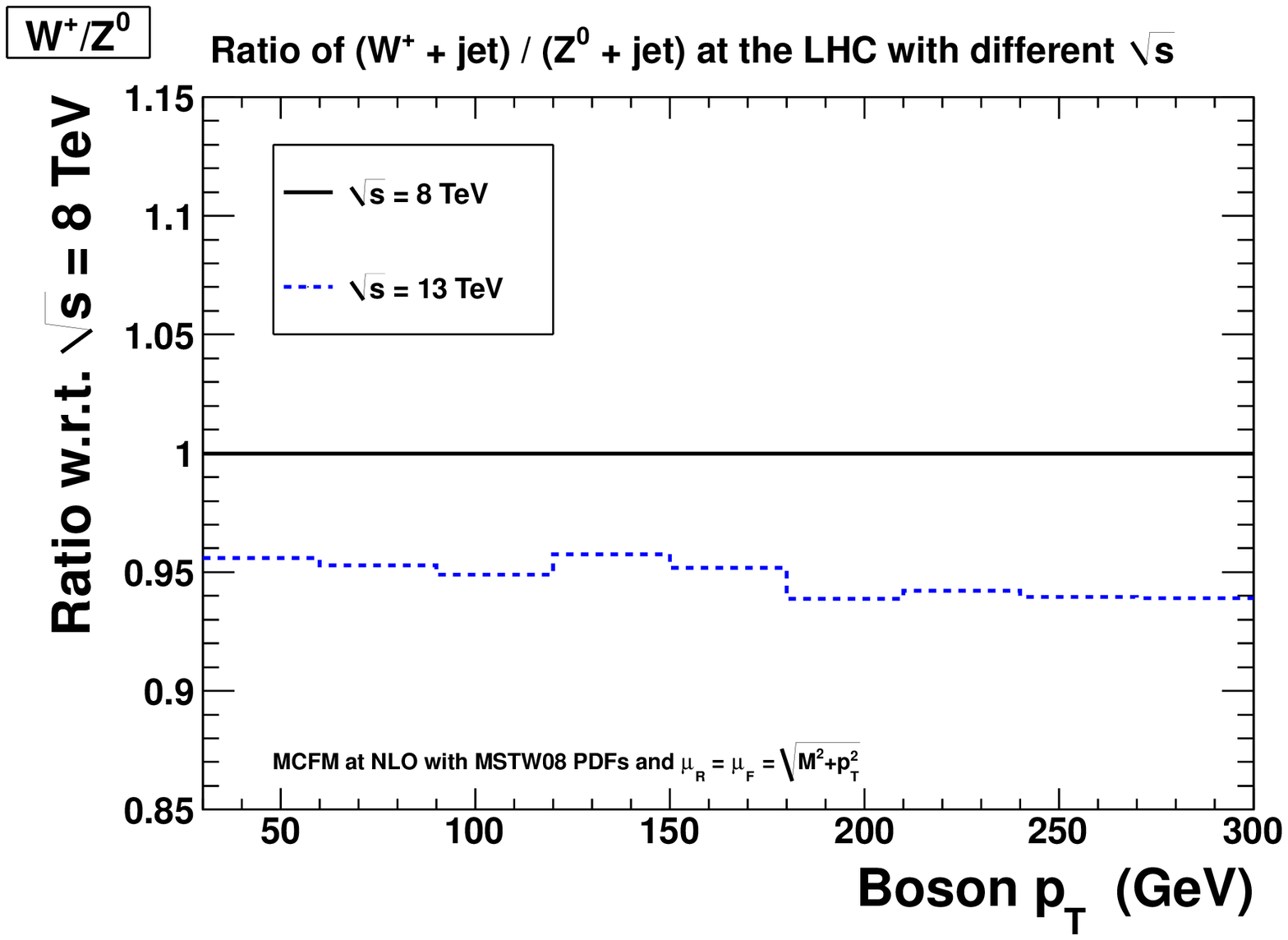}}
  \subfigure[$W^-/Z^0$]{\includegraphics[width=0.5\textwidth]{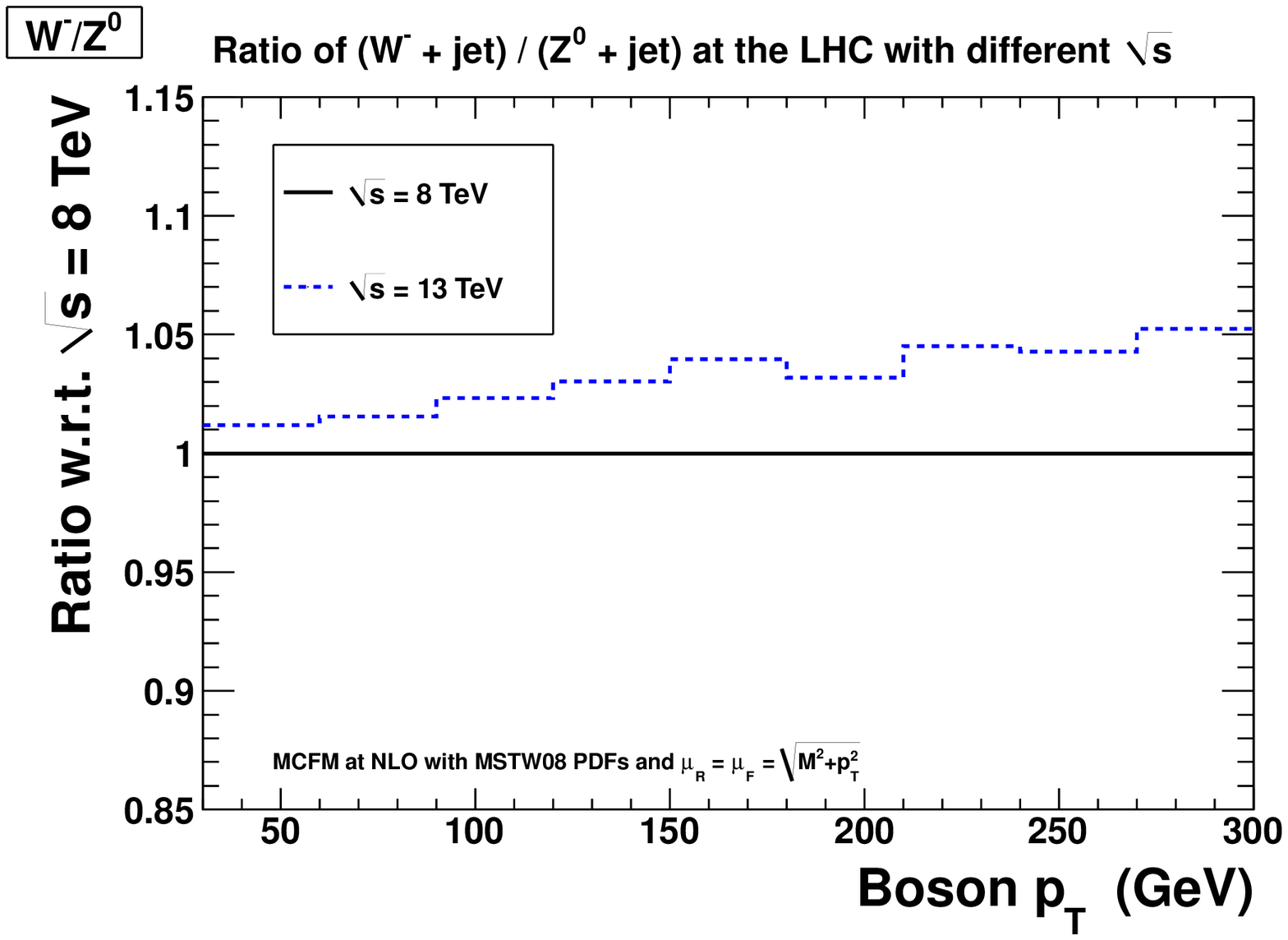}}%
  \subfigure[$W^\pm/Z^0$]{\includegraphics[width=0.5\textwidth]{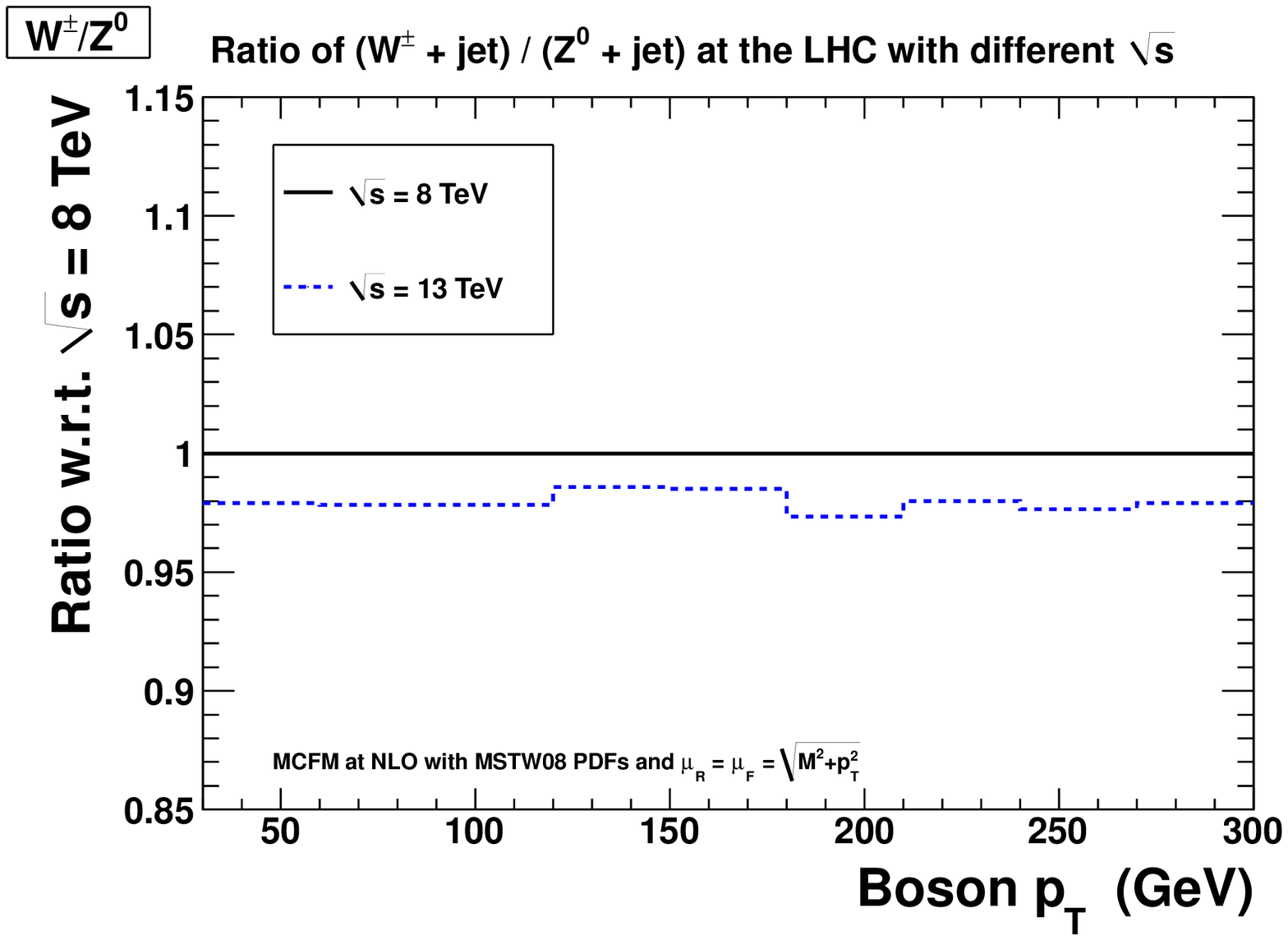}}
  \caption{Cross-section ratios for the $V$+jet process as a function of boson $p_T$ with $\sqrt{s} = 13$~TeV, normalised to the result at $\sqrt{s} = 8$~TeV, for (a)~$W^+/W^-$, (b)~$W^+/Z^0$, (c)~$W^-/Z^0$ and (d)~$W^\pm/Z^0$.}
  \label{fig:divRdsdpt13TeV}
\end{figure}
In figure~\ref{fig:divRdsdpt13TeV} we show the cross-section ratios at an LHC centre-of-mass energy of 13~TeV, normalised to the corresponding ratios at 8~TeV, as predicted by \textsc{mcfm}~\cite{Campbell:2010ff} at NLO using the best-fit MSTW 2008 NLO PDF set~\cite{Martin:2009iq}.  Increasing the LHC centre-of-mass energy will allow measurements of boson $p_T$ distributions, and their ratios, to be made at higher values of the boson $p_T$.  Recall that the two momentum fractions probed in the PDFs satisfy $x_1\,x_2=\hat{s}/s$, where $\sqrt{\hat{s}}$ and $\sqrt{s}$ are the partonic and hadronic centre-of-mass energies, respectively, and so typical $x$ values are given by $x\sim \sqrt{\hat{s}/s}$.  Therefore, increasing $\sqrt{s}$ from 8~TeV to 13~TeV with a fixed value of the boson $p_T$ decreases the typical $x$ values by a factor of $13/8$.  The $u/d$ ratio of PDFs is smaller at lower $x$ values, therefore the $W^+/W^-$ and $W^+/Z^0$ ratios are smaller at 13~TeV than at 8~TeV, while the $W^-/Z^0$ ratio is larger.  The $W^+/W^-$ ratio is most sensitive to PDFs and so is affected the most, while the $W^\pm/Z^0$ ratio is least affected, decreasing by only 2\% independent of boson $p_T$.  We expect our conclusions regarding the theoretical uncertainties on cross-section ratios at $\sqrt{s}=8$~TeV to be valid also at $\sqrt{s}=13$~TeV.  In particular, the relative size of the electroweak corrections has been found to hardly differ when the centre-of-mass energy is varied~\cite{Denner:2011vu}.

Ratios of various observables measured at different centre-of-mass energies have been studied in detail in ref.~\cite{Mangano:2012mh}.  Note in particular that the double ratio of $(W^+/W^-)_{13}/(W^+/W^-)_{8}$ measured as a function of boson $p_T$ may provide further constraints on the $u/d$ ratio of PDFs, while the double ratio $(W^\pm/Z^0)_{13}/(W^\pm/Z^0)_{8}$ is likely to have almost no theoretical SM uncertainty and hence may be sensitive at large boson $p_T$ values to potential beyond-the-SM contributions.

\section{Conclusions} \label{sec:conclusions}

We have presented a detailed study of the dependence of various ratios of $W^+$, $W^-$ and $Z^0$ cross sections as a function of the boson transverse momentum ($p_T$), and we have shown how these ratios depend on the multiplicity of associated jets and the LHC centre-of-mass energy.  We have evaluated the theoretical uncertainties from higher-order QCD corrections, including renormalisation/factorisation scale variation and dependence on matching to a parton shower, together with the choice of PDFs and associated value of $\alpha_S(M_Z^2)$, and we have discussed the potential impact of higher-order electroweak corrections.  We focused on the region of large boson $p_T\gtrsim M_{W,Z}$ where fixed-order calculations are sufficient without the need to resum large logarithms of $M_{W,Z}/p_T$ (most important for $p_T\ll M_{W,Z}$).  We find the uncertainties from higher-order QCD corrections for all cross-section ratios to be below a few percent.  This conclusion is supported by multiple evidence, such as the similarity of a fixed-order calculation at LO and NLO, the insensitivity to renormalisation/factorisation scale variation (assumed to be correlated between numerator and denominator), the stability of the ratios to different jet multiplicities, and the comparison between a fixed-order calculation (\textsc{mcfm}) with a multiparton tree-level calculation matched to a parton shower (\textsc{madgraph}+\textsc{pythia}).

The uncertainty from choice of PDFs almost completely cancels in the ratio of $W^\pm/Z^0$ which is most relevant for determining the $Z(\to\nu\bar{\nu})$+jets background from $W(\to\ell\nu)$+jets events.  We estimate that the $W^\pm/Z^0$ ratio as a function of the boson $p_T$ has a total theoretical QCD uncertainty of less than 5\%, where this estimate mainly comes from a comparison of \textsc{madgraph}+\textsc{pythia} with \textsc{mcfm} (see figure~\ref{fig:compare}).  More detailed studies would be useful to check this estimate, for example, by imposing realistic experimental cuts and using NLO calculations for large jet multiplicities, preferably with matching to a parton shower.  Alternative methods to estimate the $Z(\to\nu\bar{\nu})$+jets background carry a large statistical uncertainty in the case of $Z(\to\ell\ell)$+jets, and larger theoretical QCD uncertainty in the case of $\gamma$+jets (within 10\%~\cite{Bern:2011pa,Ask:2011xf,Bern:2012vx}), making the $W$+jets method competitive and complementary.  In particular, the use of $\gamma$+jets involves larger uncertainties due mainly to the massless photon compared to the massive $Z$ boson, but also due to the different composition of initial-state up- and down-quark contributions, together with complications arising from the need to impose photon isolation criteria and the potential inclusion of contributions from parton-to-photon fragmentation.  The largest theoretical uncertainty on the $W^\pm/Z^0$ ratio may be due to higher-order electroweak corrections, which can reach up to several percent for the $W^+/Z^0$ and $W^-/Z^0$ ratios for very large boson $p_{T} > 1$~TeV at the 14~TeV LHC~\cite{Kuhn:2007cv}, but again these corrections are much smaller than for the $\gamma/Z^0$ ratio~\cite{Kuhn:2005gv}.  A precise measurement of the $W^\pm/Z^0$ ratio would validate the theoretical predictions and would also be a good consistency check of the SM.  Assuming that the theoretical uncertainties are smaller than the statistical uncertainty, any deviations from the SM predictions may indicate the presence of new physics (see, for example, refs.~\cite{Abouzaid:2003is,Kom:2010mv,Kom:2010qs}).

We also showed that the $W^+/W^-$ ratio at large boson $p_T$ may be used to constrain PDFs by probing the up-quark to down-quark ($u/d$) ratio at larger $x$ values than the usual inclusive $W$ charge asymmetry.  Since the other theoretical uncertainties on this ratio are negligible, including those from higher-order electroweak corrections which almost completely cancel~\cite{Kuhn:2007cv}, a measurement of this ratio as a function of the boson $p_T$ could provide complementary information on PDFs to those from the old fixed-target experiments and thus be an important ingredient in a future PDF fit using only `collider' data, or only HERA and LHC data.  The boson $p_T$ distributions themselves, ${\rm d}\sigma/{\rm d}p_T$, have the potential to constrain the gluon distribution, provided that theoretical uncertainties are brought under control by inclusion of higher-order electroweak and anticipated NNLO QCD corrections.  With the currently available NLO QCD corrections where the scale uncertainties are $\mathcal{O}(10\%)$, one needs to go to very large boson $p_T\gtrsim 1$~TeV before the current global PDF uncertainties overwhelm the scale uncertainties, although the PDF sensitivity can be enhanced by moving from central to forward rapidity~\cite{Brandt:2013hoa}.  Moreover, we showed in figure~\ref{fig:PDFdsdpt} that predictions using the ABM11 PDF set can deviate from predictions using the current global PDF sets (MSTW08, CT10, NNPDF2.3) by more than 10\% even for $p_T\in[30,300]$~GeV.  With a reduction in the scale uncertainties expected from NNLO QCD corrections, $W$ and $Z$ production at large boson $p_T$ could potentially provide a cleaner constraint on the gluon distribution than either inclusive jet production or top-pair production, including in the crucial region of $x\sim M_H/\sqrt{s}$ most relevant for Higgs boson production via gluon--gluon fusion~\cite{Forte:2013wc}.

\acknowledgments

We would like to thank Zhenbin Wu for assistance and we acknowledge helpful discussions with Anwar Bhatti, John Campbell, Massimiliano Grazzini, Dan Green, Steve Mrenna, Pavel Nadolsky, Andreas Papaefstathiou and Stefano Pozzorini.  The work of S.A.M.~is in part supported by the CMS Fellowship program at Fermilab.

\bibliographystyle{JHEP}

\bibliography{wzratio.bib}

\end{document}